\documentclass[graybox]{svmult}

\usepackage{mathptmx}       
\usepackage{helvet}         
\usepackage{courier}        
\usepackage{type1cm}        
\usepackage{makeidx}         
\usepackage{graphicx}        
\usepackage{multicol}        
\usepackage[bottom]{footmisc}
\usepackage{amsmath}
\usepackage[left=4.6cm,right=4.6cm,top=5.5cm,bottom=6.5cm,bindingoffset=0cm]{geometry}

\makeindex

\begin{document}

\title*{Two Approaches to Modeling the Interaction of Small and Medium Price-Taking Traders with a Stock Exchange by Mathematical Programming Techniques}
\titlerunning{Two Approaches to Modeling the Interaction of Small and Medium Price-Taking Traders with a Stock Exchange by Mathematical Programming Techniques}

\author{Alexander S. Belenky, Lyudmila G. Egorova}

\institute{Alexander S. Belenky \at Department of Mathematics, Faculty of Economic Sciences, National Research University Higher School of Economics, Russian Federation, and Center for Engineering Systems Fundamentals, Massachusetts Institute of Technology, USA, \email{abelenky@hse.ru}
\and Lyudmila G. Egorova \at Department of Mathematics, Faculty of Economic Sciences, National Research University Higher School of Economics, Russian Federation, \email{legorova@hse.ru}}
 
\maketitle

\abstract{The paper presents two new approaches to modeling the interaction of small and medium price-taking traders with a stock exchange. In the framework of these approaches, the traders can form and manage their portfolios of financial instruments traded on a stock exchange with the use of linear, integer, and mixed programming techniques. Unlike previous authors’ publications on the subject, besides standard securities, the present publication considers derivative financial instruments such as futures and options contracts. When a trader can treat price changes for each financial instrument of her interest as those of a random variable with a known (for instance, a uniform) probability distribution, finding an optimal composition of her portfolio turns out to be reducible to solving an integer programming problem. When the trader possesses no particular information on the probability distribution of the above-mentioned random variable for financial instruments of her interest but can estimate the areas to which the prices of groups of financial instruments are likely to belong, a game-theoretic approach to modeling her interaction with the stock exchange is proposed. In antagonistic games modeling the interaction in this case, finding the exact value of the global maximin describing the trader’s guaranteed financial result in playing against the stock exchange, along with the vectors at which this value is attained, is reducible to solving a mixed programming problem. Finding the value of an upper bound for this maximin (and the vectors at which this upper bound is attained) is reducible to finding a saddle point in an auxiliary antagonistic game on disjoint polyhedra, which can be done by solving linear programming problems forming a dual pair.}

\section{Introduction}
\label{sec:1}

Stock exchanges as markets of a special structure can be viewed as economic institutions whose functioning  affects both the global economy and economic developments in every country. This fact contributes to a great deal of attention to studying the stock exchange behavior, which has been displayed for years by a wide spectrum of experts, especially by financiers, economists, sociologists, psychologists, politicians, and mathematicians. What becomes known as a result of their studies, what these experts can (and wish to) explain and interpret from findings of their studies to both interested individuals and society as a whole to help them understand how the stock exchanges work, and how good (or bad) these explanations are makes a difference. Indeed, economic issues and policies, the financial stability and the financial security of every country, and the financial status of millions of individuals in the world who invest their personal money in sets of financial instruments traded in stock exchanges are affected by the stock exchange behavior. The existing dependency of so many ``customers'' on the above-mentioned ability (or inability) of the experts to provide trustworthy explanations of this behavior makes advances in developing tools for quantitatively analyzing the work of stock exchanges important for both the financial practice and economic science. 

These tools seem indispensable, first of all, for specialists in economics and finance, since they let them a) receive, process, analyze, and interpret available information on the behavior of both stock exchanges and their participants, b) form, test, and analyze both scientific and experience-based hypotheses on the stock exchange  behavior, along with mathematical models for its description, and c) study, evaluate, and generalize the experience of successful traders. However, since the quality of the above analyses heavily affects financial decisions of so many individuals whose well-being substantially depends on the quality of decisions on forming and managing their portfolios, it is clear that developing the tools presents interest for a sizable number of these individuals as well, especially if the tools are easy to operate, are widely available, and the results of the tools' work are easy to understand. 

Developing such tools for quantitatively studying the financial behavior and strategies of price-taking traders and those of any groups of them presents particular interest, since these strategies and this behavior, in fact, a) determine the behavior of stock exchanges, b) reflect both the state of the global economy and that of every country in which particular stock exchanges function, and c) let one draw and back up conclusions on the current investors' mood. However, the development of this kind of the tools requires substantial efforts from researchers to make the tools helpful in studying particular characteristics  attributed to stock exchanges, for instance, regularities of the dynamics of financial instruments depending on the financial behavior of so-called ``bulls'' and ``bears'' [Lin et al. 1994]. The same is true for studying the reaction of particular stock exchanges in particular countries on forming financial ``bubbles,'' on crushes of particular stocks, and on financial and general economic crises, especially taking into account large volumes of the available data, interdependencies of particular ingredients of this data, and the probabilistic nature of the data.    

Three questions on tools for quantitatively studying the financial behavior and strategies of price-taking traders are in order: 1) can the above-mentioned tools be developed in principle, and if yes, what can they help analyze, 2) who and how can benefit from their development, and 3) is there any need for developing such tools while so many different tools for studying stock exchanges have already been developed (and have been recognized at least by the scientific community)? 

1. Every stock exchange is a complicated system whose behavior is difficult to predict, since this behavior depends on a) decisions made by its numerous participants, b) political and economic situations and tendencies both in the world and in particular countries, c) breakthroughs in science and technology, and d) environmental issues associated with the natural anomalies and disasters that may affect agriculture, industry, and people's everyday life. However, there are examples of global systems having a similar degree of complexity whose behavior has been studied and even successfully forecast. Weather, agricultural systems, electoral systems, certain kinds of service systems, including those supplying energy (electricity, water, and gas), and particular markets, where certain goods are traded, can serve as such examples. Indeed, for instance, the dynamics of the land productivity with respect to particular agricultural crops in a geographic region, which substantially depends on both local weather and human activities relating to cultivating the crops and is quite difficult to study, has been successfully researched. The dynamics of changing priorities of the electorate in a country, which  substantially depends on the political climate both there and in the world, as well as on actions undertaken by candidates on the ballot and their teams to convince the voters to favor these candidates, is successfully monitored and even predicted in the course of, for instance, U.S. presidential election campaigns, despite the obvious complexity of studying it. The dynamics of energy consumption by a resident in a region, which depends on the financial status of this resident and her family, on the climate and local weather in the region, on the time of the day, on her life style and habits, etc., is successfully forecast in calculating parameters of regional electrical grids though its formalized description presents substantial difficulties.

While there are obvious similarities in the dependency of all the above-mentioned global systems on the nature and human behavior, the conventional wisdom suggests that developing any universal decision-support systems applicable to studying and analyzing these systems from any common, uniform positions is hardly possible. However, the authors believe that certain regularities detected in studying these systems [Belenky 1981] can successfully be applied in studying and analyzing stock exchanges and financial strategies of their participants [Belenky \& Egorova 2015].

At the same time, one should bear in mind that according to [Kahneman 2011; Penikas \& Proskurin 2013], with all the tools available to world financial analysts, they correctly predict the behavior of financial instruments approximately in 50\% of the cases. This may suggest that either the tools adequately describing the stock exchange regularities have not so far been developed, or not everything in stock exchanges can be predicted with a desirable accuracy in principle though the existing tools seem helpful for understanding regularities underlying the tendencies of the stock exchange behavior. In any case, it seems that the tools allowing one to analyze the ``potential'' of a price-taking trader and the impact of her decisions on both the composition of her portfolio of financial instruments traded in a stock exchange and on the behavior of this stock exchange as a whole are needed the most.

2. Economists, financial analysts, and psychologists are direct beneficiaries of developing tools for quantitatively studying the financial behavior and decision-making strategies of price-taking traders, whereas these traders themselves are likely to benefit from developing these tools at least indirectly, by using results of the studies that can be undertaken by the above-mentioned direct beneficiaries. A set of mathematical models describing the process of making investment decisions by price-taking traders and software implementing existing or new techniques for solving problems formulated with the use of these models, along with available statistical data reflecting the stock exchange behavior, should constitute the core of the tools for analyzing the psychology of making decisions by price-taking traders possessing abilities to divine the market price dynamics with certain probabilities. One should expect that the use of the tools by price-taking traders for improving their decisions (no matter whether such an improvement can be attained by any of them) is likely to change the behavior of every stock exchange as a whole, and this is likely to affect both the economy of a particular country and the global economy. 

3. The financial theory in general and financial mathematics in particular offer models describing the financial behavior of price-taking traders assuming that these traders are rational and make their decisions in an attempt to maximize their utility functions under a reasonable estimate of both the current and the future market status. In all these models, including the Black-Scholes model for derivative investment instruments [Crack 2014] and those developed by Markowitz  for the stocks [Markowitz 1952], their authors assume that the trader possesses information on the probability distribution of the future prices of the financial instruments (at least for those being of interest to the trader). The use of these models leads to solving quite complicated mathematical problems that can in principle be solved only approximately, and even approximate solutions can be obtained only for a limited number of the variables that are to be taken into consideration. Since the parameters of the above-mentioned probability distributions are also known only approximately, the question on how correctly these models may reflect the real financial behavior of the traders in their everyday work with financial instruments, and to what extent these models are applicable as tools for quantitatively analyzing the work of any stock exchange as a whole seems to remain open.

The present paper discusses a new approach to developing a package of mathematical tools for quantitatively analyzing the financial behavior of small and medium price-taking traders (each possessing the ability to divine future price values of certain financial instruments traded in any particular stock exchange) by means of integer, mixed, and linear programming techniques (the latter being the most powerful techniques for solving optimization problems). It is assumed that each such trader forms her portfolio of only those financial instruments of her interest traded in a stock exchange for which the above ability has been confirmed by the preliminary testing that the trader undergoes using the publicly available data on the dynamics of all the financial instruments traded there. Once the initial trader's portfolio has been formed, at each moment, the trader gets involved in making decisions on which financial instruments from the portfolio (and in which volumes) to sell and to hold, as well as on which financial instruments traded in the stock exchange (and in which volumes) to buy to maximize the portfolio value. The paper concerns the decisions that a price-taking trader might make if she had tools for analyzing the dynamics of financial instruments being of her interest (at the time of making these decisions) in the following two situations: 

a) The trader possesses information that allows her to treat the price value changes of each financial instrument as those of a random variable with a known (to her) probability distribution, and 

b) no information on the probability distribution of the above-mentioned random variable is available to or can be obtained by the trader though she can estimate the areas in which the price values of groups of financial instruments from her portfolio, considered as components of  vectors in  finite-dimensional spaces, are likely to belong. 

It is shown that in the first situation, the deployment of one of the two mathematical models, proposed in the paper, allows the trader to reduce the problem of finding an optimal composition of her portfolio to solving an integer programming problem. In the second situation, the use of the other model allows the trader to find its optimal investment strategy as that of a player in a two-person game on sets of disjoint player strategies, analogous to the game with the nature, in which the payoff function is a sum of a linear and a bilinear function of two vector variables. It is proven that in the second situation, finding an optimal investment strategy of a trader is equivalent to solving a mixed programming problem, whereas finding an upper bound for the trader's guaranteed result in the game is reducible to calculating an equilibrium point in an auxiliary antagonistic game, which can be done by solving linear programming problems forming a dual pair. 

The structure of the rest of the paper is as follows: Section 2 briefly discusses ideas underlying the development of the tools for detecting a set of financial instruments for which a price-taking trader is able to divine their future price values. Section 3 addresses the problem of forming an optimal portfolio of securities by a price-taking trader assuming that the trader knows only the range within which the price value of each particular financial security of her interest (considered as that of a uniformly distributed random variable) changes. Section 4 presents a game model for finding strategies of a price-taking trader with respect to managing her portfolio when the trader cannot consider price values of financial securities of her interest as random variables with known probability distributions. In this case, finding a global maximum of the minimum function describing the guaranteed financial result for a trader associated with her decision to buy, hold, and sell financial securities is equivalent to solving a mixed programming problem, whereas finding an upper bound for this guaranteed result is reducible to finding Nash equilibrium points in an auxiliary antagonistic game on polyhedra of disjoint player strategies. Finding these equilibrium points is, in turn, reducible to solving linear programming problems forming a dual pair. Section 5 addresses the problem of forming and managing traders' investment portfolios that include derivative financial instruments. Section 6 provides a numerical example illustrative of using the game model, presented in Section 4, in calculating optimal investment strategies of a trader in forming a new portfolio of financial securities traded in a stock exchange. Section 7 presents concluding remarks on the problems under consideration in the paper.

A brief review of publications on modeling the financial behavior of small and medium price-taking traders in a stock exchange, along with examples illustrative of their behavior in managing existing portfolios of financial securities traded in a stock exchange, can be found in the author’ publication [Belenky \& Egorova 2016].

\section{ Detecting the ability of a price-taking trader to divine future price values of a financial instrument and to succeed in using this ability in a standard and in a margin trading }
\label{sec:2}

The ability of a trader to divine the price value dynamics of a set of particular financial instruments traded in a stock exchange matters a great deal in forming her optimal portfolio. However, even for a person gifted in divining either future values of any time series in general or only those of time series describing the dynamics of particular financial instruments with a probability exceeding 50\%, it is clear that this ability as such may turn out to be insufficient for successfully trading securities either in a long run or even in a short period of time. Thus, tools for both detecting the ability of a potential trader to divine the values of the share prices of, for instance, securities from a particular set of securities with a probability exceeding 50\% and testing this ability (from the viewpoint of a final financial result that the trader may expect to achieve by trading corresponding financial securities within any particular period of time) are needed. These tools should help the potential trader develop confidence in her ability to succeed by trading particular financial securities and evaluate whether this ability is safe to use in trading with risks associated with the margin trading at least under certain leverage rates, offered by brokers working at a stock exchange. It seems obvious that, apparently, no tools can guarantee in principle that the results that they produce are directly applicable in a real trading. However, they may a) give the interested person (i.e., a potential trader) the impression on what she should expect by embarking the gamble of trading in stock exchanges, and b) advise those who do not display the ability to succeed in trading financial securities (either in general or in a margin trading with a particular leverage) to abstain from participating in these activities. 

The above-mentioned tools for testing the ability of a trader to divine the upward and downward directions of changing the value of the share price of a financial security and those for evaluating possible financial results of trading this security in a particular stock exchange consist of two separate parts. The first part (for testing the trader's ability to divine) is a software complex in which a) the trader is offered a time series describing the dynamics of the price value of the security share for a particular financial security that interests her, and  b) her prediction made at the endpoint of a chosen segment of the time series is compared with the real value of the share price of this financial security at the point next to that endpoint. It is clear that to estimate the probability of the random event consisting of correctly predicting this value of the share price, one should first find the frequency of correct answers offered by the trader (provided the trials are held under the same conditions) and make sure that the outcome of each trial does not depend on the outcomes of the other trials (i.e., that the so-called Bernoulli scheme [Feller 1991] of conducting the trails takes place). If these conditions are met, one can calculate the frequency of this event as a ratio of the correct predictions to the total number of trials, and this ratio can be considered as an estimate of the probability under consideration [Feller 1991]. A possible approach to making the trials independent of each other and to securing the same conditions for the trials may look as follows: One may prepare a set of randomly chosen segments of the time series having the same length and let the trader make her prediction at the endpoint of each segment proceeding from the observation of the time series values within the segment. 

The second part of the testing tools (for estimating possible final financial results of trading a particular security with a detected probability to divine the directions of changing the value of the share price of this security) is also a software complex in which trading experiments can be conducted. For instance, the trader can be offered a chance to predict the direction of changing the value of the share price of a security at any consequent set of moments (at which real value of the share price  of the security constitute a time series) and to choose the number of shares that she wishes to trade (to buy or to sell) at each moment from the set. By comparing the results of the trader's experiments with the real values of the share price of securities from the sets (time series segments) of various lengths at the trader's choice, she concludes about her potential to succeed or to fail in working with the security under consideration. Finally, the second complex allows the trader to make sure that at each testing step (i.e., at the moment $t$) of evaluating financial perspectives of working with each particular financial security of her interest, the probability with which the trader divines the value of the share price of this financial security at the moment $t$+1 does coincide with the one detected earlier (or at least is sufficiently close to it). This part of the software is needed to avoid unjustified recommendations on including a particular financial security in the trader's portfolio if for whatever reasons, the above coincidence (or closeness) does not take place.

\section{Finding optimal trader's strategies of investing in standard financial securities. Model 1. The values of financial securities are random variables with uniform probability distributions }
\label{sec:3}

In considering the financial behavior of a price-taking trader who at the moment $t$ wants to trade financial instruments that are traded in a particular stock exchange, two situations should be analyzed.

Situation 1.

 The trader does not possess any financial instruments at the moment $t$ while possessing a certain amount of cash that can be used both for buying  financial instruments and for borrowing them from a broker (to sell the borrowed financial instruments short).

Situation 2.

The trader has a portfolio of financial instruments, along with a certain amount of cash, and she tries to increase the value of her portfolio by selling and buying  financial instruments of her interest, as well as by borrowing them from the brokers  (to sell the borrowed financial instruments short). 

To simplify the material presentation and to avoid the repetition of parts of the reasoning to follow, in Model 1, which is studied  in Section 3, Situation 2 is considered first. Moreover, it is assumed that the trader's portfolio consists of financial securities only;  cases in which derivative financial instruments are parts of the trader's portfolio are considered in Section 5. Remark 1 at the end of Section 3 explains how the model developed for finding the best investment strategy of the trader in Situation 2 (Model 1) can be used in finding such a strategy in Situation 1.

\subsection {Notation }

Let 

$N=\{1,2,...,n\}$ be a set of (the names of) financial securities comprising the portfolio of a trader that are traded in a particular stock exchange and interest the trader;

$t_0<...<t< t+1<t+2<...$ be a set of the time moments at which the trader adopts decisions on changing the structure of her portfolio; 

$m_t$ be the amount of cash that the trader possesses at the moment $t$; 

$W_t$ be the total value of the trader's portfolio at the moment $t$ (in the form of cash and financial securities), i.e., the trader's welfare at the moment $t$;
 
$s_{i,t}$ be the spot value of the share price of financial security $i$ at the moment $t$, i.e., the price at which a share of financial security $i$ is traded at the moment $t$ at the stock exchange under the conditions of an immediate financial operation; 

$v_{i,t}$ be the (non-negative, integer) number of shares of financial security $i$ that the trader possesses at the moment $t$. 

The following four assumptions on how the trader makes decisions on changing her portfolio at the moment $t$ seem natural: 

1) The trader possesses a confirmed (tested) ability of estimating the probability $p_i$ with  which the future value of the share price of financial security $i$ may change at the moment $t+1$ in a particular direction for each $i \in \overline{1,n}$, i.e., the ability to predict whether this value will increase or will not increase. (See some of the details further in Section 4.)

2) At each moment $t$ (from the above set of moments), the trader can divide  the set of financial securities $N$ into three subsets $I_t^+$, $I_t^-$, $I_t^0$  for which $N=I_t^+ \cup I_t^- \cup I_t^0$, and $I_t^+ \cap I_t^-=\emptyset$ , $I_t^- \cap I_t^0=\emptyset $, $I_t^+ \cap I_t^0=\emptyset $, where 

$I_t^+$ is the set of financial securities on which the trader is confident that the values of their share prices will increase at the moment $t+1$ (so she intends to buy securities from this set at the moment $t$), 

$I_t^-$ is the set of financial securities on which the trader is confident that the values of their share prices will decrease at the moment $t+1$ (so she intends to sell securities from this set at the moment $t$),

$I_t^0$ is the set of financial securities on which the trader is confident that the values of their share prices will not change at the moment $t+1$ or will change insignificantly (so she does not intent to either sell or buy securities from this set at the moment $t$).

3) For buying financial securities from the set $I_t^+$ , the trader can spend both the available cash and the money to be received as a result of selling financial securities from the set $I_t^-$ at the moment $t$, as well as finances that the trader can borrow from any lenders (if such finances are available to the trader). Analogously, for selling financial securities from the set $i \in I_t^-$, the trader may use her own reserves of this security (of the size $v_{i,t}$), as well as to borrow a certain number of shares of this security from a broker to open a short position  (if this service is available to the trader from the broker);

4) The trader does not undertake any actions with financial securities from the set $I_t^0$.

To simplify the mathematical formulation of problems to be considered in this section of the paper in the framework of Model 1, in the reasoning to follow, it is assumed that the trader a) works only with shares and bonds as financial securities (called standard securities further in this paper), and b) puts only market orders, i.e., those that can be implemented immediately, at the spot market prices.

\subsection {The statement and mathematical formulation of the problem of finding a trader's optimal investment strategy  }

Let at the moment $t$, the trader have a portfolio of standard securities $v_{i,t},i \in \overline{1,n}$ and a certain amount of cash $m_t$ so that her welfare at the moment $t$ equals $W_t=m_t + \sum_{i=1}^n {v_{i,t} s_{i,t}}$. The problem of finding optimal investment strategies of the trader consists of choosing the numbers of shares of securities  $x_{i,t}^+$ (integers) from the set $I_t^+$  to buy (about which the trader expects the increase of the values of their  share prices at the moment $t+1$), the numbers of shares of securities  $x_{i,t}^-$ (integers) from the set $I_t^-$ in her current portfolio to sell (about which the trader expects the decrease of the values of their share prices at the moment $t+1$), and the numbers of shares of securities  $z_{i,t}^-$ (integers) from the set $I_t^-$ to sell, which are to be borrowed from a broker at the value of the share price equaling $s_{i,t}$ to open a short position at the moment $t$ with the return of these securities to the broker at the moment $t+1$ at the share  price value $s_{i,t+1}$ (for which the trader expects the inequality $s_{i,t} > s_{i,t+1}$ to hold). 

The welfare that the trader expects to have at the moment $t+1$ thus equals 
$$
W_{t+1}=\sum_{i \in I_t^0}{v_{i,t} s_{i,t+1} }+\sum_{i \in I_t^+} {(v_{i,t}+x_{i,t}^+ )  s_{i,t+1}} + \sum_{i \in I_t^-} {(v_{i,t}-x_{i,t}^-)  s_{i,t+1}} +
$$
$$
+ \left(m_t-\sum_{i \in I_t^+} {x_{i,t}^+  s_{i,t}} + \sum_{i \in I_t^-} {x_{i,t}^-  s_{i,t}} + \sum_{i \in I_t^-} {z_{i,t}^- (s_{i,t}-s_{i,t+1} )} \right), 
$$
where the first three summands determine a part of the trader's welfare formed by the value of the securities from her portfolio at the moment $t+1$, and the last summand determines the amount of cash remaining after the finalization of all the deals on buying and selling securities by the moment $t+1$, including the return of the borrowed securities to the broker. 

The (positive or negative) increment of the trader's welfare that she expects to attain at the moment $t+1$ compared with that at the moment $t$ after the completion of all the transactions equals 
$$
\triangle W_{t+1} = \sum_{i \in I_t^0} {v_{i,t} (s_{i,t+1}-s_{i,t} ) } + \sum_{i \in I_t^+} {(v_{i,t}+x_{i,t}^+ ) (s_{i,t+1}-s_{i,t} )} + 
$$
$$
+ \sum_{i \in I_t^-} {(v_{i,t}-x_{i,t}^- ) (s_{i,t+1}-s_{i,t} )} + \sum_{i \in I_t^-} {z_{i,t}^- (s_{i,t}-s_{i,t+1} )}. 
$$

Here, $v_{i,t}$, $s_{i,t}$, $m_t$, $i \in I_t^+$, $I_t^-$ are known real numbers (the numbers $v_{i,t}$ are integers), and the numbers $s_{i,t+1}$, $i \in I_t^+$, $I_t^-$ are the values of random variables. Further, it is assumed that the values of the share prices of securities $i,j \in N$ at the moment $t+$1 are independent random variables.

The trader conducts her transactions taking into consideration the following constraints: 

1) the numbers of shares of financial securities bought, sold, and borrowed are integers, 

2) $x_{i,t}^-$, the number of shares of security $i$ sold from the trader's portfolio, cannot exceed the available number of shares $v_{i,t}$ of this security that the trader possesses at the moment $t$, i.e., the inequalities
$$
x_{i,t}^- \le v_{i,t}, i \in I_t^-,
$$
hold (one should notice that if the trader plans to sell any  number of shares of security $i$ that exceeds $v_{i,t}$, then she borrows the number of shares of this security $z_{i,t}^-$ from a broker to open a short position to sell security $i$ in the number of shares additional to the number $v_{i,t}$),

3) the limitations on the accessibility to the capital to borrow while using a credit with a particular (credit) leverage cannot be exceeded; these limitations may be described, for instance, by the inequality
$$
\sum_{i \in I_t^+} {x_{i,t}^+  s_{i,t}} + \sum_{i \in I_t^-} {z_{i,t}^- s_{i,t} } - \left( m_t + \sum_{i \in I_t^-} {x_{i,t}^-  s_{i,t} } \right) \le k_t \left(m_t+ \sum_{i=1}^n {v_{i,t} s_{i,t}}\right).
$$

Here, $k_t$  is the size of the credit leverage, the first two summands on the left hand side of this inequality represent the sum of the total expenses bore by the trader at the moment $t$ (that are associated with buying securities in the market and with the trader's debt to the brokers who lent her securities from the set $I_t^-$ to open a short position). The third summand (on the left hand side of the above inequality) reflects the amount of cash that the trader will possess as a result of selling securities from the set $I_t^-$ that the trader has as part of her own portfolio at the moment $t$. The right hand side of the inequality reflects the maximal amount of money (that is assumed to be) available to the trader for borrowing with the credit leverage of the size $k_t$, and this amount depends on the total amount of capital that the trader possesses at the moment $t$ before she makes any of the above-mentioned transactions. One should bear in mind that particular mathematical relations reflecting the limitations on the accessibility of a particular trader to the capital to borrow may vary, and such relations much depend on the situation in the stock exchange at the moment $t$ and on the ability of the trader to convince particular brokers to lend her securities and particular lenders to lend her cash (or both).

It is also assumed that in making investment decisions at the moment $t$, the trader proceeds from the value $\alpha$ of a threshold, determining whether to make transactions in the stock exchange in principle. That is, she makes the transactions if the inequality $W_{t+1} \ge \alpha W_t$ holds, meaning that the trader tends to keep the level of the ratio of her welfare at every moment compared with that at the previous moment not lower than a particular value $\alpha$ of the threshold, $\alpha > 0$. 

\subsection {Transforming the problem of finding an optimal investment strategy of a trader into an integer programming problem }

Let at the moment $t$, the trader be able to estimate $s_{i,t+1}^{max}$ and $s_{i,t+1}^{min}$, the boarders of a segment  to which the values of the share price  of security $i \in I_t^+ \cup I_t^-$ will belong  at the moment $t+1$  (based upon either the previous data or any fundamental assumptions on the dynamics of the value of the share price that this security may have). If the trader can make no assumptions on a particular probability distribution of the  values of the share price that security $i$ may have within these boarders, it is natural to consider that these values  change upwards and downwards (with respect to the value $s_{i,t}$) as continuous random variables $u$ and $v$ uniformly distributed on the segments $[s_{i,t},s_{i,t+1}^{max}]$ and $[s_{i,t+1}^{min},s_{i,t}]$, respectively, with the probability distribution densities 
\begin{eqnarray} 
f_1 (u)= \left\{
\begin{aligned}
\frac{1}{s_{i,t+1}^{max}-s_{i,t} }, if  \ u \in (s_{i,t}, s_{i,t+1}^{max} ], \\
0, if \  u \notin (s_{i,t}, s_{i,t+1}^{max}],  \nonumber
\end{aligned}
\right.  \\
f_2 (v)= \left\{
\begin{aligned}
\frac{1}{s_{i,t}-s_{i,t+1}^{min} }, if  \ v \in (s_{i,t+1}^{min}, s_{i,t}], \\
0, if \  v \notin (s_{i,t+1}^{min}, s_{i,t}].  \nonumber
\end{aligned}
\right.
\end{eqnarray} 

Thus, if the trader assumes that the value of the share price  of security $i$ will increase at the moment $t+1$ compared with its current value, i.e., that the inequality $s_{i,t+1}>s_{i,t}$ will hold, then the expectation of the value of the share price  that this security will have at the moment $t+1$ equals $Ms_{i,t+1}=\frac{s_{i,t} + s_{i,t+1}^{max}}{2}$. On the contrary, if the trader assumes that this value of the share price will decrease at the moment $t+1$, i.e., that the inequality $s_{i,t+1}<s_{i,t}$ will hold, then the expectation of the value of the share price that security $i$ will have at the moment $t+1$ equals $Ms_{i,t+1}=\frac{s_{i,t+1}^{min}+s_{i,t}}{2}$. Finally, if the trader cannot make either assumption about the value of the share price  that security $i$ will have at the moment $t+1$, it is natural to consider that the value of the share price of this security will not change, i.e., that the equality $s_{i,t+1}=s_{i,t}$ will hold. 

If at the moment $t$, the trader expects with the probability $p_i$ that the value of the share price of security $i$ will increase at the moment $t+1$, i.e., that the inclusion $i \in I_t^+$ will hold (event $A_1$), then the conditional expectation of the value of the share price that this security will have at the moment $t+1$ assumes the value $\frac{s_{i,t} + s_{i,t+1}^{max}}{2}$ with the probability $p_i$. Otherwise, two events are possible at the moment $t+1:$ a) the value of the share price of security $i$ at the moment $t+1$ will decrease (event $A_2$), and b) the value of the share price of security $i$ at the moment $t+1$ will remain equal to the one at the moment $t$ (event $A_3$), and it is natural to assume that these two events are equally possible, i.e., each of them may occur with the probability $\frac{1-p_i}{2}$. 

Thus, the conditional expectation of the value of the share price that security $i \in I_t^+$ will have at the moment $t+1$ can be 
calculated proceeding from the probabilities of the three mutually exclusive events $A_1, A_2,A_3$, reflected  in Table 1.

\begin{table}
\caption{The values of the conditional expectation  $M(s_{i,t+1}/A_k),i \in I_t^+,k\in\overline {1,3}$}
\label{tab:1}      
\begin{tabular}{p{3cm}|p{1.5cm}|p{1.5cm}|p{1.5cm}}
\hline\noalign{\smallskip}
$M(s_{i,t+1}/A_k),i \in I_t^+$ & $\frac{s_{i,t} + s_{i,t+1}^{max}}{2}$ & $\frac{s_{i,t+1}^{min}+s_{i,t}}{2}$ & $s_{i,t}$  \\
\noalign{\smallskip}\svhline\noalign{\smallskip}
\hline \\
$P(A_k)$ & $p_i$  & $\frac{1-p_i}{2}$ & $\frac{1-p_i}{2}$ \\
\noalign{\smallskip}\hline\noalign{\smallskip}
\end{tabular}
\end{table}

If at the moment $t$, the trader expects with the probability $p_i$ that the value of the share price of security $i$ will decrease at the moment $t+1$, i.e., that the inclusion $i \in I_t^-$ will hold, then the reasoning similar to the previous one allows one to conclude that the expectation of the value of the share price that security $i \in I_t^-$ will have at the moment $t+1$  can be calculated proceeding from the probabilities of the three incompatible events $B_1,B_2,B_3$, reflected  in Table 2.

\begin{table}
\caption{The values of the conditional expectation $M(s_{i,t+1}/B_k),i \in I_t^-,k\in\overline {1,3}$}
\label{tab:2}      
\begin{tabular}{p{3cm}|p{1.5cm}|p{1.5cm}|p{1.5cm}}
\hline\noalign{\smallskip}
$M(s_{i,t+1}/B_k),i \in I_t^-$ & $\frac{s_{i,t+1}^{min}+s_{i,t}}{2}$ & $\frac{s_{i,t} + s_{i,t+1}^{max}}{2}$ &  $s_{i,t}$  \\
\noalign{\smallskip}\svhline\noalign{\smallskip}
\hline \\
$P(B_k)$ & $p_i$  & $\frac{1-p_i}{2}$ & $\frac{1-p_i}{2}$ \\
\noalign{\smallskip}\hline\noalign{\smallskip}
\end{tabular}
\end{table}

Finally, if the trader expects with the probability $p_i$ that for security $i$ the inclusion $i \in I_t^0$ will hold at the moment $t+1$, the conditional expectation of the value of the share price that security $i \in I_t^0$ will have at the moment $t+1$ can be calculated proceeding from the probabilities of the three incompatible events $C_1,C_2,C_3$, reflected in Table 3.

\begin{table}
\caption{The values of the conditional expectation $M(s_{i,t+1}/C_k),i \in I_t^0,k\in\overline {1,3}$}
\label{tab:3}      
\begin{tabular}{p{3cm}|p{1.5cm}|p{1.5cm}|p{1.5cm}}
\hline\noalign{\smallskip}
$M(s_{i,t+1}/C_k),i \in I_t^0$ &  $s_{i,t}$ & $\frac{s_{i,t+1}^{min}+s_{i,t}}{2}$ & $\frac{s_{i,t} + s_{i,t+1}^{max}}{2}$   \\
\noalign{\smallskip}\svhline\noalign{\smallskip}
\hline \\
$P(C_k)$ & $p_i$  & $\frac{1-p_i}{2}$ & $\frac{1-p_i}{2}$ \\
\noalign{\smallskip}\hline\noalign{\smallskip}
\end{tabular}
\end{table}

Thus, in the above three cases for security $i$ to belong to one of the three subsets of the set $N$, the expectations $Ms_{i,t+1}$ are calculated as follows [Feller 1991]: 
$$
Ms_{i,t+1} = p_i  \frac{s_{i,t} + s_{i,t+1}^{max}}{2} + \frac{1-p_i}{2}  \frac{s_{i,t+1}^{min} + s_{i,t}}{2} + \frac{1-p_i}{2} s_{i,t} , i \in I_t^+,
$$
$$
Ms_{i,t+1} = p_i  \frac{s_{i,t+1}^{min} + s_{i,t}}{2}  + \frac{1-p_i}{2} \frac{s_{i,t} + s_{i,t+1}^{max}}{2}  + \frac{1-p_i}{2} s_{i,t} , i \in I_t^-,
$$
$$
Ms_{i,t+1} = p_i s_{i,t}  + \frac{1-p_i}{2}  \frac{s_{i,t+1}^{min} + s_{i,t}}{2}  + \frac{1-p_i}{2} \frac{s_{i,t} + s_{i,t+1}^{max}}{2}  , i \in I_t^0,
$$

Certainly, generally, the trader can make any particular assumptions on the regularities that  probability distributions of the future values of the share prices may have for securities from the sets $I_t^+$ and $I_t^-$  at the moment $t+1$ (for instance, that these distributions will be normal).  Such  assumptions may let her more accurately calculate the expectations of the values of these share prices using the same logic that was employed  under the assumption on the uniform distributions of these values.

To calculate an optimal trader's strategy of changing her portfolio at the moment $t$, one should choose the value of the  threshold $\alpha$ and formulate the problem of finding such a strategy as, for instance, that of maximizing the expectation of the portfolio value increment, provided all the constraints associated with this choice hold. In the simplest case of such a formulation, one can assume that a) the trader deals with and is interested in only those securities that are present in her portfolio at the moment $t$, 
b) she may buy securities only from the set $I_t^+$, and she may sell securities only from the set $I_t^-$, and c) the trader does not make any transactions with securities from the set $I_t^0$ (see assumption 4 on page 10). Then,  this maximization problem can be  formulated, for instance,  as follows [Belenky \& Egorova 2015]:

\begin{eqnarray}
M[\triangle W_{t+1}] = \sum_{i \in I_t^0} {v_{i,t} (Ms_{i,t+1}-s_{i,t} ) } + \sum_{i \in I_t^+} {(v_{i,t}+x_{i,t}^+ ) (Ms_{i,t+1}-s_{i,t} )} + \\  \nonumber
+ \sum_{i \in I_t^-} {(v_{i,t}-x_{i,t}^- ) (Ms_{i,t+1}-s_{i,t} )} + \sum_{i \in I_t^-} {z_{i,t}^- (s_{i,t}-Ms_{i,t+1} )} \to max.  \nonumber
\end{eqnarray}
$$
\sum_{i \in I_t^0}{v_{i,t} Ms_{i,t+1} } + \sum_{i \in I_t^+} {(v_{i,t}+x_{i,t}^+ ) Ms_{i,t+1}} + \sum_{i \in I_t^-} {(v_{i,t}-x_{i,t}^-) Ms_{i,t+1}} + 
$$
$$
+ \left( m_t-\sum_{i \in I_t^+} {x_{i,t}^+  s_{i,t}} + \sum_{i \in I_t^-} {x_{i,t}^-  s_{i,t}} + \sum_{i \in I_t^-} {z_{i,t}^- (s_{i,t}-Ms_{i,t+1} )} \right) \ge \alpha \left( m_t + \sum_{i=1}^n {v_{i,t} s_{i,t}} \right),  
$$
$$
\sum_{i \in I_t^+} {x_{i,t}^+  s_{i,t}} + \sum_{i \in I_t^-} {z_{i,t}^- s_{i,t} } - \left( m_t + \sum_{i \in I_t^-} {x_{i,t}^-  s_{i,t} } \right) \le k_t \left(m_t+ \sum_{i=1}^n {v_{i,t} s_{i,t}}\right),
$$
$$
x_{i,t}^- \le v_{i,t}, i \in I_t^-,
$$
where $x_{i,t}^+$, $i \in I_t^+$, $x_{i,t}^-$, $i \in I_t^-$, $z_{i,t}^-$, $i \in I_t^-$ are integers. This problem is an  integer programming one in which $x_{i,t}^+$, $i \in I_t^+$, $x_{i,t}^-$, $i \in I_t^-$, and $z_{i,t}^-$, $i \in I_t^-$ are the variables.

Generally, a) the set of standard securities $\tilde N$ (which contains $N$ as a subset) that are of the trader's interest  may include those that are not necessarily present in her portfolio at the moment $t$, and b) the trader proceeds from the estimates of $Ms_{i,t+1}$ for all the securities from the set $\tilde N$ and makes decisions of changing the composition of her portfolio based upon the values of the differences $Ms_{i,t+1}-s_{i,t}$ for all of these securities (so that assumption 4 on page 10 does not hold).

Let the trader divide the whole set $\tilde N$  of standard securities that interest her at the moment $t$ into the subsets $\tilde I_t^+$, $\tilde I_t^-$, and $\tilde I_t^0$, where $\tilde I_t^+$ is a set of standard securities for which the trader believes with the probability $p_i>0.5$ that the share price values that these securities will have at the moment $t+1$ will increase,  $\tilde I_t^-$ is a set of standard securities for which the trader believes with the probability $p_i>0.5$ that the share price values that these securities will have at the moment $t+1$ will decrease, and $\tilde I_t^0$ is a set of standard securities for which the trader believes with the probability $p_i>0.5$ that the share price values that these securities will have at the moment $t+1$ will not change.

Let the trader know the boarders of the segment $[s_{i,t+1}^{min},s_{i,t+1}^{max}]$ within which the value of $s_{i,t+1},$
$i \in \tilde I_t^+ \cup \tilde I_t^-$ will change at the moment $t+1$ while the trader can make no assumptions on a particular probability distribution that the value of $s_{i,t+1}$, considered as that of a random variable, may have (within these borders). Then, as before, it seems natural to assume that this value changes upwards as a continuous random variable $u$ uniformly distributed on the segment $[s_{i,t}, s_{i,t+1}^{max}]$ and changes downwards as a continuous random variable $v$ distributed uniformly on the segment $[s_{i,t+1}^{min},s_{it}]$. The latter assumption allows one to calculate the expectations $Ms_{i,t+1}$ in just the same manner this was done earlier for standard securities from the set $N$. 

First, consider standard securities that the trader may be interested in buying, including securities with particular names that some of the standard securities in her portfolio have. Let $\hat{I}_t^+ \subset \tilde I_t^+$,$\hat{I}_t^- \subset \tilde I_t^-$, and $\hat{I}_t^0 \subset \tilde I_t^0$ be the sets of
standard securities for which the differences $Ms_{i,t+1}-s_{i,t}$ are strictly positive. If at least one of the three sets $\hat{I}_t^+$,$\hat{I}_t^-$, and $\hat{I}_t^0$ is not empty, the trader may consider buying new standard securities from the set $\hat{I}_t^+\cup \hat{I}_t^- \cup \hat{I}_t^0$ at the moment $t$. 

Second, consider standard securities  that are already in the trader's portfolio at the moment $t$. Let $\tilde {I}_t^+ (av) \subset \tilde I_t^+$, $\tilde {I}_t^- (av) \subset \tilde I_t^-$, $\tilde {I}_t^0 (av) \subset \tilde I_t^0$ be the sets of names of the standard securities that the trader possesses at the moment $t$, and let $v_{i,t}$ be the number of shares of standard security $i$, $i \in {\tilde I}̂_t^+(av)\cup \tilde {I}̂_t^-(av) \cup {\tilde I}̂_t^0(av)$. Let $\hat{I}_t^+ (av)\subset {\tilde I}_t^+(av)$, $\hat{I}_t^-(av)\subset {\tilde I}_t^-(av)$,  and $\hat{I}_t^0(av)\subset {\tilde I}_t^0(av)$  be the sets  of $i$ for which the differences $Ms_{i,t+1}-s_{i,t}$ are strictly positive. 

It is clear that the trader may consider a) holding the standard securities  from the sets $\hat{I}_t^+ (av)$, $\hat{I}_t^-(av)$, and $\hat{I}_t^0(av)$, and b) selling all the standard securities from the sets $\tilde I_t^+(av) \setminus \hat{I}_t^+(av)$, $\tilde I_t^-(av) \setminus \hat{I}_t^-(av)$, and $\tilde I_t^0(av) \setminus \hat{I}_t^0(av)$ and  borrowing   standard securities from these sets from brokers. Since the trader believes that selling standard securities, in particular,  from the sets $\tilde I_t^+(av) \setminus \hat{I}_t^+ (av)$, $\tilde I_t^-(av) \setminus \hat{I}_t^- (av)$, and $\tilde I_t^0(av) \setminus \hat{I}_t^0 (av)$ short leads to receiving the money that can be spent, particularly, for buying new standard securities  from the sets $\hat{I}_t^+$, $\hat{I}_t^-$, and $\hat{I}_t^0$ (provided these sets are not empty), the trader needs to find an optimal investment strategy of changing her portfolio. This problem can be formulated as follows:

\begin{eqnarray}
M[\triangle W_{t+1}] =   \sum_{i \in \hat I_t^+} {x_{i,t}^+( Ms_{i, t+1}-s_{i,t})}+  \sum_{i \in \hat I_t^-} {x_{i,t}^+( Ms_{i, t+1}-s_{i,t})} +\\  \nonumber 
 \sum_{i \in \hat I_t^0} {x_{i,t}^+( Ms_{i, t+1}-s_{i,t})} 
+ \sum_{i \in \hat{I}_t^+ (av)} { v_{it}(Ms_{i,t+1}-s_{i,t} )} +  \sum_{i \in \hat{I}_t^- (av)} { v_{it}(Ms_{i,t+1}-s_{i,t} )}+ \\ \nonumber
 \sum_{i \in \hat{I}_t^0 (av)} { v_{it}(Ms_{i,t+1}-s_{i,t} )} 
+\sum_{i \in (\tilde I_t^+\setminus \hat{I}_t^+)\cup (\tilde I_t^-\setminus \hat{I}_t^-) \cup (\tilde I_t^0\setminus \hat{I}_t^0)} {z_{i,t}^- (s_{i,t}-Ms_{i,t+1} )} \to max.  \nonumber
\end{eqnarray}
$$
 \sum_{i \in \hat I_t^+} {x_{i,t}^+ Ms_{i, t+1} }+  \sum_{i \in \hat I_t^-} {x_{i,t}^+Ms_{i, t+1} } + \sum_{i \in \hat I_t^0} {x_{i,t}^+M s_{i, t+1} } +
$$
$$
+ \sum_{i \in \hat{I}_t^+ (av)} { v_{i,t}Ms_{i,t+1}} +  \sum_{i \in \hat{I}_t^- (av)} { v_{i,t}Ms_{i,t+1}}+  \sum_{i \in \hat{I}_t^0 (av)} { v_{i,t}Ms_{i,t+1})} +
$$
$$
+ \left( m_t- \sum_{i \in \hat I_t^+} {x_{i,t}^+ s_{i, t} }-  \sum_{i \in \hat I_t^-} {x_{i,t}^+s_{i, t} } - \sum_{i \in \hat I_t^0} {x_{i,t}^+ s_{i, t} } \right)+
$$
$$
+\sum_{i \in \tilde I_t^+(av) \setminus \hat{I}_t^+(av)} {v_{i,t} s_{i, t} }+ \sum_{i \in \tilde I_t^-(av) \setminus \hat{I}_t^-(av)} {v_{i,t} s_{i, t} }+\sum_{i \in \tilde I_t^0(av) \setminus \hat{I}_t^0(av)} {v_{i,t} s_{i, t} }+
$$
$$
+\sum_{i \in (\tilde I_t^+\setminus \hat{I}_t^+)\cup (\tilde I_t^-\setminus \hat{I}_t^-) \cup (\tilde I_t^0\setminus \hat{I}_t^0)} {z_{i,t}^- (s_{i,t}-Ms_{i,t+1} )}\ge \alpha \left( m_t + \sum_{i\in  \tilde I_t^+(av)\cup\tilde I_t^-(av)\cup\tilde I_t^0(av)} {v_{i,t} s_{i,t}} \right), 
$$
$$
 \sum_{i \in \hat I_t^+} {x_{i,t}^+ s_{i, t} }+  \sum_{i \in \hat I_t^-} {x_{i,t}^+s_{i, t} } + \sum_{i \in \hat I_t^0} {x_{i,t}^+ s_{i, t} } +\sum_{i \in (\tilde I_t^+\setminus \hat{I}_t^+)\cup (\tilde I_t^-\setminus \hat{I}_t^-) \cup (\tilde I_t^0\setminus \hat{I}_t^0)} {z_{i,t}^- s_{i,t}}-
$$
$$
-\left( m_t+\sum_{i \in \tilde I_t^+(av) \setminus \hat{I}_t^+(av)} {v_{i,t} s_{i, t} }+ \sum_{i \in \tilde I_t^-(av) \setminus \hat{I}_t^-(av)} {v_{i,t} s_{i, t} }+\sum_{i \in \tilde I_t^0(av) \setminus \hat{I}_t^0(av)} {v_{i,t} s_{i, t} }\right) \le 
$$
$$
\le k_t \left(m_t+ \sum_{i\in  \tilde I_t^+(av)\cup\tilde I_t^-(av)\cup\tilde I_t^0(av)} {v_{i,t} s_{i,t}}\right),
$$
where $x_{i,t}^+, i\in  \tilde I_t^+\cup\tilde I_t^-\cup\tilde I_t^0,$ are the numbers of shares of securities from the set $\tilde I_t^+\cup\tilde I_t^-\cup\tilde I_t^0$ that are bought at the moment $t$. 

As before, the (expected) increment of the trader's welfare is calculated as the difference between the expected trader's welfare at the moment $t+1$ as a result of buying  and selling securities in the stock exchange and her welfare at the moment $t$ (with respect to the activities related to the interaction with the stock exchange). That is, at the moment $t+1$, the expected trader's welfare is a sum of a) the expected value of new securities bought at the moment $t$, b) the expected value of securities from her portfolio that have been held since the moment $t$, c) the amount of cash remaining at the moment $t+1$ after spending a part of cash that is available at the moment $t$ for buying new securities and receiving cash as a result of selling securities from the set  $i \in (\tilde I_t^+\setminus \hat{I}_t^+)\cup (\tilde I_t^-\setminus \hat{I}_t^-) \cup (\tilde I_t^0\setminus \hat{I}_t^0)$, and d) the amount of cash expected to be received as a result of selling short securities borrowed from brokers.

This  problem is also an  integer programming one in which $x_{i,t}^+$, $i \in \hat I_t^+\cup \hat I_t^- \cup \hat I_t^0$, and $z_{i,t}^-$, $i \in (\tilde I_t^+\setminus \hat{I}_t^+)\cup (\tilde I_t^-\setminus \hat{I}_t^-) \cup (\tilde I_t^0\setminus \hat{I}_t^0)$ are the variables.

Both problem (1) and problem (2) can be solved exactly, with the use of software for solving integer programming problems, if the number of the variables allows one to solve this problem in an acceptable time. 

As is known, in solving applied integer programming problems, integer variables are often considered as continuous ones, i.e., a relaxation of the problem is solved instead of the initial integer programming problem, and all the non-integer components of the solution are rounded-off [Yudin \& Yudin 2009] in line with any methodology. Such a transformation is mostly used when the constraints in the integer programming problem have the form of inequalities (which is the case in the problem under consideration). One should notice that the problem of rounding-off non-integer solutions in relaxed linear programming problems (with respect to the initial integer programming ones) and an approach to estimating the accuracy of this rounding-off are discussed in scientific publications, in particular, in [Asratyan \& Kuzyurin 2004]. 

Thus, the conditions for the variables to be integer are to be replaced in problem (1) with those of non-negativity for the volumes of securities to be bought, sold, and borrowed 
$$
x_{i,t}^+ \ge 0, i \in I_t^+, \  x_{i,t}^- \ge 0, i \in I_t^-, \  z_{i,t}^- \ge 0, i \in I_t^-,
$$
which transforms the above-formulated integer programming problem into a linear programming one. Analogously, for problem (2), conditions for the variables to be integer are to be replaced with those of non-negativity
$$
x_{i,t}^+\ge 0, i \in \hat I_t^+\cup \hat I_t^- \cup \hat I_t^0, \  z_{i,t}^-\ge 0, i \in (\tilde I_t^+\setminus \hat{I}_t^+)\cup (\tilde I_t^-\setminus \hat{I}_t^-) \cup (\tilde I_t^0\setminus \hat{I}_t^0).
$$
Adding these conditions to the system of constraints of problem (2) transforms this problem into a linear programming one.

As one can easily notice, the system of constraints of problem (1) and that of problem (2) are substantially different. Particularly, there are no inequalities of the kind $x_{i,t}^- \le v_{i,t}, i \in I_t^-$ in the system of constraints of problem (2). Under the assumptions made in formulating problem (1), the trader may or may not sell all the standard securities from the set $I_t^-$. On the contrary,  in problem (2), the suggested division of the set $\tilde N$ into the subsets implies that the trader will sell all the standard securities from the set $(\tilde I_t^+(av) \setminus \hat{I}_t^+(av))\cup (\tilde I_t^-(av) \setminus \hat{I}_t^-(av)) \cup (\tilde I_t^0(av) \setminus \hat{I}_t^0(av))$. 

Though, at first glance, the  decision to buy any securities from the sets $\hat{I}_t^- \cup \hat{I}_t^0$  may seem counterintuitive, one should bear in mind that the trader's division of the set $N$ into the three subsets $I_t^+$, $I_t^-$, $I_t^0$ , for which $N=I_t^+ \cup I_t^- \cup I_t^0$ and $I_t^+ \cap I_t^-=\emptyset$ , $I_t^- \cap I_t^0=\emptyset $, $I_t^+ \cap I_t^0=\emptyset $, is purely intuitive. This division is not based on any mathematical analysis of either directions of potential changes in which  the share price values of particular securities may move or on any numerical relations among the probabilities with which these moves may take place and the limits within which the changes are possible.  In contrast,  the division of the set $\tilde N$  of standard securities that interest the trader  at the moment $t$ into the subsets $\tilde I_t^+$, $\tilde I_t^-$, and $\tilde I_t^0$ and dealing only with those securities $i$ from this set for which the differences $Ms_{i,t+1}-s_{i,t}$ are strictly positive are a result of such an analysis. In the framework of this analysis, solving problem (2)  may, in fact, be viewed as a means for testing the trader's intuition with respect to her ability to properly choose the set of securities to consider for potential transactions.

{\it Example 1.} Consider security $A$ from the set $X_t^+$, whose current share price value (at the moment $t$) equals 10.00 US dollars. Let the trader expect that at the moment $t+1$, the share price value of security $A$ a) will be between 10.00 US dollars and 12.00 US dollars with the probability 0.6, b) will be between 2.00 US dollars and 10.00 US dollars with the probability 0.2, and c) will remain equal to 10.00 US dollars (i.e., will remain unchanged) with the probability 0.2.

Then using the above formulae for calculating the expectation of the share price value for a security from the set $I_t^+$, one can easily be certain that the expectation of the share price value of security $A$ at the moment $t+1$ equals 9.80 US dollars, i.e., contrary to the trader’ initial analysis, the expectation of the share price value of this security will decrease. 

{\it Example 2.} Consider security $B$ from the set $X_t^-$, whose current share price value (at the moment $t$) equals 100.00 US dollars. Let the trader expect that at the moment $t+1$, the share price value of security $B$ a) will be between 90.00 US dollars and 100.00 US dollars with the probability 0.6, b) will be between 100.00 US dollars and 160.00 US dollars with the probability 0.2, and c) will remain equal to 100.00 US dollars (i.e., will remain unchanged) with the probability 0.2.

Then using the above formulae for calculating the expectation of the share price value for a security from the set $I_t^-$, one can easily be certain that the expectation of the share price value of security $B$ at the moment $t+1$ equals 103.00 US dollars, i.e., contrary to the trader’ initial analysis, the expectation of the share price value of this security will increase.

Remark 1. It is clear that finding an optimal investment strategy of the trader in Situation 1, one should add the equalities $v_{i,t}=0$, $i \in \overline{1,n}$ and $x_{i,t}^-=0, \ i\in I_t^-$  to the system of constraints of problem (1) and set $I_t^+(av)=\emptyset, \  I_t^-(av)=\emptyset,  \ I_t^0(av)=\emptyset$  in the system of constraints of problem (2). Also, one should bear in mind that in the formulation of problems (1) and  (2), it is assumed that the value of the money at which  securities are sold at the moment $t$ remains unchanged at the moment $t+1$. However, if this is not the case, it is easy to reformulate problem (2) taking into consideration the difference in this value.

\section{Finding optimal trader's strategies of investing in standard financial securities. Model 2: The trader can numerically estimate only the areas in which the values of the share prices of all the securities that interest her may change}
\label{sec:4}

Let $N$ be a set of (names of) standard securities that interest a trader at the moment $t$. Further, let us assume that at the moment $t$, a trader can choose 1) a set of securities $I_t^+ \subseteq N$ whose share price values (as she believes) will increase at the moment $t+1$ compared with their share price values at the moment $t$, and 2) a set of securities $I_t^- \subseteq N$ whose share price values  (as she believes) will decrease at the moment $t+1$ compared with those at the moment $t$. Finally, let the trader correctly forecast  that the share price values of securities from the set $I_t^+$ will increase with the probability $p^+>0.5$ (so that  the share price values of securities from the set $I_t^+$ will not increase with the probability $1-p^+$). Analogously, let the trader correctly forecast  that the share price values of securities from the set $I_t^-$ will decrease  with the probability $p^->0.5$  (so that the share price values  of securities from the set $I_t^-$  will not decrease with the probability $1-p^-$). 

If a) the set of securities $N$  also contains standard securities forming the set $I_t^0=N \setminus (I_t^+ \cup I_t^- )$, b) she believes that the share price values of securities forming this set may change at the moment $t+1$, and c)  she does not have any assumptions on  the direction in which the share price values of these securities will change at the moment $t+1$, it seems natural to assume that both the increasing and not increasing of the share price values of these securities are equally possible with the probability 0.5. It is natural to assume that $I_t^+\cap I_t^-=\emptyset$, $I_t^-\cap I_t^0=\emptyset$, and $I_t^+\cap I_t^0=\emptyset$.

Let

1) $x_t=(x_t^+,x_t^-,x_t^0 ) $ be the vector of volumes (numbers of shares) of securities from the set $N$ that the trader intends to buy  and to sell at the moment $t$ (based on her beliefs), where $x_t^+ \in X_t^+ \subset R_+^{|I_t^+|}$ is the vector of volumes (numbers of shares) of such securities  from the set $I_t^+$, $x_t^- \in X_t^- \subset R_+^{|I_t^-|}$  is the vector of volumes (numbers of shares) of such  securities  from the set $I_t^-$, and $x_t^0 \in X_t^0 \subset R_+^{|I_t^0|}$ is the vector of volumes (numbers of shares) of such securities from the set $I_t^0$;

2) $y_{t+1}=(y_{t+1}^+,y_{t+1}^-,y_{t+1}^0) \in Y_{t+1}^+ \times Y_{t+1}^- \times Y_{t+1}^0 \subset R_+^{|I_t^+ |+|I_t^- |+|I_t^0 |}$ be the vector whose components are the values of the share prices of securities from the set $N$ at the moment $t+1$ if the trader correctly determines directions in which  the values of the share prices of these securities may change, where $y_{t+1}^+ \in Y_{t+1}^+ \subset R_+^{|I_t^+|}$ is the vector whose components are the values of the share prices of securities from the set $I_t^+$ at the moment $t+1$ if the trader correctly determines directions in which these values of the share prices may change (with the probability $p^+>0.5$), $y_{t+1}^- \in Y_{t+1}^- \subset R_+^{|I_t^-|}$  is the vector whose components are the values of the share prices of securities from the set $I_t^-$ at the moment $t+1$ if the trader correctly determines directions in which these values of the share prices may change (with the probability $p^->0.5$), and $y_{t+1}^0 \in Y_{t+1}^0\subset R_+^{|I_t^0|}$  is the vector whose components are the values of the share prices of securities from the set $I_t^0$ at which they will be available in the stock exchange at the moment $t+1$, if the trader correctly determines the areas in which these values of the share prices may change (with the probability $p^0=0.5$); 

3) $z_{t+1}=(z_{t+1}^+,z_{t+1}^-,z_{t+1}^0) \in Z_{t+1}^+ \times Z_{t+1}^- \times Z_{t+1}^0 \subset R_+^{|I_t^+ |+|I_t^- |+|I_t^0 |}$ be the vector whose components are the values of the share prices of securities from the set $N$ at the moment $t+1$ if the trader incorrectly determines directions in which the values of the share prices of these securities may change, where $z_{t+1}^+ \in Z_{t+1}^+\subset R_+^{|I_t^+|}$ is the vector whose components are the values of the share prices of securities from the set $I_t^+$ at the moment $t+1$ if the trader incorrectly determines  directions in which these values of the share prices may change (with the probability $1-p^+$), $z_{t+1}^- \in Z_{t+1}^-\subset R_+^{|I_t^-|}$ is the vector whose components are the values of the share prices of securities from the set $I_t^-$ at the moment $t+1$ if the trader incorrectly determines directions in which these values of the share prices may change (with the probability $1-p^-$), and $z_{t+1}^0 \in Z_{t+1}^0\subset R_+^{|I_t^0|}$  is the vector whose components are the values of the share prices of securities from the set $I_t^0$ at which they will be available in the stock exchange at the moment $t+1$ if the trader incorrectly determines the areas  in which these values of the share prices may change (with the probability $1-p^0=0.5$).

Throughout Section 4, the optimality  of the strategy to be exercised at the moment $t$ is understood in the sense of maximizing the value of the trader' s portfolio at the moment $t+1$. 

As mentioned earlier (see Section 3), the trader may consider finding an optimal investment strategy in two situations: a) in forming a new portfolio (Situation 1), and b) in changing a composition of the existing portfolio (Situation 2). Unlike in Section 3, for Model 2, Situation 1 is considered first, and based upon the analysis of the results obtained for Situation 1, Situation 2 is considered.

Situation 1.

Let the trader have no securities in her portfolio at the moment $t$ while $N$ is a set of (names of) standard securities that interest the trader at the moment $t$. As before, let $N=I_t^+\cup I_t^-\cup I_t^0$, where all the three sets have the same meaning as described earlier, at the beginning of Section 4, and let $|N|=|I_t^+|+|I_t^-|+|I_t^0|=n$. 

It is obvious that if the trader does not have any securities in her portfolio at the moment $t$, she can only either buy securities (by investing cash that she possesses at the moment $t$) or borrow money or securities or both  (under certain conditions offered by potential lenders or/and brokers that the trader views  to be acceptable) and use the borrowed money (or/and the money to be received as a result of selling the borrowed securities short)  to invest it in  securities from the set $N$. With respect to Situation 1, the vectors $x_t^-$ and $x_t^0$ should be understood as volumes of those  securities (from the set $I_t^-\cup I_t^0$)  that are the only securities that the trader may eventually consider to borrow from  brokers  to sell these securities  short to receive the above-mentioned cash.  However, at the moment $t$, the trader also has a certain amount of cash (see the description of the underlying conditions of Situation 1 at the beginning of Section 3).  So her major problem is to find the best variant of investing all the cash that she can afford to invest in securities (i.e., in buying securities) at the moment  $t$ in such a manner that the value of  her portfolio of securities, which is formed as a result of this investment, will be maximal at the moment $t+1$. 

Thus, in Situation 1, all the three sets  $X_t^+$, $X_t^-$, $X_t^0$ are those from which the trader may buy securities at the moment $t$, and the trader forms all these three sets at any moment $t$ at her own discretion, proceeding from her financial abilities at the moment $t$.  One should also emphasize that if the trader decides to borrow securities from a broker to sell them short (provided the broker offers such a transaction to the trader), and the trader can choose which particular securities to borrow within financial limitations agreed upon by both parties, she will do this  in the course of forming the above three sets.   

It is clear that if at the moment $t$, the trader were absolutely sure that the share price values of all the securities from the set $I_t^+$ would only increase, the share price values of all securities from the set $I_t^-$ would only decrease, and if the set $I_t^0$ were empty, then she would invest all the  cash available at the moment $t$ in securities from the set $I_t^+$. The trader would certainly not buy any securities from the set $I_t^-$ though she would borrow securities from this set to sell them short at the moment $t$ (provided such a transaction is offered to her by any broker or brokers) and to invest the money received (from this selling)  in securities from the set $I_t^+$ by adding the money received to all the cash available to the trader at the moment $t$ (for the purpose of investing in standard securities). As one will have a chance to be certain, mathematically, the corresponding problem is a particular case of the problem under consideration in this section of the  paper. 

 However, in the rest of this section,  it is assumed that the trader believes that a) the share price values of each of securities from the set $I_t^+$ may increase only with a certain probability $p^+$, whereas these values may decrease with the probability $1-p^+$, and b) there is a non-empty set $I_t^0$ of securities for each of which its share price value may increase or decrease with the same probability $p^0=0.5$. Analogously,  the trader believes that the share price values of securities from the set $I_t^-$ may decrease also only with a certain probability $p^-$, whereas they may  increase with the probability $1-p^-$. 

Examples at the end of Section 3 are illustrative of such relations between the values of the probabilities $p^+ (p^+>0.5)$,  $p^- (p^->0.5)$,  and those of coordinates of the vectors from the set $X_t^+\cup X_t^-$ that the expectations of the share price values of some securities from the set $I_t^+$ at the moment $t+1$ will be lower than their current values (i.e., those at the moment $t$), whereas the expectations of the  share price values of some securities from the set $I_t^-$ at the moment $t+1$ will exceed their current values. The same reasons are applicable to the set $I_t^0$ as well, which explains the trader's interest to securities from this set. 

While it seems quite clear how the trader may form the sets $X_t^+$ and $X_t^0$, one may raise a natural question: what should be considered as the set $X_t^-$ in Situation 1? The trader does not possess any securities at the moment $t$ at all, and she assumes that if she possessed securities from the set $I_t^-$, she would certainly have sold at least some of them trying to protect her welfare. When the optimality  of the strategy to be exercised at the moment $t$ is understood in the sense of maximizing the value of the trader's portfolio at the moment $t+1$ (which is the case under consideration in this paper),  at least three approaches to what the trader may view as the set $X_t^-$ are possible.

$\it Approach \ 1.$ Taking into consideration the above-mentioned examples at the end of Section 3, the trader considers spending a part of the available (her own) cash for buying securities from the set $I_t^-$ at the moment $t$ at the share price values of these securities that exist at the moment $t$ (while she has no access to credits in any form), and possible (feasible) variants of allocating this money among these securities determine the set $X_t^-$. The determination of the set $X_t^-$ should be followed by making a decision on which securities from the set $I_t^-$ (or from its subset) and in which volumes to buy to maximize the trader portfolio's value at the moment $t+1$. (In contrast, choosing particular volumes of securities (to buy and to sell) proceeding from  a particular vector  of them in the set $X_t^-$ corresponds to considering Situation 2 in which the trader already has securities from the set $X_t^-$ in her portfolio.)

$\it Approach \ 2.$ The trader can get a cash credit at the moment $t$ on certain acceptable conditions, and she is to return this credit at the moment $t+1$ or later (possibly, with some interest in both cases). Once again, taking  into consideration the above-mentioned examples at the end of Section 3, the trader may consider spending a part of this credit for buying some securities from the set $I_t^-$ at the moment $t$ in an attempt to maximize the trader portfolio's value at the moment $t+1$ in just the same way this is done under Approach 1. 

$\it Approach \ 3.$ At the moment $t$, the trader borrows securities from a subset of the set $I_t^-$ from a broker to sell the borrowed securities short at the moment $t$; however,  she is to return the borrowed securities to the broker (possibly, along with a certain percentage of the cost of the borrowed securities to be paid to the broker for using them  as a form of a credit from this broker) later than at the moment $t+1$. This move is based  on the hope that at the time of returning the borrowed  securities, their share price values will be lower than those at which these securities were borrowed.  The trader uses the money received as a result of selling the borrowed securities  short for buying securities from the set $N$. Here, as under Approaches 1 and 2, the trader's aim is to maximize the value of her portfolio at the moment $t+1$, and  securities to borrow are chosen from among those from the set $I_t^-$ that are offered by the broker. The trader is interested in borrowing such securities from the broker whose share price values at the moment $t$ would allow her to sell these securities at the maximal possible amount of money to be added to the trader's cash (that she  can afford to spend for buying securities at the moment $t$). This borrowing is done with the aim of spending all the cash (that the trader can spend for buying securities from the whole set $N$ at the moment $t$) to maximize the trader portfolio's value at the moment $t+1$.

Thus, under any of these three approaches, one may consider that at the moment $t$, the trader has a certain amount of cash that she can spend in forming her portfolio in such a manner that this portfolio would have the maximal market value at the moment $t+1$.  (Here, some strategies of allocating a portion of the cash available at the moment $t$ for buying some securities to be returned to the broker (or to the brokers) later than at the moment $t+1$ can be exercised.) It is the allocation of this available cash either among securities from the set $I_t^+$ or among securities from the set $I_t^+\cup I_t^-\cup I_t^0$ that determines the set $X_t^-$.

Let us first consider Situation 1 assuming that one of the above three approaches to determining the set $X_t^-$ is used, which means that no matter what particular approach is employed, taking into consideration examples at the end of Section 3, the trader chooses which securities to buy from all the three sets $I_t^+$, $I_t^-$, and $I_t^0$ to form her portfolio. 

Further, let at each moment $t$ the trader proceed from the existence of linear constraints of the balance kind imposed on the components of the vector $x_t$, including bilateral constraints-inequalities imposed on each component of each of the three vectors forming the vector $x_t$. It is natural to assume that these constraints, which, in particular,  reflect the trader's financial status at the moment $t$, are compatible.  The presence of such constraints allows one to consider, for instance, that the sets $X_t^+$ (the set of feasible values of the vector $x_t^+$), $X_t^-$ (the set of feasible values of the vector $x_t^-$) , and $X_t^0$ (the set of feasible values of the vector $x_t^0$) are  formed by the vectors  from subsets of  convex polyhedra $M_t^+\subset  R_+^{|I_t^+|}$, $M_t^-\subset R_+^{|I_t^-|}$, and $M_t^0\subset R_+^{|I_t^0|}$, respectively. In this case, each of these three polyhedra  is described by a system of compatible linear constraints binding variables forming vectors from the corresponding space only, and the above-mentioned subset of the polyhedron  is determined by the requirement for all the coordinates of the vectors from this subset to be non-negative integers so that a)   the above mentioned subsets take the form $M_t^+=\{x^+_t \in R_+^{|I_t^+|}:B^+_t x^+_t \ge d^+_t, x^+_t \in Q_+^{|I_t^+|} \}$, $M_t^-=\{x^-_t \in R_+^{|I_t^-|}:B^-_t x^-_t \ge d^-_t, x^-_t \in Q_+^{|I_t^-|} \}$, and $M_t^0=\{x^0_t \in R_+^{|I_t^0|}:B^0_t x^0_t \ge d^0_t, x^0_t \in Q_+^{|I_t^0|} \}$, where $B^+_t , B^-_t , B^0_t $ are matrices and $ d^+,  d^-,  d^0$ are vectors of corresponding sizes, and $Q_+^{k}$ is a direct product of $k$ sets of the set of all non-negative integers $Q_+$, and b) $X_t$, a set of feasible values of the vectors $x_t=(x_t^+, x_t^-, x_t^0)$, has the form $X_t=M_t^+\times M_t^-\times M_t^0$. According to the assumptions on the bilateral constraints-inequalities, the sets $M_t^+, M_t^-,$ and $M_t^0$ are either subsets of the corresponding parallelepipeds or coincide with them. 

However, generally, the sets of feasible values $X_t^+$ , $X_t^-$ , and $X_t^0$  may be determined by a set of linear equations and inequalities binding together  the variables being coordinates of all the vectors $x_t^+$, $x_t^-$, and $x_t^0$ so that $X_t$ (the set of feasible values of the vectors $x_t=(x_t^+,x_t^-,x_t^0)$), may have  the form $X_t=M_t=\{x_t \in R_+^n:B_t x_t \ge d_t, x_t \in Q_+^n \}\subset M_t^+\times M_t^-\times M_t^0$, where each of these three sets is non-empty and contains the zero vector. Analogously, it is natural to assume that each of the sets $Y_{t+1}^+$, $Y_{t+1}^-$,$Y_{t+1}^0$ and $Z_{t+1}^+$, $Z_{t+1}^-$, $Z_{t+1}^0$ is a (non-empty) convex polyhedron, since the values of the share prices of securities from the set $N$ are non-negative, real numbers bounded from above. Finally, let the trader believe that at each moment $t$, the directions in which the values of the share prices of securities from the set $N$ may change are ``connected'' within each of the three sets $I_t^+$, $I_t^-$, and $I_t^0$. Here, this ``connection'' is understood in the sense  that  the values of the share prices of all the securities from the set  $I_t^+$ will change in one and the same direction at the moment $t+1$, and the same is true for the values of the share prices of all the securities from each of the two sets  $I_t^-$ and $I_t^0$. Also, let the trader believe that the share price values within each of the sets $Y_{t+1}^+$, $Y_{t+1}^-$,$Y_{t+1}^0$ and $Z_{t+1}^+$, $Z_{t+1}^-$, $Z_{t+1}^0$ change independently of those in the other five sets.

At each moment, one may view the interaction between the trader and the stock exchange in Situation 1 as an antagonistic game between them. In this game, a strategy of the trader is to choose a) how many shares of securities from the sets $I_t^+$, $I_t^-$, and $I_t^0$ should be bought at the moment $t$, and b) how many shares of securities from the set $N$ to borrow from a broker to sell them short at the moment $t$ (see Remark 3 at the end of Section 4) with  the intent of both types of the transactions to form her portfolio with the maximum possible value at the moment $t+1$. The stock exchange's strategy in this game is ``to choose'' the values of the share prices of securities from the set $N$ the most unfavorably to the trader. This game can be viewed to be analogous to the game with the nature in which ``the nature'' (the stock exchange in the game under consideration) may offer the trader the most unfavorable combinations of the values of the share prices that securities from the sets $I_t^+$, $I_t^-$, and $I_t^0$ may assume at the moment $t+1$ (while the trader chooses the volumes of security shares to buy from each of these three sets at the moment $t$). These combinations (of the share price values) are chosen  in the form of vectors from the (non-empty) convex polyhedra $Y_{t+1}^+$, $Y_{t+1}^-$,$Y_{t+1}^0$ and $Z_{t+1}^+$, $Z_{t+1}^-$, $Z_{t+1}^0$, and (as mentioned earlier) vectors from these convex polyhedra are chosen independently of each other. 

The structure of this game allows one to find an optimal trader's strategy by solving a mixed programming problem. Finding an upper bound of the trader's guaranteed result in this game can be done by solving linear programming problems forming a dual pair [Belenky 1981].

{\it {\bf Theorem.} There exists an antagonistic game describing the interaction between the trader and the stock exchange at each moment $t$, and this game  is the one  on (non-empty) sets of disjoint player strategies one of which is 
$X_t=M_t=\{x_t \in R_+^n:B_t x_t \ge d_t, x_t \in Q_+^n \}\subseteq M_t^+ \times M_t^- \times M_t^0$,  and the other is  $\theta_{t+1}=\{w_{t+1} \in R_+^{2n}:A_t w_{t+1} \ge b_t\}$ with the bilinear payoff function $\langle x_t,D_t w_{t+1} \rangle$, where 
$$
D_t=\begin{pmatrix} D^{|I_t^+|}(p^+) & D^{|I_t^+|}(1-p^+) & 0 & 0 & 0 & 0 \\ 0 & 0 & D^{|I_t^-|}(p^-) & D^{|I_t^-|}(1-p^-) &  0 & 0 \\ 0 & 0 & 0 & 0 & D^{|I_t^0|}(\frac{1}{2}) & D^{|I_t^0|}(\frac{1}{2})   \end{pmatrix},
$$
$x_t=(x_t^+,x_t^-,x_t^0 ) \in  X_t$, $w_{t+1}=(w_{t+1}^+,w_{t+1}^-,w_{t+1}^0) \in \theta_{t+1}= \theta_{t+1}^+ \times \theta_{t+1}^- \times \theta_{t+1}^0$, $D_t$ is a $(|I_t^+|+|I_t^-|+|I_t^0|) \times 2(|I_t^+|+|I_t^-|+|I_t^0|)$ matrix, $D^{|I|}(x)$ is a diagonal matrix of the size $|I|$ all whose elements on the main diagonal equal $x$, $X_t$ is  a  set of the trader's strategies, $\theta_{t+1}$ is a set of the stock exchange strategies, $\theta_{t+1}^+=Y_{t+1}^+ \times Z_{t+1}^+$, $\theta_{t+1}^-=Y_{t+1}^- \times Z_{t+1}^-$, $\theta_{t+1}^0=Y_{t+1}^0 \times Z_{t+1}^0$ are (non-empty) convex polyhedra,  $w_{t+1}^+=(y_{t+1}^+,z_{t+1}^+) \in \theta_{t+1}^+$, $w_{t+1}^-=(y_{t+1}^-,z_{t+1}^-) \in \theta_{t+1}^-$, $w_{t+1}^0=(y_{t+1}^0,z_{t+1}^0) \in \theta_{t+1}^0$ are vectors, $Q_+$ is the set of all non-negative, integer numbers, $Q_+^n$ is a direct product of $n$ sets $Q_+$, and the payoff function is maximized with respect to the vector $x$ and is minimized with respect to the vector $w_{t+1}$. In this game, an optimal trader's strategy is the one at which the maximin of the payoff function of the game is attained, and finding the exact value of this maximin is reducible to solving a mixed programming problem. Finding an upper bound of this maximin is reducible to solving linear programming problems forming a dual pair [Belenky \& Egorova 2015].}

\begin{proof}

 Let us first assume that the set of trader's strategies $ X_t$ is  a direct product of the three subsets of vectors with all integer components from disjoint polyhedra $ M_t^+$, $M_t^-$, and $M_t^0$, i.e., $X_t= X_t^+ \times X_t^- \times X_t^0= M_t^+ \times M_t^- \times M_t^0$  in the  spaces $R_{+}^{|I_t^+|}$, $R_{+}^{|I_t^-|}$, and $R_{+}^{|I_t^0|}$, respectively, where $M_t^+=\{x_t \in R_+^{|I_t^+|}:B^+_t x^+_t \ge d^+_t, x_t \in Q_+^{|I_t^+|} \}$,  $M_t^-=\{x^-_t \in R_+^{|I_t^-|}:B^-_t x^-_t \ge d^-_t, x^-_t \in Q_+^{|I_t^-|} \}$, and $M_t^+=\{x^0_t \in R_+^{|I_t^0|}:B^0_t x^0_t \ge d^0_t, x^0_t \in Q_+^{|I_t^0|} \}$.

1. Let us consider securities forming the set $I_t^+$ at the moment $t$. If the trader correctly forecast directions in which the values of the share prices of securities from this set may change, then a) by buying securities from the set $I_t^+$ in the volumes (numbers of shares) being components of the vector $x_t^+$, and b) by expecting the values of the share prices of these securities at the moment $t+1$ to be components of the vector $y_{t+1}^+$, the trader would hope to invest the money available to her at the moment $t$ in such a manner that would maximize the value of the part of her portfolio (associated with securities from the set $I_t^+$) at the moment $t+1$. Here, the trader's best investment strategy in the game with the stock exchange (with ``the nature'') with respect to securities from the set $X_t^+$ consists of choosing such volumes of securities from the set $I_t^+$ to buy that can be found by solving the problem 
$$
\min_{y_{t+1}^+ \in Y_{t+1}^+ } {\langle x_t^+,y_{t+1}^+ \rangle} \to \max_{x_t^+ \in X_t^+}⁡.
$$

If the trader did not  correctly forecast the directions in which the values of the share prices of securities from the set $I_t^+$ may change, i.e., if the values of the share prices of securities from the set $I_t^+$ did not increase at the moment $t+1$, the best investment strategy of the trader in her game with the stock exchange with respect to securities from the set $X_t^+$ would be determined by solutions to the problem
$$
\min_{z_{t+1}^+ \in Z_{t+1}^+ } {\langle x_t^+,z_{t+1}^+ \rangle} \to \max_{x_t^+ \in X_t^+}⁡.
$$

Since the trader correctly forecasts the directions in which the values of the share prices of securities from the set $I_t^+$ may change only with the probability $p^+$, the worst financial result of the trader's choice of the volumes of securities from the set $I_t^+$ to be bought at the moment $t$, i.e., the worst financial result of choosing the vector $x_t^+ \in X_t^+$ at the moment $t$ by the trader, can be viewed as a discrete random variable taking the values $\min_{y_{t+1}^+ \in Y_{t+1}^+ } {\langle x_t^+,y_{t+1}^+ \rangle}$ and $\min_{z_{t+1}^+ \in Z_{t+1}^+ } {\langle x_t^+,z_{t+1}^+ \rangle}$ with the probabilities $p^+$ and $1-p^+$, respectively. It is clear that an optimal trader's strategy in the case under consideration may consist of choosing a vector $x_t^+ \in X_t^+$ that maximizes the expectation of this discrete random variable. If this is the case, the optimal trader's strategy  is found by solving the problem
$$
p^+ \min_{y_{t+1}^+ \in Y_{t+1}^+ } {\langle x_t^+,y_{t+1}^+ \rangle} + (1-p^+) \min_{z_{t+1}^+ \in Z_{t+1}^+ } {\langle x_t^+,z_{t+1}^+ \rangle} \to \max_{x_t^+ \in X_t^+}⁡.
$$

One can easily be certain that the equality 
$$
\max_{x_t^+ \in X_t^+} {\left[ p^+ \min_{y_{t+1}^+ \in Y_{t+1}^+ } {\langle x_t^+,y_{t+1}^+ \rangle} + (1-p^+) \min_{z_{t+1}^+ \in Z_{t+1}^+ } {\langle x_t^+,z_{t+1}^+ \rangle} \right]} =
$$
$$
\max_{x_t^+ \in X_t^+} {\left[ \min_{y_{t+1}^+ \in Y_{t+1}^+ } {\langle x_t^+,D^{|I_t^+|}(p^+) y_{t+1}^+ \rangle} + \min_{z_{t+1}^+ \in Z_{t+1}^+ } {\langle x_t^+,D^{|I_t^+|}(1-p^+) z_{t+1}^+ \rangle} \right]}
$$
holds, and since the vectors $y_{t+1}^+$ and $z_{t+1}^+$ from the sets $Y_{t+1}^+$ and $Z_{t+1}^+$ are chosen independently of each other, the following equalities also hold: 
$$
\max_{x_t^+ \in X_t^+} {\left[ \min_{y_{t+1}^+ \in Y_{t+1}^+ } { \langle x_t^+,D^{|I_t^+|}(p^+) y_{t+1}^+ \rangle} + \min_{z_{t+1}^+ \in Z_{t+1}^+ } {\langle x_t^+,D^{|I_t^+|}(1-p^+) z_{t+1}^+ \rangle} \right]} =
$$
$$
\max_{x_t^+ \in X_t^+} {\left[ \min_{(y_{t+1}^+,z_{t+1}^+) \in Y_{t+1}^+ \times Z_{t+1}^+} {\langle x_t^+,D^{|I_t^+|}(p^+) D^{|I_t^+|}(1-p^+) (y_{t+1}^+,z_{t+1}^+) \rangle} \right]} =
$$
$$
\max_{x_t^+ \in X_t^+} {\left[ \min_{w_{t+1}^+ \in \theta_{t+1}^+ } {\langle x_t^+,D^{2|I_t^+|}(p^+, 1-p^+) w_{t+1}^+ \rangle} \right]},
$$
where $w_{t+1}^+= (y_{t+1}^+,z_{t+1}^+)$, $\theta_{t+1}^+=Y_{t+1}^+ \times Z_{t+1}^+$, $D^{2|I_t^+|}(p^+, 1-p^+)=D^{|I_t^+|}(p^+) D^{|I_t^+|} \\ (1-p^+)$, $D^{|I_t^+|}(p^+)$ is a $|I_t^+ | \times |I_t^+ |$ diagonal matrix all whose elements on the main diagonal equal $p^+$, $D^{|I_t^+|}(1-p^+)$ is a $|I_t^+ | \times |I_t^+ |$ diagonal matrix all whose elements on the main diagonal equal $1-p^+$, and $D^{2|I_t^+|}(p^+, 1-p^+)$ is a $|I_t^+ | \times 2|I_t^+ |$ matrix formed by accessing the matrix $D^{|I_t^+|}(1-p^+)$ to the matrix $D^{|I_t^+|}(p^+)$ from the right.

2. Let us consider securities forming the set $I_t^-$ at the moment $t$. If the trader correctly forecast  directions in which the values of the share prices of securities from this set may change, then a) by buying securities from the set $I_t^-$ in the volumes (numbers of shares) being components of the vector $x_t^-$, and b) by expecting the values of the share prices of these securities at the moment $t+1$ to be components of the vector $y_{t+1}^-$, the trader would hope to invest the money available to her at the moment $t$ in such a manner that would maximize the value of the part of her portfolio (associated with securities from the set $I_t^-$) at the moment $t+1$. Here, the trader's best investment strategy in the game with the stock exchange  with respect to securities from the set $X_t^-$ consists of choosing such volumes (numbers of shares) of securities to buy that can be found by solving the problem
$$
\min_{y_{t+1}^- \in Y_{t+1}^- } {\langle x_t^-,y_{t+1}^- \rangle} \to \max_{x_t^- \in X_t^-}⁡.
$$

If the trader did not correctly forecast directions in which the values of the share prices of securities from the set $I_t^-$ may change, i.e., if the values of the share prices of securities from the set $I_t^-$ did not decrease at the moment $t+1$, the best investment strategy of the trader in her game with the stock exchange with respect to securities from the set $X_t^-$ would be determined by solutions to the problem
$$
\min_{z_{t+1}^- \in Z_{t+1}^- } {\langle x_t^-,z_{t+1}^- \rangle} \to \max_{x_t^- \in X_t^-}⁡.
$$
The reasoning analogous to the one presented in part 1 of this Proof lets one write the expression for the expectation of the worst financial result of the trader's decision to buy securities from the set $I_t^-$ in the volumes (numbers of shares) determined by the vector $x_t^-$ in the form
$$
\min_{w_{t+1}^- \in \theta_{t+1}^- } { \langle x_t^-,D^{2|I_t^-|}(p^-, 1-p^-) w_{t+1}^- \rangle}.
$$

Under the assumption on the optimality of the trader's strategy that was made with respect to securities from the set $I_t^+$, one can be certain that the trader tries to maximize her expected financial result associated with choosing the vector $x_t^- \in X_t^-$ by solving the problem
$$
\max_{x_t^- \in X_t^-} {\left[ \min_{w_{t+1}^- \in \theta_{t+1}^- } {\langle x_t^-,D^{2|I_t^-|}(p^-, 1-p^-) w_{t+1}^- \rangle} \right]},
$$
where $w_{t+1}^-= (y_{t+1}^-,z_{t+1}^-)$, $\theta_{t+1}^-=Y_{t+1}^- \times Z_{t+1}^-$, $D^{2|I_t^-|}(p^-, 1-p^-)=D^{|I_t^-|}(p^-) D^{|I_t^-|} \\ (1-p^-)$, $D^{|I_t^-|}(p^-)$ is a $|I_t^- | \times |I_t^- |$ diagonal matrix all whose elements on the main diagonal equal $p^-$, $D^{|I_t^-|}(1-p^-)$ is a $|I_t^- | \times |I_t^- |$ diagonal matrix all whose elements on the main diagonal equal $1-p^-$, and $D^{2|I_t^-|}(p^-, 1-p^-)$ is a $|I_t^- | \times 2|I_t^- |$ matrix formed by accessing the matrix $D^{|I_t^-|}(1-p^-)$ to the matrix $D^{|I_t^-|}(p^-)$ from the right.

3. Let us consider securities forming the set $I_t^0$ at the moment $t$ for which the trader determines the  direction in which the values of their share prices at the moment $t+1$ may change with the probability $p^0=1/2$. The best investment strategy of the trader in her game with the stock exchange with respect to securities from the set $I_t^0$ would be to choose the volumes (numbers of shares) of securities from this set that are determined by solutions to the problems
$$
\min_{y_{t+1}^0 \in Y_{t+1}^0 } {\langle x_t^0,y_{t+1}^0 \rangle} \to \max_{x_t^0 \in X_t^0}⁡.
$$
and
$$
\min_{z_{t+1}^0 \in Z_{t+1}^0 } {\langle x_t^0,z_{t+1}^0 \rangle} \to \max_{x_t^0 \in X_t^0}⁡.
$$

A reasoning similar to that presented in parts 1 and 2 of this Proof allows one to write the expression for the expectation of the financial result associated with choosing (buying) by the trader the volumes (numbers of shares) of securities from the set $I_t^0$ (being components of the vector $x_t^0$) in the form 
$$
\min_{w_{t+1}^0 \in \theta_{t+1}^0 } {\langle x_t^0,D^{2|I_t^0|}(p^0, 1-p^0) w_{t+1}^0 \rangle},
$$
Under the same assumption on the optimality of the trader's strategy that was made with respect to securities from the set $I_t^+$ and $I_t^-$, the trader tries to maximize this minimum by choosing the vector $x_t^0 \in X_t^0$ as a vector component of a solution to the problem
$$
\max_{x_t^0 \in X_t^0} {\left[\min_{w_{t+1}^0 \in \theta_{t+1}^0 } {\langle x_t^0,D^{2|I_t^0|}(p^0, 1-p^0) w_{t+1}^0 \rangle} \right]},
$$
where $w_{t+1}^0= (y_{t+1}^0,z_{t+1}^0)$, $\theta_{t+1}^0=Y_{t+1}^0 \times Z_{t+1}^0$, $D^{2|I_t^0|}(p^0, 1-p^0)=D^{|I_t^0|}(p^0) D^{|I_t^0|}(1-p^0)$, $D^{|I_t^0|}(p^0)$ is a $|I_t^0| \times |I_t^0|$ diagonal matrix all whose elements on the main diagonal equal $p^0$, $D^{|I_t^0|}(1-p^0)$ is a $|I_t^0| \times |I_t^0|$ diagonal matrix all whose elements on the main diagonal equal $1-p^0$, and $D^{2|I_t^0|}(p^0, 1-p^0)$ is a $|I_t^0| \times 2|I_t^0|$ matrix formed by accessing the matrix $D^{|I_t^0|}(1-p^0)$ to the matrix $D^{|I_t^0|}(p^0)$ from the right.

4. Since the financial results of choosing the volumes (numbers of shares) of securities from the sets $I_t^+$, $I_t^-$, and $I_t^0$ are random variables (since the trader forecasts the directions in which the values of their share prices at the moment $t+1$ will change within the polyhedra $Y_{t+1}^+$, $Y_{t+1}^-$,$Y_{t+1}^0$ and $Z_{t+1}^+$, $Z_{t+1}^-$, $Z_{t+1}^0$ only with certain probabilities), the expectations of the worst compound financial result is a sum of the above three expectations [Feller 1991]. 

Let  the matrix $D_t$ have the form
$$
D_t=\begin{pmatrix} D^{2|I_t^+|}(p^+, 1-p^+) & 0 & 0 \\ 0 & D^{2|I_t^-|}(p^-, 1-p^-) & 0 \\ 0 & 0 & D^{2|I_t^0|}(p^0, 1-p^0)   \end{pmatrix} =
$$
$$
=\begin{pmatrix} D^{|I_t^+|}(p^+) & D^{|I_t^+|}(1-p^+) & 0 & 0 & 0 & 0 \\ 0 & 0 & D^{|I_t^-|}(p^-) & D^{|I_t^-|}(1-p^-) &  0 & 0 \\ 0 & 0 & 0 & 0 & D^{|I_t^0|}(\frac{1}{2}) & D^{|I_t^0|}(\frac{1}{2})   \end{pmatrix},
$$
while  $x_t=(x_t^+,x_t^-,x_t^0)$ belongs to the set $X_t$, and  $w_{t+1}=(w_{t+1}^+,w_{t+1}^-,w_{t+1}^0)$ belongs to the convex polyhedron $\theta_{t+1}=\theta_{t+1}^+ \times \theta_{t+1}^- \times \theta_{t+1}^0$.  Further, let linear inequalities describing the convex polyhedron  $\theta_{t+1}=\{w_{t+1} \in R_+^{2n}:A_t w_{t+1} \ge b_t \}$ be compatible so that  $A_t,B_t$ are matrices, and $b_t,d_t$ are vectors of corresponding dimensions, whose elements are formed by the coefficients of the above two  compatible systems of  linear equations and inequalities. Then, when the trader chooses a particular vector $x_t$ from the set $X_t$, the expectation of the compound worst financial result determined by this choice can be calculated as
$$
\min_{w_{t+1} \in \theta_{t+1}} \langle x_t,D_t w_{t+1} \rangle.
$$

5. Let now $X_t=M_t=\{x_t \in R_+^n:B_t x_t \ge d_t, x_t \in Q_+^n \}\subset M_t^+ \times M_t^- \times M_t^0$,  where   $B_t$ is a matrix of a general structure, not necessarily corresponding to the structure of the set $M_t=M_t^+ \times M_t^- \times M_t^0$ as a direct product of subsets of the three polyhedra from the spaces $R_{+}^{|I_t^+|}$, $R_{+}^{|I_t^-|}$, and $R_{+}^{|I_t^0|}$, respectively. This means that the system of linear equations and inequalities in the description of the set $M_t$ contains at least one that binds together components of all the three vectors $x_t^+$, $x_t^-$, and $x_t^0$. 

 Let the trader choose the vector $x_t=(x_t^+, x_t^-, x_t^0 )\in X_t$. Depending on in which direction the share price values of securities from the sets $I_t^+$, $I_t^-$, and $I_t^0$ may change, the trader may obtain the following worst financial results:

1)  $\min_{y_{t+1}^+ \in Y_{t+1}^+ } {\langle x_t^+,y_{t+1}^+ \rangle} $ or  $\min_{z_{t+1}^+ \in Z_{t+1}^+ } {\langle x_t^+,z_{t+1}^+ \rangle} $ for securities from the set $I_t^+$,

2)  $\min_{y_{t+1}^- \in Y_{t+1}^- } {\langle x_t^-,y_{t+1}^- \rangle} $ or  $\min_{z_{t+1}^- \in Z_{t+1}^- } {\langle x_t^-,z_{t+1}^- \rangle} $   for securities from the set  $I_t^-$, 

3) $\min_{y_{t+1}^0 \in Y_{t+1}^0} {\langle x_t^0,y_{t+1}^0 \rangle} $ or  $\min_{z_{t+1}^0 \in Z_{t+1}^0 } {\langle  x_t^0,z_{t+1}^0 \rangle}$ for securities from the set $I_t^0$.

According to the  (earlier made)  assumptions on the sets $Y_{t+1}^+$, $Y_{t+1}^-$,$Y_{t+1}^0$ and $Z_{t+1}^+$, $Z_{t+1}^-$,$Z_{t+1}^0$,  

a) non-empty convex polyhedra in each of which all the components of the vectors belonging to the sets $I_t^+$, $I_t^-$, and $I_t^0$, respectively,  change in one  and the same direction, and 

b) the direction of changing the values for all the components of the vectors  $y_{t+1}^+$, $y_{t+1}^-$,$y_{t+1}^0$ and $z_{t+1}^+$, $z_{t+1}^-$, $z_{t+1}^0$ are chosen (by the stock exchange) randomly, with the probabilities $p^+, p^-, p^0$ and $(1-p^+), (1-p^-), (1-p^0)$, respectively, independently of each other for all the components of these six vectors, 

the above six worst financial results can be viewed as the values of three random variables $\xi^+, \xi^-, \xi^0$.  

Each of these three random variables is, in turn, a discrete random variable with two possible values for each variable. That is,  the discrete random variable $\xi^+$ assumes the values $\min_{y_{t+1}^+ \in Y_{t+1}^+ } {\langle x_t^+,y_{t+1}^+ \rangle} $ and  $\min_{z_{t+1}^+ \in Z_{t+1}^+ } {\langle x_t^+,z_{t+1}^+ \rangle} $ with the probabilities $p^+$ and $1-p^+$, respectively (since, in line with assumption a),  the probability with which all the components of those vectors whose components belong to the set $I_t^+$ hit the sets $Y_{t+1}^+$ and $Z_{t+1}^+$ with the probabilities  $p^+$ and $1-p^+$, respectively). Analogously, the discrete random variable $\xi^-$ assumes two values $\min_{y_{t+1}^- \in Y_{t+1}^- } {\langle x_t^-,y_{t+1}^- \rangle} $ and  $\min_{z_{t+1}^- \in Z_{t+1}^- } {\langle x_t^-,z_{t+1}^- \rangle} $ with the probabilities  $p^-$ and $1-p^-$, respectively, whereas the discrete random variable  $\xi^0$  assumes two values  $\min_{y_{t+1}^0 \in Y_{t+1}^0} {\langle x_t^0,y_{t+1}^0 \rangle} $ and  $\min_{z_{t+1}^0 \in Z_{t+1}^0 } {\langle  x_t^0,z_{t+1}^0 \rangle}$  with the probabilities $p^o$ and $1-p^0$, respectively.

Since the expectation of the sum of the random variables $\xi^+, \xi^-, \xi^0$ equals the sum of their expectations [Feller 1991], the equality 
$$
M[\xi^+ + \xi^- +\xi^0]= p^+(\min_{y_{t+1}^+ \in Y_{t+1}^+ } {\langle x_t^+,y_{t+1}^+ \rangle}) + (1-p^+)  (\min_{z_{t+1}^+ \in Z_{t+1}^+ } {\langle x_t^+,z_{t+1}^+ \rangle})+
$$
$$
p^-(\min_{y_{t+1}^- \in Y_{t+1}^- } {\langle x_t^-,y_{t+1}^-\rangle}) + (1-p^-)  (\min_{z_{t+1}^- \in Z_{t+1}^- } {\langle x_t^-,z_{t+1}^- \rangle})+
$$
$$
p^0(\min_{y_{t+1}^0 \in Y_{t+1}^0 } {\langle x_t^0,y_{t+1}^0\rangle}) + (1-p^0)  (\min_{z_{t+1}^0 \in Z_{t+1}^0 } {\langle x_t^+,z_{t+1}^0 \rangle})
$$
holds, which, in line with the notation from the formulation of the Theorem, takes the form
$$
M[\xi^+ + \xi^- +\xi^0]=\min_{w_{t+1} \in \theta_{t+1}} \langle x_t,D_t w_{t+1} \rangle 
$$
for any $x_t\in X_t$.

6. It seems natural to consider that the best trader's choice of the vector $x_t$ is the vector at which the maximin
$$
\max_ {x_t \in X_t} \left[ \min_{w_{t+1} \in \theta_{t+1}} \langle x_t,D_t w_{t+1} \rangle \right]
$$
is attained. Though all the components of the vector $x_t$  are integers, the same logic that was applied in [Belenky 1981] in finding the maximum of the minimum function similar to the above one (but with all the components of the vector variable under the maximum sign assuming non-negative, real values) allows one to be certain that the equality 
$$
\max_ {x_t \in X_t} \left[ \min_{w_{t+1} \in \theta_{t+1}} \langle x_t,D_t w_{t+1} \rangle \right] = \max_ {x_t \in X_t} \left[ \max_{z_{t+1} \in \{z_{t+1} \ge 0  \  \   : z_{t+1} A_t \le x_tD_t \}} \langle b_t,z_{t+1} \rangle \right]
$$
holds. Indeed, since the set $\theta_{t+1}=\{w_{t+1} \in R_+^{2n}:A_t w_{t+1} \ge b_t \}$ is a (non-empty) convex polyhedron so that the linear function $\langle x_t,D_t w_{t+1} \rangle$ attains its minimum on this convex polyhedron for any $x_t \in X_t$, the set $\{z_{t+1} \ge 0:z_{t+1} A_t \le x_t D_t \}$, which is a set of feasible solutions to the linear programming problem that is dual to the problem $\min_{w_{t+1} \in \theta_{t+1}} \langle x_t,D_t w_{t+1} \rangle$, is nonempty  for any $x_t \in X_t$ [Yudin \& Golshtein 1965]. Thus,   in both problems
$$
\langle x_t,D_t w_{t+1} \rangle \to \min_{w_{t+1} \in \theta_{t+1}}
$$
and 
$$
\langle b_t,z_{t+1} \rangle \to \max_{z_{t+1} \in \{z_{t+1} \ge 0 :z_{t+1} A_t \le x_tD_t \}} ,
$$
which form a dual pair of linear programming problems for any $x_t \in X_t$, the goal functions attain their extreme values at certain points of the sets $\theta_{t+1}=\{w_{t+1} \in R_+^{2n}:A_t w_{t+1} \ge b_t \}$ and $\{z_{t+1} \ge 0 : z_{t+1} A_t \le x_tD_t \}$,  respectively, for every $x_t \in X_t$ due to the duality theorem of linear programming [Yudin \& Golshtein 1965]. Thus, the equality 
$$
\min_{w_{t+1} \in \theta_{t+1}} \langle x_t,D_t w_{t+1} \rangle = \max_{z_{t+1} \in \{z_{t+1} \ge 0 : z_{t+1} A_t \le x_tD_t \}} \langle b_t,z_{t+1} \rangle
$$
holds for every $x_t \in X_t$, and since the set $X_t$ is finite, the equality 
$$
\max_ {x_t \in X_t} \left[ \min_{w_{t+1} \in \theta_{t+1}} \langle x_t,D_t w_{t+1} \rangle \right] = \max_ {x_t \in X_t} \left[ \max_{z_{t+1} \in \{z_{t+1} \ge 0 : z_{t+1} A_t \le x_tD_t \}} \langle b_t,z_{t+1} \rangle \right]
$$
also holds, which means that the equality 
$$
\max_ {x_t \in X_t} \left[ \min_{w_{t+1} \in \theta_{t+1}} \langle x_t,D_t w_{t+1} \rangle \right] = \max_ {\{x_t \in R_+^n:B_t x_t \ge d_t, x_t \in Q_+^n \} } \left[ \max_{z_{t+1} \in \{z_{t+1} \ge 0 : z_{t+1} A_t \le x_tD_t \}} \langle b_t,z_{t+1} \rangle \right],
$$
where $Q_+$ is a set of all non-negative, integer numbers, and $Q_+^n$ is a direct product of $n$ sets $Q_+$, holds. This means that the value
$$
\max_ {x_t \in M_t} \left[ \min_{w_{t+1} \in \theta_{t+1}} \langle x_t,D_t w_{t+1} \rangle \right] 
$$
can be found by solving the problem
$$
\langle b_t,z_{t+1} \rangle  \to \max_ { \{ (x_t,z_{t+1}) \in R_+^n \times R_+^m: B_t x_t \ge d_t, z_{t+1} A_t \le x_t D_t, x_t \in Q_+^n \} },
$$
where $m$ is the number of rows in the matrix $A_t$, which is a mixed programming problem. 

7. It is clear that if the numbers of securities in the sets $I_t^+$, $I_t^-$, and $I_t^0$ are large, solving this problem may present considerable difficulties. At the same time, since the values of components of the vector $x_t$ usually substantially exceed 1, one can consider these numbers as non-negative, real ones, solve the problem of finding 
$$
\max_ {x_t \in \tilde M_t} \left[ \min_{w_{t+1} \in \theta_{t+1}} \langle x_t,D_t w_{t+1} \rangle \right] 
$$
with the vector variables $x_t$ belonging to the above-mentioned convex polyhedron $\tilde M_t=\{x_t \in R_+^n:B_t x_t \ge d_t\}$, which contains the set $M_t$ (and is described by a compatible system of linear equations and inequalities), and round off all the non-integer components of the vector $x_t$ in the solution in just the same way it was mentioned in Section 3, in considering problems (1) and (2). Thus (if the number of shares in the trader's portfolio is large), the trader may decide to calculate the above maximum of the minimum function, which is an upper bound for the number $\max_ {x_t \in  M_t} \left[ \min_{w_{t+1} \in \theta_{t+1}} \langle x_t,D_t w_{t+1} \rangle \right] $. The value of this upper bound is attained at a saddle point of an antagonistic game on the convex polyhedra $\tilde{M}_t$ and $\theta_{t+1}$ with the payoff function 
\begin{equation}
\langle x_t,D_t w_{t+1} \rangle.
\end{equation}

Let
$$
Q_t=\{(x_t, h_t) \ge 0:h_t A_t \le x_t D_t, B_t x_t \ge d_t \},
$$
$$
P_{t,t+1}=\{(w_{t+1},\pi_{t+1}) \ge 0: \pi_{t+1} B_t \le -D_t w_{t+1}, A_t w_{t+1} \ge b_t \}.
$$

Then the optimal values of the vectors $(x_t )^*$ and $(w_{t+1})^*$, forming a saddle point of function (3) on the set $\tilde{M}_t \times \theta_{t+1}$, are found as components of the solution vectors to linear programming problems 
$$
\langle b_t,h_t \rangle \to \max_{(x_t,h_t) \in Q_t }⁡,
$$
$$
\langle -d_t,\pi_{t+1} \rangle \to \min_{(w_{t+1},\pi_{t+1}) \in P_{t,t+1} }⁡,
$$
forming a dual pair.

If $\left((x_t )^*,(h_t )^*,(w_{t+1} )^*,(\pi_{t+1} )^* \right)$ is a solution of the above pair of linear programming problems, then the values of the vectors $(x_t^+ )^*$, $(x_t^- )^*$ and $(x_t^0 )^*$, where $(x_t )^*=((x_t^+ )^*,(x_t^- )^*,(x_t^0 )^* )$, are completely determined by the values of the vector $(x_t )^*$ [Belenky 1981]. The Theorem  is proved.
\qed
\end{proof}

Remark 2. As mentioned in the course of proving the Theorem, all the variables $x_t$  are integers so that the value of the maximin of the function (3) when $x_t\in \tilde M_t$--which is attained at a saddle point of the game on the sets  $\tilde{M}_t$ and $\theta_{t+1}$ with the payoff function $\langle x_t,D_t w_{t+1} \rangle$ that is maximized with respect to $x\in\tilde {M}_t$ and is minimized with respect to $w_{t+1}\in \theta_{t+1}$--is only an upper bound of the maximin of this function when $x_t\in  M_t$.  Also, as shown there, finding the exact value of this maximin is reducible to solving a mathematical programming problem with mixed variables and a linear goal function. However, it is clear that solving this mixed programming problem within an acceptable period of time may present considerable difficulties for the problems with sizes being of interest for both theoretical studies and practical calculations while solving linear programming problems in finding a saddle point of the game on $\tilde{M}_t \times \theta_{t+1}$  with the payoff function described by (3) does not present any computational difficulties in such calculations. Moreover, quickly  finding an upper bound of the maximin of the function (3) may interest small and medium price-taking traders for their practical calculations the most. Also, in theoretical studies of the interaction between a trader and a stock exchange (to which the present paper belongs), traditionally (see, for instance, the seminal publication of  Markowitz [Markowitz 1952]), volumes of shares to be bought and sold by a trader are assumed to be non-negative, real numbers (variables). Finally, generally, the coefficients in the systems of linear equations and inequalities describing the convex polyhedra that participate in the mathematical formulation of the mixed programming  problem under consideration are known only approximately. With all this in mind, the replacement of the problem of finding the exact value of the maximin of the function (3) when $x_t\in M_t$ with finding an upper bound of this value seems justifiable in practical calculations.

Situation 2. 

There are two cases to be considered in Situation 2. In the first case, the trader does not have any intent to keep particular securities   that she  possesses at the moment $t$ (either based on her own beliefs or at someone's advice), whereas in the second case, the trader has this intent with respect to particular securities. It is clear that in the first case, to estimate what portfolio would  have the maximum value at the moment $t+1$, the trader should first estimate the total cash that she would have if she sold all the securities from her portfolio at the moment $t$ proceeding from the share price values that these securities have at the moment $t$. 
Then the trader should solve the same problem that she would solve in Situation 1 in forming a portfolio a) proceeding from the total amount of cash available to her at the moment $t$, and b) taking into account that she can borrow cash and/or securities from a broker to be returned later. If the borrowed cash or securities should be returned later than at the moment $t+1$, then in the first case of Situation 2, finding the trader's best investment strategies (in the sense of maximizing the value of her portfolio at the moment $t+1$) is either reducible to solving a mixed programming problem (for finding the exact value of the maximin of the function  (3) when $x_t\in  M_t$) or to finding saddle points in an antagonistic game (for finding an upper bound of the above-mentioned maximin) that are similar to those considered in finding such strategies earlier, in Situation 1.

In the second case of  Situation 2, one can easily show that the considered game of changing the portfolio of securities is formulated as the game on the sets $M_t^+\times M_t^-\times M_t^0$ or $M_t$ and $\theta_{t+1}=\{w_{t+1} \in R_+^{2n}: A_t w_{t+1} \ge b_t \}$ of the player strategies with the payoff function $\langle x_t,D_t w_{t+1} \rangle+ \langle q,w_{t+1} \rangle$. Here, $q \in R_+^{2n}$ is a particular vector, $A_t$, $B_t$ are matrices, and $d_t$, $b_t$ are vectors of corresponding dimensions. Their elements are formed by coefficients of compatible systems of  linear equations and inequalities of the balance kind that describe sets of feasible values of the variables forming the vectors $x_t$ and $w_{t+1}$. Two subcases should  then be considered. 

In the first subcase,  the trader does not borrow any securities (from a broker) from the set $I_t^-$. 

Let  $v_t=(v_t^+,v_t^-,v_t^0) \in R_+^{|I_t^+|+|I_t^-|+|I_t^0|}$ be the vector of volumes (numbers of shares) of securities from the set $N$ that the trader has in her portfolio at the moment $t$ and would like to keep at the moment $t+1$ for whatever reasons. As in the Proof of the Theorem, let us first consider the case in which $X_t= X_t^+ \times X_t^- \times X_t^0= M_t^+ \times M_t^- \times M_t^0$  in the  spaces $R_{+}^{|I_t^+|}$, $R_{+}^{|I_t^-|}$, and $R_{+}^{|I_t^0|}$, respectively, where $M_t^+=\{x_t \in R_+^{|I_t^+|}:B^+_t x^+_t \ge d^+_t, x_t \in Q_+^{|I_t^+|} \}$,  $M_t^-=\{x^-_t \in R_+^{|I_t^-|}:B^-_t x^-_t \ge d^-_t, x^-_t \in Q_+^{|I_t^-|} \}$, and $M_t^+=\{x^0_t \in R_+^{|I_t^0|}:B^0_t x^0_t \ge d^0_t, x^0_t \in Q_+^{|I_t^0|} \}$. Then the optimal trader's strategy of choosing the volumes of securities from the set $X_t^+$ to buy is found by maximizing the expectation of the discrete random variable

$$
p^+ \min_{y_{t+1}^+ \in Y_{t+1}^+ } {\left[ \langle x_t^+,y_{t+1}^+ \rangle+ \langle v_t^+,y_{t+1}^+ \rangle \right]} + (1-p^+) \min_{z_{t+1}^+ \in Z_{t+1}^+ } {\left[ \langle x_t^+,z_{t+1}^+ \rangle + \langle v_t^+,z_{t+1}^+ \rangle \right]}, 
$$
which describes the expectation of the financial result associated with buying securities from the set $I_t^+$.

Since the equality 
{\small
$$
\max_{x_t^+ \in X_t^+} {\left[ p^+  \min_{y_{t+1}^+ \in Y_{t+1}^+ } {\left( \langle x_t^+,y_{t+1}^+ \rangle + \langle v_t^+,y_{t+1}^+ \rangle \right)} + (1-p^+) \min_{z_{t+1}^+ \in Z_{t+1}^+ } {\left( \langle x_t^+,z_{t+1}^+ \rangle 
+ \langle v_t^+,z_{t+1}^+ \rangle \right)} \right]} = 
$$ }
$$
= \max_{x_t^+ \in X_t^+} \left[ \min_{y_{t+1}^+ \in Y_{t+1}^+ } {\left( \langle x_t^+, D^{|I_t^+|}(p^+) y_{t+1}^+ \rangle + \langle p^+v_t^+,y_{t+1}^+\rangle \right)} + \right.
$$
$$
\left. \min_{z_{t+1}^+ \in Z_{t+1}^+ } {\left( \langle x_t^+, D^{|I_t^+|}(1-p^+) z_{t+1}^+ \rangle + \langle (1-p^+) v_t^+,z_{t+1}^+ \rangle \right)} \right] 
$$
holds, and the since the vectors $y_{t+1}^+$ and $z_{t+1}^+$ from the sets $Y_{t+1}^+$ and $Z_{t+1}^+$ are chosen independently of each other, the equalities 
$$
\max_{x_t^+ \in X_t^+} \left[ \min_{y_{t+1}^+ \in Y_{t+1}^+ } {\left( \langle x_t^+, D^{|I_t^+|}(p^+) y_{t+1}^+ \rangle + \langle p^+v_t^+,y_{t+1}^+ \rangle \right)} + \right.
$$
$$
\left. +\min_{z_{t+1}^+ \in Z_{t+1}^+ } {\left( \langle x_t^+, D^{|I_t^+|}(1-p^+) z_{t+1}^+ \rangle + \langle (1-p^+) v_t^+,z_{t+1}^+ \rangle \right)} \right]= 
$$
$$
= \max_{x_t^+ \in X_t^+} \left[ \min_{(y_{t+1}^+,z_{t+1}^+) \in Y_{t+1}^+ \times Z_{t+1}^+ } {\left( \langle x_t^+, D^{|I_t^+|}(p^+) y_{t+1}^+ \rangle + \langle p^+v_t^+,y_{t+1}^+ \rangle \right) + } \right.
$$
$$
\left. {+ \langle x_t^+, D^{|I_t^+|}(1-p^+) z_{t+1}^+ \rangle + \langle (1-p^+) v_t^+,z_{t+1}^+ \rangle } \right]= 
$$
$$
= \max_{x_t^+ \in X_t^+} \left[ \min_{w_{t+1}^+ \in \theta_{t+1}^+ } {\left( \langle x_t^+, D^{2|I_t^+|}(p^+,1-p^+) w_{t+1}^+ \rangle + \langle (p^+v_t^+,(1-p^+) v_t^+),w_{t+1}^+ \rangle \right) } \right], 
$$
hold.

Analogously, the maximum of the expectation of the financial result associated with selling securities from the set $I_t^-$ (owned by the trader at the moment $t$) can be written as
$$
\max_{x_t^- \in X_t^-} \left[ \min_{w_{t+1}^- \in \theta_{t+1}^- } {\left( \langle x_t^-, D^{2|I_t^-|}(p^-,1-p^-) w_{t+1}^- \rangle + \langle (p^-v_t^-,(1-p^-) v_t^-),w_{t+1}^- \rangle \right) } \right], 
$$
whereas the maximum of the expectation of the financial result associated with choosing (buying) the volumes of securities from the set $I_t^0$ can be written as
$$
\max_{x_t^0 \in X_t^0} \left[ \min_{w_{t+1}^0 \in \theta_{t+1}^0 } {\left( \langle x_t^0, D^{2|I_t^0|} \left( \frac{1}{2},\frac{1}{2} \right) w_{t+1}^0 \rangle + \langle \left(\frac{1}{2}v_t^0,\frac{1}{2} v_t^0 \right),w_{t+1}^0 \rangle \right)  } \right]. 
$$

Thus, if the trader's best strategy  of choosing the volumes of financial securities from the set $M_t$ is understood as that maximizing the expectation of the financial result of buying securities being components of the vector $x_t\in X_t^+ \times X_t^- \times X_t^0$, this strategy can be found the calculating 
$$
\max_{x_t \in X_t} \left[ \min_{w_{t+1} \in \theta_{t+1} } {\left(  \langle x_t,D_t w_{t+1} \rangle + \langle q,w_{t+1} \rangle \right)}\right],
$$
where $q=\left( (p^+ v_t^+,(1-p^+) v_t^+),(p^- v_t^-,(1-p^-) v_t^-),\left(\frac{1}{2} v_t^0,\frac{1}{2} v_t^0 \right) \right)$. 

In just the same way this was done in the course of proving the Theorem, one can be certain that this strategy remains optimal if 
$X_t= M_t=\{x_t \in R_+^n:B_t x_t \ge d_t, x_t \in Q_+^n \}\subset M_t^+ \times M_t^- \times M_t^0$,  where   $B_t$ is a matrix of a general structure, not necessarily corresponding to the structure of the set $M_t=M_t^+ \times M_t^- \times M_t^0$ as a direct product of the above-mentioned subsets of the three polyhedra from the spaces $R_{+}^{|I_t^+|}$, $R_{+}^{|I_t^-|}$, and $R_{+}^{|I_t^0|}$, respectively (see earlier in Section 4).

One can easily be certain that the equalities
$$
\max_{x_t \in M_t} \left[ \min_{w_{t+1} \in \theta_{t+1} } {\left(  \langle x_t,D_t w_{t+1} \rangle + \langle q,w_{t+1} \rangle \right)}\right]=
$$
$$
=\max_{ \{x_t \in R_+^n: B_t x_t \ge d_t, x_t \in Q_+^n \} } { \left[ \max_{ \{z_{t+1} \ge 0: z_{t+1} A_t \le x_t D_t + q \}} {\langle b_t, z_{t+1} \rangle } \right]}=
$$
$$
=\max_{ \{(x_t, \ z_{t+1}) \ge 0: \ B_t x_t \ge d_t, \ z_{t+1} A_t \le x_t D_t + q, \ x_t \in Q_+^n \}} {\langle b_t, z_{t+1} \rangle }
$$
hold for both types of the structure of the set $X_t=M_t$ so that the maximin 
$$ \max_{x_t \in M_t} \left[ \min_{w_{t+1} \in \theta_{t+1} } {\left(  \langle x_t,D_t w_{t+1} \rangle + \langle q,w_{t+1} \rangle \right)}\right]
$$ 
is found by solving a mixed programming problem of finding the maximum of the linear function $\langle b_t, z_{t+1} \rangle$ on the set $\{(x_t,z_{t+1}) \ge 0: B_t x_t \ge d_t, z_{t+1} A_t \le x_t D_t + q, x_t \in Q_+^n \}$.

In just the same way it was done in considering Situation 1, if one treats components of the vector $x_t$ as non-negative, real numbers, finding the maximin 
$$
 \max_{\{ x_t \in R_+^n: B_t x_t \ge d_t\} } \left[ \min_{w_{t+1} \in \theta_{t+1} } {\left(  \langle x_t,D_t w_{t+1} \rangle + \langle q,w_{t+1} \rangle \right)}\right],
$$
which is an upper bound of the maximin 
$$ \max_{x_t \in M_t} \left[ \min_{w_{t+1} \in \theta_{t+1} } {\left(  \langle x_t,D_t w_{t+1} \rangle + \langle q,w_{t+1} \rangle \right)}\right],
$$
is reducible to finding a saddle point in the antagonistic game  on the sets of player strategies $\tilde{M}_t$ and $\theta_{t+1}$ with the payoff function  $\langle x_t,D_t w_{t+1} \rangle + \langle q,w_{t+1} \rangle$.

A saddle point in this game can be found [Belenky 1981] by solving linear programming problems
$$
\langle b_t,h_t \rangle \to \max_{(h_t,x_t) \in Q_t }⁡,
$$
$$
\langle -d_t,\pi_{t+1} \rangle + \langle q,w_{t+1} \rangle \to \min_{(\pi_{t+1},w_{t+1}) \in P_{t,t+1} }⁡,
$$
forming a dual pair, where $Q_t=\{(h_t,x_t ) \ge 0:h_t A_t \le q + x_t D_t, B_t x_t \ge d_t \}$, and $P_{t,t+1}=\{(\pi_{t+1},w_{t+1}) \ge 0: \pi_{t+1} B_t \le -D_t w_{t+1}, A_t w_{t+1} \ge b_t \}$.

In the second subcase, the trader borrows securities from the broker to sell them at the moment $t$ to have additional cash for buying those securities at the moment $t$ whose share price values she expects to increase at the moment $t+1$ (and the trader should return the borrowed securities later than at the moment $t+1$). The only difference between this subcase and the first subcase is in the amount of cash available for buying shares of securities that interest the trader at the moment $t$, i.e., in  the parameters determining the set $M_t$.

Remark 3. One should bear in mind that both the trader's guaranteed result and its upper estimate in her game with the stock exchange determine only the trader's investment strategies at the moment $t$, and they do not determine the financial result of applying these strategies. This is the case,  since neither the goal function in the maximin problem nor the payoff function, for instance,  in game (3) (when $x_t\in \tilde M_t$) take into consideration such components of the trader's welfare at the moment $t+1$ as, for instance,  the amount of cash remaining after finalizing all the transactions associated with buying securities from the sets $I_t^+$ and $I_t^0$.  However, the above-mentioned financial result can easily be calculated based upon the solutions to the mixed programming problems and games considered for both Situation 1 and Situation 2.

One should also bear in mind that if the trader borrows securities from a broker, and she needs to return them to the broker at the moment $t+1$, other approaches to what should be chosen as the set $X_t^-$ are to be considered. The deployment of such approaches leads to a different structure of the payoff functions in the games describing the interaction of the trader with the stock exchange, including the structure of the  matrix $D_t$. One can show that in the framework of this interaction, finding corresponding maximin values or saddle points in corresponding  games can be done based on the same theoretical foundation developed in [Belenky 1981]. Certainly, in some cases, the interaction between the trader and the stock exchange is formalized in the form of maximin problems and games of more complicated structures  than those studied in Section 4; however,  their consideration lies beyond the scope of the present publication. 

Finally, one should notice that by solving either above-mentioned  problem (i.e., the problem of finding the trader's guaranteed result or that of finding its upper estimate), the trader determines which share price values she should expect to deal with at the moment $t$ with respect to all the standard securities from the set $N$. This information can be used, in particular,  in making decisions on borrowing standard securities to be returned to brokers at the moment $t+1$.

\section{Forming and managing traders' investment portfolios that include derivative financial instruments }
\label{sec:5}

In both problems formulated and studied in Sections 3 and 4 of this paper, no derivative financial instruments (for instance, no futures contracts and no options contracts) were considered. However, one can show that at least in some simple cases of operating with futures and options contracts, models underlying the mathematical formulations of these two problems can be used for solving the problem of forming and managing a trader's investment portfolio which includes derivative financial instruments, futures contracts, and options contracts being the most popular ones among those usually present in portfolios of small and medium price-taking traders. 

\subsection {Model 1. The price values of the futures and options contracts that interest a trader are random variables with uniform probability distributions }

First, let us consider a trader who at the moment $t$, besides standard securities, has some futures contracts as the only derivative financial instruments and plans to buy new futures contracts and to sell some (or all the) futures contracts from her portfolio, i.e., let us consider Situation 2 first (see the beginning of Section 3). 

 For the sake of definiteness, let us first consider futures contracts for supplying (buying) financial instruments, and  for the sake of simplicity, let a) commodities be the only financial instruments that are the underlying assets of the futures contracts that interest the trader, and b) all the expenses associated with storing the commodity being the subject of  futures contract $j$ (since the moment $t$ of buying the contract and  until its expiration date at the moment $t+1$) be reflected in the number $c_{j,t+1}$. Let $v_{j,t+1}$ be the number of futures contracts $j$  with the  price equaling  $K_{j,t+1}+c_{j,t+1}$ that the trader buys at the moment $t$, where $K_{j,t+1}$ is calculated according to well-known formulae, proceeding from both the spot price of the underlying asset of the futures contract at the moment $t$ and the expected risk-free interest rate at the moment $t+1$. 

Consider a trader who at the moment $t$ buys $v_{j,t+1}$ futures contracts $j$ for supplying (buying), for instance, a particular volume of a particular commodity (being the subject of the futures contract) at the moment $t+1$. If the trader  buys these futures contracts with the intent to hold them until the expiration date, she expects to receive a profit from this transaction, for instance, by selling $v_{j,t+1}$ contracts at the moment $t+1$, and the amount of the (expected) profit equals $\Delta_{j.t+1}=v_{j,t+1} \left( s_{j,t+1}-K_{j,t+1}-c_{j,t+1} \right)$. Here,  $s_{j,t+1}$ is  the expected market price of 
 futures contract $j$ at the moment $t+1$.  This (expected) profit is positive,  i.e.,  the inequality  $\Delta_{j.t+1}>0$  holds, if the value of $s_{j,t+1}$  at the moment $t+1$ exceeds $K_{j,t+1}+c_{j,t+1}$.  The trader  will sustain a loss, and  the amount of the loss equals $|\Delta_{j.t+1}|$ if the inequality $\Delta_{j.t+1}<0$ holds. 

An analogous reasoning  is applicable to considering the selling of the same number of futures contracts $j$; however, in this case, a profit is attained if $\Delta_{j.t+1}^0=v_{j,t+1} \left(K_{j,t+1} +c_{j,t+1}-s_{j,t+1} \right)$ is positive. In both cases,  futures contract $j$ with the expiration at the moment $t+1$ that the trader possesses at the moment $t$ can be sold at the moment $t$ as a standard security at the price $K_{j,t+1}+c_{j,t+1}$. 

Let the trader divide  the whole set of futures contracts that interest her at the moment $t$  into the subsets $J_t^+$, $J_t^-$, and $J_t^0$, where $J_t^+$ is a set of futures contracts for which the trader believes with the probability $p_j>0.5$ that the values of the prices that the underlying assets of these contracts will have  at the moment $t+1$ will increase, $J_t^-$ is a set of futures contracts for which the trader believes with the probability $p_j>0.5$ that the values of the prices that the underlying assets of these contracts will have  at the moment $t+1$ will decrease, and $J_t^0$ is a set of futures contracts for which the trader believes with the probability $p_j>0.5$ that the values of the prices that the underlying assets of these contracts will have  at the moment $t+1$ will not change.

Remark 4. To simplify the notation, it is assumed that two different futures contracts for the same commodity with the expiration at the moment $t+1$ have different indices $j$  in each of the sets $J_t^+$, $J_t^-$, and $J_t^0$  if either the volumes of the commodity stipulated in these futures contracts  that  is the subject of the futures contracts (i.e., the sizes of these futures contracts) or the future  prices of these contracts at the moment $t+1$ (or both) are different. One should bear in mind that from any  practical viewpoint, the trader may sell all (or some of) the $v_{j,t+1}$ futures contracts for supplying (buying) the commodity (being the underlying asset of futures contract $j$) before the expiration date of these contracts depending on the dynamics of the market price for the commodity between the moment $t$ and the expiration date $t+1$. However, the consideration of such an action lies beyond the scope of this paper, since it is assumed here (see the beginning of Section 3) that the trader adopts decisions on forming and managing her portfolio of financial instruments at the moments $t$ and $t+1$ as at two consecutive moments, no matter  how much time may (physically) be between these two moments. So it is assumed that at the moment of adopting decisions on forming and managing her portfolio of financial instruments (i.e., at the moment $t$), the trader estimates her potential profit/loss as a result of these decisions with respect to the moment $t+1$. (Here it is assumed that the dynamics of the prices for the futures contracts on commodity $j$ and those for this commodity as such have the same directions (i.e., both increase or not increase), which is usually the case in stock exchange markets.)

Let the trader know  the boarders of the segment $[s_{j,t+1}^{min},s_{j,t+1}^{max}]$ within which the value of $s_{j,t+1},$
$j \in J_t^+ \cup J_t^-$ will change at the moment $t+1$ (or let her believe that the boarders of this segment will be such) while the trader can make no assumptions on a particular probability distribution that the value of  $s_{j,t+1}$, considered as that of a random variable, may have (within these borders). 

Consider first new futures contracts for supplying (buying) commodities that the trader may be interested in buying at the moment $t$, including futures contracts with particular names (indices) from the sets  $J_t^+$, $J_t^-$, and $J_t^0$, that some of the futures contracts in her portfolio have.

For each contract $j$ from these sets of futures contracts, the trader should estimate the value $\Delta_{j.t+1}$, and only if $s_{j,t+1}$ is such that the inequality $\Delta_{j.t+1}>0$ holds, may the trader consider to deal with this futures contract. Thus,  only the futures contracts (from among those of interest to the trader) for which the expectation of the value of $s_{j,t+1}$ at the moment $t+1$ within the segment $[s_{j,t+1}^{min},s_{j,t+1}^{max}]$ exceeds $h_j(t,t+1)=K_{j,t+1}+c_{j,t+1}$ may deserve her attention. The reasoning identical to that from Section 3 allows the trader a) to assume that the values of  $s_{j,t+1}$ are those of continuous random variables $u$ and $v$ and that these random variables are uniformly distributed on the segments $[s_{j,t+1}^{min},h_j(t,t+1)]$ and $[h_j(t,t+1),s_{j,t+1}^{max}]$, respectively, where the inequalities $s_{j,t+1}^{min}<h_j(t,t+1) < s_{j,t+1}^{max}$ hold, and b) to calculate the expectation of $s_{j,t+1}$ using the formulae identical to those from Section 3.

Thus,  the profit/loss that the trader should expect to receive/sustain at the moment $t+1$ as a result of  buying $v_{j,t+1}$ futures contracts $j$ at the moment $t$ (with the expiration at the moment $t+1$) is a  random variable, and the expectation of this random variable (i.e., that of the trader's profit/loss) can be calculated as follows:

a) $M_f FinRes_{j,t+1}(v_{j,t+1})=p_j \left( v_{j,t+1} \left[ \frac{h_j(t,t+1)+ s_{j,t+1}^{max}}{2}-h_j(t,t+1) \right] \right)+$
{\small
$$
+\frac{1-p_j}{2} \left( v_{j,t+1} \left[ \frac{s_{j,t+1}^{min}+h_j(t,t+1)}{2}-h_j(t,t+1) \right] \right)+\frac{1-p_j}{2} \left( v_{j,t+1} \left[h_j(t,t+1)-h_j(t,t+1) \right] \right), 
$$ }
which is a linear function of the volume $v_{j,t+1}$, if the trader believes that for futures contract $j$, the inclusion $j \in J_t^+$ will hold at the moment $t+1$ with the probability $p_j$, the inclusions $j \in J_t^-$ and $j \in J_t^0$ are equally possible, and she buys $v_{j,t+1}$  futures contracts $j$ for supplying the commodity that is the subject of futures contract $j$,

b) $M_f FinRes_{j,t+1}(v_{j,t+1})=p_j \left( v_{j,t+1} \left[ \frac{s_{j,t+1}^{min}+h_j(t,t+1)}{2}-h_j(t,t+1) \right] \right)+$
{\small
$$
+\frac{1-p_j}{2} \left( v_{j,t+1} \left[ \frac{h_j(t,t+1)+ s_{j,t+1}^{max}}{2} -h_j(t,t+1) \right] \right)+\frac{1-p_j}{2} \left( v_{j,t+1} \left[h_j(t,t+1)-h_j(t,t+1) \right] \right), 
$$ }
which is a linear function of the volume $v_{j,t+1}$, if the trader believes that for futures contract $j$, the inclusion $j \in J_t^-$ will hold at the moment $t+1$ with the probability $p_j$, the inclusions $j \in J_t^+$ and $j \in J_t^0$ are equally possible, and  she buys $v_{j,t+1}$  futures contracts $j$  for supplying the commodity that is the subject of futures contract $j$, and

c) $M_f FinRes_{j,t+1}(v_{j,t+1})=p_j \left( v_{j,t+1} \left[h_j(t,t+1)-h_j(t,t+1) \right] \right) +$
{\small
$$
+\frac{1-p_j}{2} \left( v_{j,t+1} \left[ \frac{s_{j,t+1}^{min}+h_j(t,t+1)}{2}-h_j(t,t+1) \right] \right)
$$
$$+\frac{1-p_j}{2} \left( v_{j,t+1} \left[ \frac{h_j(t,t+1)+ s_{j,t+1}^{max}}{2} -h_j(t,t+1) \right] \right), 
$$}which is a linear function of the volume $v_{j,t+1}$, if the trader believes that for futures contract $j$, the inclusion $j \in J_t^0$ will hold at the moment $t+1$ with the probability $p_j$, the inclusions $j \in J_t^+$ and $j \in J_t^-$ are equally possible, and she buys $v_{j,t+1}$  futures contracts $j$  for  supplying the commodity that is the subject of  futures contract $j$. (One can easily  notice that the third summand in the expression for $M_f FinRes_{j,t+1}(v_{j,t+1})$ in cases a) and b)  and the first summand in that of case c) equal 0, since if  $s_{j,t+1}=K_{j,t+1}+c_{j,t+1}$, the equality $s_{j,t+1}- h_j(t,t+1)=0$ holds for $j\in {J}_t^+\cup {J}_t^- \cup {J}_t^0$ at the moment $t$. However, the above three expressions seem more descriptive since they help draw attention to the comparison of the expectation of the market price that futures contract $j$ will have at the moment $t+1$  and the price of this futures contract at the moment $t$.)

It is clear that the trader may be interested in including  in her portfolio a futures contract for buying a particular commodity   at the moment $t$  only if the expectation of the profit at the moment $t+1$ associated with buying this contract at the moment $t$ is positive.  One should notice that, generally, more exact estimates of the profit/loss can be obtained by an interested trader if more information on the structure of the random variable $s_{j,t+1}$ is available, for instance, if the trader may assume that this random variable is normally distributed on the segment $[s_{j,t},s_{j,t+1}^{max}]$ [Feller 1991].

Let $\hat{J}_t^+ \subset J_t^+$,$\hat{J}_t^- \subset J_t^-$, and $\hat{J}_t^0 \subset J_t^0$ be the sets of futures contract names  (indices) for which the expectations $M_f FinRes_{j,t+1}(v_{j,t+1})$  are strictly positive (according to the estimates of the expected profit/loss from the transactions with them, calculated with the use of the above formulae, that are planned to be made at the moment $t$). (For futures contracts with normally distributed values of  $s_{j,t+1}$ on the segments  $[s_{j,t+1}^{min},s_{j,t+1}^{max}]$, one may also consider these sets to be chosen in such a manner that the likely intervals for the profit/loss values are calculated based on this information [Feller 1991].) If at least one of the three sets $\hat{J}_t^+$,$\hat{J}_t^-$, and $\hat{J}_t^0$ is  not empty, the trader may consider buying new futures contracts from the set $\hat{J}_t^+\cup \hat{J}_t^- \cup \hat{J}_t^0$ at the moment $t$ (see Remark 4). 

Consider now the futures contracts that are already in the trader's portfolio at the moment $t$, and let $c_{j,t+1}$ have the same meaning as it does for futures contracts (for supplying (buying) commodities) under consideration (see the beginning of Section 5.1) that are bought at the moment $t$.

Let $J_t^+ (av) \subset J_t^+$, $J_t^- (av) \subset J_t^-$, $J_t^0 (av) \subset J_t^0$ be the sets of the names (indices) of the futures contract that the trader possesses at the moment $t$, and let $v_{j,t}$ be the number of contracts $j$, $j \in {J}̂_t^+(av)\cup {J}̂_t^-(av) \cup {J}̂_t^0(av)$. Unlike in determining which new futures contracts to buy, to determine which of the futures contracts that the trader possesses at the moment $t$ should be held, the trader should first estimate the expectations of the values of $s_{j,t+1}$ for  contracts from all the three sets $J_t^+ (av)$, $J_t^- (av)$, and $J_t^0 (av)$ at the moment $t+1$. The expectations of these values are calculated according to the formulae
$$
Ms_{j,t+1} = p_j \frac{ s_{j,t} + s_{j,t+1}^{max}}{2} + \frac{1-p_j}{2} \frac{s_{j,t+1}^{min} + s_{j,t}}{2} + \frac{1-p_i}{2} s_{j,t}, j\in J_t^+ (av),
$$
$$
Ms_{j,t+1} = p_j \frac{s_{j,t+1}^{min} + s_{j,t}}{2} + \frac{1-p_j}{2} \frac{s_{j,t} + s_{j,t+1}^{max}}{2} + \frac{1-p_j}{2} s_{j,t}, j\in J_t^- (av),
$$
$$
Ms_{j,t+1} = p_js_{j,t} + \frac{1-p_j}{2} \frac{s_{j,t+1}^{min} + s_{j,t}}{2} + \frac{1-p_j}{2} \frac{s_{j,t} + s_{j,t+1}^{max}}{2} , j \in J_t^0 (av),
$$
where $s_{j,t}$ is the price value of futures contract $j$ that the trader already possesses at the moment $t, \ j\in {J}̂_t^+(av)\cup {J}̂_t^-(av) \cup {J}̂_t^0(av)$, which are completely identical to the ones from Section 3 (for standard  securities). 

Let $\hat{J}_t^+ (av)\subset {J}_t^+(av)$, $\hat{J}_t^-(av)\subset {J}_t^-(av)$, and $\hat{J}_t^0(av)\subset {J}_t^0(av)$ be the sets of names $j$ for which the differences $Ms_{j,t+1}-s_{j,t}$ are strictly positive. It is clear that the trader may consider  a) holding the futures contracts from the sets $\hat{J}_t^+ (av)$, $\hat{J}_t^-(av)$, and $\hat{J}_t^0(av)$, and b) selling all the futures contracts from the sets $J_t^+(av) \setminus \hat{J}_t^+(av)$, $J_t^-(av) \setminus \hat{J}_t^-(av)$, and $J_t^0(av) \setminus \hat{J}_t^0(av)$ and not borrowing from brokers any futures contracts from these sets. Since selling  futures contracts from the sets $J_t^+(av) \setminus \hat{J}_t^+ (av)$, $J_t^-(av) \setminus \hat{J}_t^- (av)$, and $J_t^0(av) \setminus \hat{J}_t^0 (av)$ leads to receiving the money that can be spent, particularly, for buying new futures contracts from the sets $\hat{J}_t^+$, $\hat{J}_t^-$, and $\hat{J}_t^0$, the trader needs to find an optimal investment strategy of changing her portfolio that includes both standard securities and futures contracts under consideration (as the only derivative financial instruments). 

In just the same way it is done in dealing with standard securities, the trader may decide to borrow futures contracts from a broker to sell them short (i.e., to open a short position) if she believes that the prices of particular futures contracts will decrease at the moment $t+1$. Based upon the estimates of the expectations of the prices that the futures contracts of her interest may have at the moment $t+1$, the trader may decide to borrow futures contracts from the set $( J_t^+\setminus \hat{J}_t^+)\cup ( J_t^-\setminus \hat{J}_t^-) \cup ( J_t^0\setminus \hat{J}_t^0)$.

Let $z_{j,t}, \ j \in ( J_t^+\setminus \hat{J}_t^+)\cup ( J_t^-\setminus \hat{J}_t^-) \cup ( J_t^0\setminus \hat{J}_t^0)$ be the number of futures contracts that the trader borrows from a broker (or from brokers), and  let $\triangle \hat W_{t+1}$ be the estimated increment of the value of the trader's welfare (at the moment $t+1$ with respect to the moment $t$). If the trader considers a strategy of changing her portfolio at the moment $t$ to be optimal if it maximizes the expectation of the random variable $\triangle \hat W_{t+1}$ as a result of choosing appropriate standard securities and futures contracts to buy, to sell, to hold, and to borrow, she should solve the problem 

\begin{equation}
\begin{split}
M[\triangle \hat W_{t+1}] &=   \sum_{i \in \hat I_t^+} {x_{i,t}^+( Ms_{i, t+1}-s_{i,t})}+  \sum_{i \in \hat I_t^-} {x_{i,t}^+( Ms_{i, t+1}-s_{i,t})} + \sum_{i \in \hat I_t^0} {x_{i,t}^+( Ms_{i, t+1}-s_{i,t})} +\\  
& + \sum_{i \in \hat{I}_t^+ (av)} { v_{it}(Ms_{i,t+1}-s_{i,t} )} +  \sum_{i \in \hat{I}_t^- (av)} { v_{it}(Ms_{i,t+1}-s_{i,t} )}+  \\ 
&+\sum_{i \in \hat{I}_t^0 (av)} { v_{it}(Ms_{i,t+1}-s_{i,t} )} +\sum_{i \in (\tilde I_t^+\setminus \hat{I}_t^+)\cup (\tilde I_t^-\setminus \hat{I}_t^-) \cup (\tilde I_t^0\setminus \hat{I}_t^0)} {z_{i,t}^- (s_{i,t}-Ms_{i,t+1} )}+\\
& + \sum_{j \in \hat{J}_t^+}{M_f FinRes_{j,t+1} (v_{j,t+1})}+ \sum_{j\in \hat{J}_t^-}{M_f FinRes_{j,t+1} (v_{j,t+1}) } + \\
& + \sum_{j\in \hat{J}_t^0}{M_f FinRes_{j,t+1} (v_{j,t+1})} + \sum_{j \in \hat{J}_t^+(av)}{v_{j,t}(Ms_{j,t+1}-h_j(t,t+1))} + \\
& + \sum_{j\in \hat{J}_t^-(av)}{v_{j,t}(Ms_{j,t+1}-h_j(t,t+1))}+ \sum_{j\in \hat{J}_t^0(av)}{v_{j,t}(Ms_{j,t+1}-h_j(t,t+1))} +\\
& +\sum_{j \in (J_t^+\setminus \hat{J}_t^+)\cup ( J_t^-\setminus \hat{J}_t^-) \cup ( J_t^0\setminus \hat{J}_t^0)} {z_{j,t}^- (h_j(t,t+1)-Ms_{j,t+1} )}\to max, \\ 
&  \sum_{i \in \hat I_t^+} {x_{i,t}^+ Ms_{i, t+1} }+  \sum_{i \in \hat I_t^-} {x_{i,t}^+Ms_{i, t+1} } + \sum_{i \in \hat I_t^0} {x_{i,t}^+M s_{i, t+1} } + \\
\end{split}
\end{equation}
\begin{align*}
\begin{split}
& + \sum_{i \in \hat{I}_t^+ (av)} { v_{it}Ms_{i,t+1}} +  \sum_{i \in \hat{I}_t^- (av)} { v_{it}Ms_{i,t+1}}+  \sum_{i \in \hat{I}_t^0 (av)} { v_{it}Ms_{i,t+1}} + \\
& + \sum_{j \in \hat{J}_t^+}{v_{j,t+1} Ms_{j,t+1}}+ \sum_{j\in \hat{J}_t^-}{v_{j,t+1}Ms_{j,t+1}} + \sum_{j\in \hat{J}_t^0}{v_{j,t+1}Ms_{j,t+1}} +\\
& + \sum_{j\in \hat J_t^+ (av)}{v_{j,t}s_{j,t+1}}+ \sum_{j\in \hat J_t^-(av)}{{v_{j,t}s_{j,t+1}}}+ \sum_{j\in \hat J_t^0(av)}{v_{j,t}s_{j,t+1}}+\\
& + \left( m_t- \sum_{i \in \hat I_t^+} {x_{i,t}^+ s_{i, t} }-  \sum_{i \in \hat I_t^-} {x_{i,t}^+s_{i, t} } - \sum_{i \in \hat I_t^0} {x_{i,t}^+ s_{i, t} -\sum_{j\in \hat J_t^+ }{v_{j,t+1}h_j(t,t+1)}- \sum_{j\in \hat J_t^-}{{v_{j,t+1}h_j(t,t+1)}}- } \right.\\
& - \left. \sum_{j\in \hat J_t^0}{v_{j,t+1}h_{j}{t, t+1}} \right) +\sum_{i \in \tilde I_t^+(av) \setminus \hat{I}_t^+(av)} {v_{i,t} s_{i, t} }+ \sum_{i \in \tilde I_t^-(av) \setminus \hat{I}_t^-(av)} {v_{i,t} s_{i, t} }+\sum_{i \in \tilde I_t^0(av) \setminus \hat{I}_t^0(av)} {v_{i,t} s_{i, t} }+\\
&  +\sum_{i \in (\tilde I_t^+\setminus \hat{I}_t^+)\cup (\tilde I_t^-\setminus \hat{I}_t^-) \cup (\tilde I_t^0\setminus \hat{I}_t^0)} {z_{i,t}^-(s_{i,t}-Ms_{i,t+1} )} + \\
& + \sum_{j \in ( J_t^+\setminus \hat{J}_t^+)\cup ( J_t^-\setminus \hat{J}_t^-) \cup ( J_t^0\setminus \hat{J}_t^0)} {z_{j,t}^- (h_j(t,t+1)-Ms_{j,t+1} )}+  \\ 
&\left.  +\sum_{j\in J_t^+(av)\setminus \hat{J}_t^+(av)}{v_{j,t}s_{j,t}}+\sum_{j\in J_t^-(av)\setminus \hat{J}_t^-(av)}{v_{j,t}s_{j,t}}+\sum_{j \in J_t^0(av) \setminus \hat{J}_t^0(av)} {v_{i,t} s_{j,t} }
\ge \right.  \alpha \Biggl( m_t   \\
&\left.  \left. + \sum_{i\in  \tilde I_t^+(av)\cup\tilde I_t^-(av)\cup\tilde I_t^0(av)} {v_{i,t} s_{i,t}} + \sum_{j\in J_t^+ (av)}{v_{j,t}s_{j,t}}+ \sum_{j\in J_t^-(av)}{{v_{j,t}s_{j,t}}}+ \sum_{j\in J_t^0(av)}{v_{j,t}s_{j,t}}\right), \right. \\
&  \sum_{i \in \hat I_t^+} {x_{i,t}^+ s_{i, t} }+  \sum_{i \in \hat I_t^-} {x_{i,t}^+s_{i, t} } + \sum_{i \in \hat I_t^0} {x_{i,t}^+ s_{i, t} } +\sum_{i \in (\tilde I_t^+\setminus \hat{I}_t^+)\cup (\tilde I_t^-\setminus \hat{I}_t^-) \cup (\tilde I_t^0\setminus \hat{I}_t^0)} {z_{i,t}^- s_{i,t}}+\\
& +\sum_{j \in (J_t^+\setminus \hat{J}_t^+)\cup ( J_t^-\setminus \hat{J}_t^-) \cup ( J_t^0\setminus \hat{J}_t^0)} {z_{j,t}^- h_j(t,t+1)}+ \sum_{j\in \hat{J}_t^+}{v_{j,t+1}h_j(t,t+1)}+ \sum_{j\in \hat{J}_t^-}{{v_{j,t+1}h_j(t,t+1)}}+ \\
& + \sum_{\hat{j} \in \hat{J}_t^0} {v_{j,t+1}h_j(t,t+1)} - \left( m_t + \sum_{i \in \tilde I_t^+(av) \setminus \hat{I}_t^+(av)} {v_{i,t} s_{i, t} }+ \sum_{i \in \tilde I_t^-(av) \setminus \hat{I}_t^-(av)} {v_{i,t} s_{i, t} }+  \right.  \\
& \left. + \sum_{i \in \tilde I_t^0(av) \setminus \hat{I}_t^0(av)} {v_{i,t} s_{i, t} } + \sum_{j\in J_t^+ (av) \setminus \hat{J}_t^+ (av)}{v_{j,t}h_j(t,t+1)} + \sum_{j\in J_t^-(av) \setminus \hat{J}_t^-(av)}{{v_{j,t}h_j(t,t+1)}} + \right.  \\
& + \sum_{j\in J_t^0(av) \setminus \hat{J}_t^0(av)}{v_{j,t}h_j(t,t+1)} \Biggr) \le k_t \left(m_t+ \sum_{i\in  \tilde I_t^+(av)\cup\tilde I_t^-(av)\cup\tilde I_t^0(av)} {v_{i,t} s_{i,t}} + \sum_{j\in J_t^+ (av)}{v_{j,t}s_{j,t}}\right. \\
& +\sum_{j\in J_t^-(av)}{{v_{j,t}s_{j,t}}}+ \sum_{j\in J_t^0(av)}{v_{j,t}s_{j,t}} \Biggr) ,\\
\end{split}
\end{align*}
where $x_{i,t}^+\ge 0, i \in \hat I_t^+\cup \hat I_t^- \cup \hat I_t^0, \  z_{i,t}^-\ge 0, i \in (\tilde I_t^+\setminus \hat{I}_t^+)\cup (\tilde I_t^-\setminus \hat{I}_t^-) \cup (\tilde I_t^0\setminus \hat{I}_t^0)$, $z_{j,t}^-\ge 0, j\in ( J_t^+\setminus \hat{J}_t^+)\cup ( J_t^-\setminus \hat{J}_t^-) \cup ( J_t^0\setminus \hat{J}_t^0)$,   $v_{j,t+1}\ge 0, j \in \hat{J}̂_t^+\cup \hat{J}̂_t^-\cup \hat{J}̂_t^0$ are integer variables,  $v_{j,t}, \ j \in {J}̂_t^+(av)\cup {J}̂_t^-(av) \cup {J}̂_t^0(av)$  is the number of futures contracts $j$ that the trader possesses at the moment $t$ (a constant parameter), and $v_{j,t}(Ms_{j,t+1}-s_{j,t}), \ j \in \hat {J}̂_t^+(av)\cup \hat {J}̂_t^-(av) \cup \hat {J}̂_t^0(av)$ is  the expectation of the financial result associated with holding futures contract $j$ from the set $\hat {J}̂_t^+(av)\cup \hat {J}̂_t^-(av) \cup \hat {J}̂_t^0(av)$, which  the trader possesses at the moment $t$. Here, both $Ms_{j,t+1}, \ j\in \hat J_t^+\cup \hat J_t^- \cup \hat J_t^0$ (constant parameters) and $M_f FinRes_{j,t+1} (v_{j,t+1})$ (linear functions of the variables $v_{j,t+1}$) are calculated according to the above formulae. 

One should bear in mind the difference between  the notation used in describing  sets of the names (indices)  of standard securities and sets of the names (indices) of futures contracts, in particular, in problem (4). That is, the whole set of standards securities   that interest the trader at the moment $t$ is the set $\tilde I_t^+\cup \tilde I_t^-\cup \tilde I_t^0$, whereas the whole set of futures contracts that interest the trader at the moment $t$ is the set $J_t^+\cup J_t^-\cup J_t^0$ (the absense of the sign $\tilde J$ in the description of the second set). 

Problem (4) is an integer programming one, and in just the same way as  problems (1) and (2) (see Section 3), particular variants of this problem can be solved exactly, with the use of software for solving integer programming problems (if the total number of integer variables in them is such that one can solve these problems in an acceptable time). Otherwise, this problem can be transformed into a linear programming problem by  replacing the integer variables with continuous ones, in just the same way it was  described in Section 3. Once this  linear programming problem has been solved,  the values of these continuous variables in the solution to this problem that are non-integer should be rounded  off. 

A problem of the same kind can be formulated for futures contracts for selling commodities.  It is clear that if at the moment $t$, the future price  of futures contract $j$ on selling a particular commodity at the moment $t+1$  equals $K_{j,t+1}+c_{j,t+1}$, the trader should estimate the value $\Delta_{j.t+1}^0$ (see the definition of $\Delta_{j.t+1}^0$ earlier in Section 5.1). Only if $s_{j,t+1}$ is such that the inequality  $\Delta_{j.t+1}^0>0$ holds, may the trader consider to deal with this futures contract.

Finally, if futures contracts of both kinds are in the trader's portfolio, she can divide the set of the futures contracts of her interest first into two parts: a) for the futures contracts for buying commodities, and b) for the futures contracts for selling commodities. Then, she can divide each of these two parts into the three subsets $J_t^+$, $J_t^-$, and $J_t^0$ and form the system of constraints for the problem of finding optimal trader's investment strategies  with respect to  the set of futures contracts for buying commodities and that with respect to the set of futures contracts for selling commodities. After that, she should combine both systems of constraints taking into account that a) there should be only one constraint on the ratio of the value of the trader's welfare at the moment $t+1$ and the value of that at the moment $t$, and b) there should be only a combined amount of cash (that the trader possesses at the moment $t$) that determines the value of the credit leverage available to the trader.  The combined system of constraints is to be the one in the integer programming problem to be formulated to cover this case, and this problem  is similar in the structure to problem (4). 

One should mention that all the above reasoning and the formulae remain true for futures contracts with respect to financial instruments other than commodities. However, if there is no cost for storing the underlying assets of these contracts (or any other costs associated with managing the underlying assets of these futures contracts), the equalities $c_{j,t+1}=0$ should hold for contract $j$ that may belong to a corresponding set of such futures contracts. 

Remark 5. Similar to how this was done in Section 3 (see Remark 1), to find an optimal investment strategy of the trader in Situation 1 (when the trader does not have any standard securities or any futures contracts in her portfolio at the moment $t$), in addition to the equalities $I_t^+(av)=\emptyset, \  I_t^-(av)=\emptyset$, and $ I_t^0(av)=\emptyset$,  one should set $J_t^+ (av)=\emptyset, \  J_t^- (av)=\emptyset$, and   $J_t^0 (av)=\emptyset$ in the system of constraints of problem (4). Also, one should mention that all the constant summands are present in the goal function of problem (4) only for the sake of the tractability of its structure with respect to the corresponding expressions parts of which are used in the representation of this function. Further, one should mention that though in the variables $s_{j,t+1}$ and $s_{i,t+1}$, the indices $j\in J_t^+\cup J_t^- \cup J_t^0$ and $i\in \tilde N$ (see Section 3) are subsets of the set of natural numbers, this does not constitute any confusion. That is, since in the general description,  the difference is clear due to that in the first index in the pair (in each of these variables), in any particular calculations with the use of software packages, the letters describing the variables are to be chosen in line with the requirements that these packages have. Finally, as in considering problems (1) and (2), one should bear in mind that in the formulation of problem (4), it is assumed that the value of the money at which some of the securities and the futures contracts are sold at the moment $t$ remains unchanged at the moment $t+1$. However, if this is not the case, it is easy to reformulate problem (4) taking into consideration the difference in this value.  

Consider now a trader who at the moment $t$, besides standard securities, has some options contracts in her portfolio as the only derivative financial instruments and plans to buy new options contracts and to sell some (or all the) options contracts that are in her possession (i.e., as before, consider Situation 2 first). 

In just the same way it was done for futures contracts, for the sake of definiteness, let us first consider options contracts for supplying (buying) financial instruments (call options contracts), and  for the sake of simplicity, let a) commodities be the only financial instruments that are the underlying assets of the options contracts that interest the trader, and b) the option premium equal  $\gamma_{l,t+1}$. Let at the moment  $t$, the trader buy call options contract $l$ for supplying (buying), for instance,  a particular volume of a particular commodity (being the subject of this call options contract) at the moment $t+1$. Further, let  $v_{l,t+1}$ be the number of call options contracts $l$ with the strike price $K_{l,t+1}$ that the trader buys at the moment $t$. (It is assumed that the seller of call option contract $l$ calculates the strike price $K_{l, t+1}$ taking into consideration all the expenses associated, for instance, with storing the whole volume of the commodity (stipulated in call option contract $l$) that is to be provided at the moment $t+1$).

At the time of buying call option contract $l$, the trader estimates the profit that she would receive from this transaction (if this call option contract were executed; see also Remark 4 earler in Section 5.1 regarding a similar situation with futures contracts), and the amount of the profit equals $\Delta'_{l,t+1}=v_{l,t+1}\left( s_{l,t+1}-K_{l,t+1}-\gamma_{l,t+1} \right)$. Here,  $s_{l,t+1}$ is the expected market price  of  the call options contract. This (expected) profit is positive, i.e.,  the inequality  $\Delta'_{l.t+1}>0$  holds,  if the value of $s_{l,t+1}$  at the moment $t+1$ exceeds $K_{l,t+1}+\gamma_{l,t+1}$.  The trader will sustain a loss, and the amount of the loss equals $\min (|\Delta'_{l.t+1}|, \gamma_{l,t+1}v_{l,t+1}) $ if the inequality $\Delta'_{l.t+1}<0$ holds.

An analogous reasoning  is applicable in considering the selling of financial instruments in the form of options contracts (put options contracts)   $l$ in the number of $v_{l,t+1}$ such contracts. However, in this case, the profit is attained if 
$$\Delta_{l.t+1}^{'0}=v_{l,t+1} \left(K_{l,t+1} +\gamma_{l,t+1}-s_{l,t+1}\right)$$ 
is positive.  In both cases,  options contract $j$ with the expiration at the moment $t+1$ that the trader possesses at the moment $t$ can be sold at the moment $t$ as a standard security at the price $K_{l,t+1}+\gamma_{l,t+1}$.

As before (in considering the futures contracts as the only derivative financial instruments that the trader possesses at the moment $t$), let the trader divide the whole set of the options contracts into the  subsets $L_t^+$, $L_t^-$, and $L_t^0$, where each of these three sets has the same meaning with respect to the options contracts as does each of the corresponding sets $J_t^+$, $J_t^-$, and $J_t^0$ with respect to the futures contracts. That is, let the trader believe with the probability $p_l>0.5$ that the values of the prices that the underlying assets of the contracts from the set $L_t^+$ will have  at the moment $t+1$ will increase, the values of the prices that the underlying assets of the contracts from the set $L_t^-$  will have  at the moment $t+1$ will decrease, and the values of the prices that the underlying assets of the contracts from the set $L_t^0$  will have  at the moment $t+1$ will not change. Also, as in considering futures contracts (see Remark 4) , to simplify the notation, it is assumed that two different options contracts with the expiration at the moment $t+1$ for the same commodity have different names (indices) $l$  in each of the sets $L_t^+$, $L_t^-$, and $L_t^0$  if either the volumes of the commodity stipulated in these options contracts that is the underlying asset of the options contracts (i.e., the sizes of these options contracts) or  the strike prices of these contracts or both are different. 

Let the trader know the boarders of the segment $[s_{l,t+1}^{min},s_{l,t+1}^{max}]$ within which the value of $s_{l,t+1}$ for  options contract $l \in L_t^+ \cup L_t^-$ will change at the moment $t+1$ (or let her believe that these boarders will be such) while the trader can make no assumptions on a particular probability distribution that the value of $s_{l,t+1}$, considered as that  of a random variable, may have (within these borders). As in the above case with the futures contracts as the only derivative financial instrument in the trader's portfolio, the  trader should estimate the value $h_l(t,t+1)=K_{l,t+1}+\gamma_{l,t+1}$, and it is clear that only if $s_{l,t+1}$ is such that the inequality $\Delta'_{l.t+1}>0$ for options contract $l$ holds, may the trader consider to deal with this call option contract.  Thus,  only the  options contracts (from among those being of interest to the trader) for which the expectation of the value of $s_{l,t+1}$ within the segment $[s_{l,t+1}^{min},s_{l,t+1}^{max}]$ exceeds $K_{l,t+1}+\gamma_{l,t+1}$ may deserve her attention. As in the case of the futures contracts, the trader may assume that the values of $s_{l,t+1}$  are those of continuous random variables $\tilde u$ and $\tilde v$ and that these random variables  are uniformly distributed on the segments $[s_{l,t+1}^{min},h_l(t,t+1)]$ and $[h_l(t,t+1), s_{l,t+1}^{max}]$, respectively, where $h_l(t,t+1)=K_{l,t+1}+\gamma_{l,t+1}$, and the inequalities  $s_{l,t+1}^{min} <h_l(t,t+1)<  s_{l,t+1}^{max}$ hold.
The same reason is applicable to put options contract $l$, which may deserve the trader's attention only if the value of $s_{l, t+1}$ within the segment $s_{l,t+1}^{min} <h_l(t,t+1)<  s_{l,t+1}^{max}$ does not exceed $K_{l,t+1}+\gamma_{l,t+1}$.

Thus, the profit/loss that the trader should expect to receive/sustain as a result of  buying $v_{l,t+1}$ options contracts $l$  at the moment $t$ (with the expiration at the moment $t+1$) is a random variable. However,  the expectations of the final financial result, i.e., those of the profit/loss that the trader should expect to receive/sustain at the moment $t+1$ as a result of  buying $v_{l,t+1}$ options contracts $l$ from the set $L_t^+ \cup L_t^- \cup L_t^0$ at the moment $t$ (with the expiration at the moment $t+1$), are calculated differently compared with those for futures contracts $j$ from the set $J_t^+ \cup J_t^- \cup J_t^0$.  That is, these expectations  for the options contracts are calculated as follows:

a) $M_{op} FinRes_{l,t+1}(v_{j,t+1})=$
$$
=p_l \left[ \max{\left( \left( v_{l,t+1} \left[\frac{h_l(t,t+1)+ s_{l,t+1}^{max}}{2}-h_l(t,t+1)\right] \right), -\gamma_{l,t+1} v_{l,t+1} \right) } \right] + 
$$
$$
+\frac{1-p_l}{2} \left[ \max{\left( \left( v_{l,t+1} \left[\frac{s_{l,t+1}^{min}+ h_l(t,t+1)}{2}-h_l(t,t+1) \right]\right), -\gamma_{l,t+1} v_{l,t+1} \right) } \right] +
$$
$$
+\frac{1-p_l}{2} \left[ \max{\left( \left( v_{l,t+1} \left[h_l(t,t+1)-h_l(t,t+1)], -\gamma_{l,t+1} v_{l,t+1} \right]\right),  \right) } \right],              
$$
which is a function of the volume $v_{l,t+1}$,  if the trader believes that for options contract $l$, the inclusion $l \in L_t^+$ will hold at the moment $t+1$ with the probability $p_l$, the inclusions $l \in L_t^-$ and $l \in L_t^0$ are equally possible, and she buys $v_{l,t+1}$  contracts $l$ as call option contracts,

b) $M_{op} FinRes_{l,t+1}(v_{j,t+1})=$
$$
=p_l \left[ \max{\left( \left( v_{l,t+1} \left[h_l(t,t+1)-\frac{s_{l,t+1}^{min}+ h_l(t,t+1)}{2} \right]\right), -\gamma_{l,t+1} v_{l,t+1} \right) } \right] +
$$
$$
+\frac{1-p_l}{2} \left[ \max{\left( \left( v_{l,t+1} \left[h_l(t,t+1)-\frac{h_l(t,t+1)+ s_{l,t+1}^{max}}{2} \right]\right), -\gamma_{l,t+1} v_{l,t+1} \right) } \right] +
$$
$$
+\frac{1-p_l}{2} \left[ \max{\left( \left( v_{l,t+1} \left[h_l(t,t+1)-h_l(t,t+1)  \right]\right), -\gamma_{l,t+1} v_{l,t+1} \right) } \right],
$$
which is a function of the volume $v_{l,t+1}$, if the trader believes that for options contract  $l$, the inclusion $l \in L_t^-$ will hold at the moment $t+1$ with the probability $p_l$, the inclusions $l \in L_t^+$ and $l \in L_t^0$ are equally possible, and she buys $v_{l,t+1}$ contracts $l$ as  put option contracts,

c) $M_{op} FinRes_{l,t+1}(v_{j,t+1})=$
$$
=p_l \left[ \max{\left( \left( v_{l,t+1} \left[h_l(t,t+1)-h_l(t,t+1) \right] \right), -\gamma_{l,t+1} v_{l,t+1} \right) } \right] +
$$
$$
+\frac{1-p_l}{2} \left[ \max{\left( \left( v_{l,t+1} \left[\frac{s_{l,t+1}^{min}+ h_l(t,t+1)}{2}-h_l(t,t+1) \right]\right), -\gamma_{l,t+1} v_{l,t+1} \right) } \right] +
$$
$$
+\frac{1-p_l}{2} \left[ \max{\left( \left( v_{l,t+1} \left[\frac{h_l(t,t+1)+ s_{l,t+1}^{max}}{2}-h_l(t,t+1) \right]\right), -\gamma_{l,t+1} v_{l,t+1} \right) } \right],
$$
which is a function of the volume $v_{l,t+1}$,  if the trader believes that for options contract  $l$, the inclusion $l \in L_t^0$ will hold at the moment $t+1$ with the probability $p_l$, the inclusions $l \in L_t^+$ and $l \in L_t^-$ are equally possible, and she buys $v_{l,t+1}$ contracts $l$ as call options contracts,

d) $M_{op} FinRes_{l,t+1}(v_{j,t+1})=$
$$
=p_l \left[ \max{\left( \left( v_{l,t+1} \left[h_l(t,t+1)-h_l(t,t+1) \right] \right), -\gamma_{l,t+1} v_{l,t+1} \right) } \right] +
$$
$$
+\frac{1-p_l}{2} \left[ \max{\left( \left( v_{l,t+1} \left[h_l(t,t+1)-\frac{s_{l,t+1}^{min}+ h_l(t,t+1)}{2} \right]\right), -\gamma_{l,t+1} v_{l,t+1} \right) } \right] +
$$
$$
+\frac{1-p_l}{2} \left[ \max{\left( \left( v_{l,t+1} \left[h_l(t,t+1)-\frac{h_l(t,t+1)+ s_{l,t+1}^{max}}{2} \right]\right), -\gamma_{l,t+1} v_{l,t+1} \right) } \right],
$$
which is a function of the volume $v_{l,t+1}$,  if the trader believes that for options contract  $l$, the inclusion $l \in L_t^0$ will hold at the moment $t+1$ with the probability $p_l$, the inclusions $l \in L_t^+$ and $l \in L_t^-$ are equally possible, and she buys $v_{l,t+1}$  contracts $l$ as put options contracts,

Similarly to futures contracts, one can easily  notice that the third summand in the expression for $M_{op} FinRes_{l,t+1}(v_{j,t+1})$ in cases a) and b)  and the first summand in that of cases c) and d) equal 0, since if  $s_{l,t+1}=K_{l,t+1}+\gamma_{l,t+1}$, the equality $s_{l,t+1}- h_l(t,t+1)=0$ holds for $j\in {L}_t^+\cup {L}_t^- \cup {L}_t^0$ at the moment $t$.

It is clear that the trader may consider  buying options contracts for particular commodities  only if at least the expectations of the profits from the corresponding transactions are positive.  In just the same way it was mentioned for futures contracts, more exact estimates of the profit/loss values associated with these transactions - if, for instance,  the values of $s_{l,t+1}$ are normally distributed  on the segments $[s_{l,t+1}^{min},s_{l,t+1}^{max}]$ - can be applied by the interested trader to be more certain about the likely intervals  within which the profit/loss values caused by the above-mentioned transactions (associated with changing the values of $s_{l,t+1}$) may be at the moment $t+1$.

As in the case of futures contracts, one should mention that all the above reasoning and the formulae remain true for options contracts with respect to their underlying assets  other than commodities. However, if there is no cost for storing the underlying assets of these contracts (or any other costs associated with managing the underlying assets of these options contracts), this should be reflected in the strike prices for a corresponding set of options contracts.

One can easily be certain that the trader should solve an integer programming problem that is  identical in the structure to problem (4), formulated for the futures contracts, to find an optimal investment strategy of changing her portfolio that includes both standard securities and options contracts as the only derivative financial instruments. Indeed, though, formally, the functions describing the expected financial results associated with buying options contracts are the maximum functions of the numbers of contracts from the corresponding sets, the relation between the functions under the maximum sign depends  on the relation between the value of the premium and the difference between the expected market price and the strike price of each particular contract. Thus, the calculation of this difference based upon  the boarders within which the corresponding random variables ($\tilde u$ and $\tilde v$) change lets one determine which of the two linear functions under the maximum sign coincides with the maximum function, and this function is, in fact, the maximum function for all the values of the numbers of the options contracts under consideration. Certainly, the reduction to the above-mentioned integer programming problem is the case as long as the trader considers an investment strategy to be optimal if this strategy  maximizes the expectation of the increment of the trader welfare's value at the moment $t+1$.

In just the same way this  was done for the futures contracts that are present in the trader's portfolio as the only derivative financial instruments, in Situation 1, besides setting $\tilde I_t^+(av)=\emptyset$, $\tilde I_t^-(av)=\emptyset$, and $\tilde I_t^0(av)=\emptyset$, one should set  $L_t^+ (av)=\emptyset, \  L_t^- (av)=\emptyset$, and $L_t^0 (av)=\emptyset$ in the system of constraints of this integer programming problem, where  $L_t^+ (av), \  L_t^- (av)$, and $ L_t^0 (av)$ have the same meaning for the options contracts as do $J_t^+ (av), \  J_t^- (av)$, and  $J_t^0 (av)$ for the futures contracts, respectively.

Moreover, if besides standard securities, the trader deals with both futures contracts and options contracts, an integer programming problem similar in the structure to problem (4) can be formulated and solved for finding an optimal investment strategy of changing the trader's portfolio in this case for both Situation 1 and Situation 2 (see the beginning of Section 3) for any types of financial instruments being underlying assets of these futures and options contracts. .

\subsection {Model 2. The trader can numerically estimate only the areas in which price values of all the futures and options contracts of her interest may change.}

Unlike in Section 5.1, similar to how this was done in Section 4, first consider a trader who at the moment $t$ a) does not have any financial instruments in her portfolio, and b) has some amount of cash and considers buying financial instruments, including derivative ones,  or/and borrowing them from a broker to open short positions. 

To begin with, let the trader be interested in dealing with futures contracts only. As in Section 5.1, for the sake of definiteness, let us consider only futures contracts with commodities being their underlying assets. Thus, let the price of a futures contract $j$ at the moment $t+1$ equal $K_{j,t+1}+c_{j,t+1}$, where both $K_{j,t+1}$ and $c_{j,t+1}$ have the same definition as in Section 5.1.

As in considering Model 1 from Section 5.1, let the trader divide the whole set of the futures contracts that interest her into the subsets ${(J_t^f )}^+$, ${(J_t^f )}^-$, and ${(J_t^f )}^0$, which have the same meaning as do the subsets $J_t^+$, $J_t^-$, and $J_t^0$ in Model 1, respectively. However, unlike in Model 1, let the trader be  only able to  assume that a) the values of all the prices that the underlying assets of the futures contracts from the set ${(J_t^f )}^+$ will have at the moment $t+1$ will increase (with the probability ${(p^f )}^+ > 0.5$), b) the values of all the prices that the underlying assets  of the futures contracts from the set ${(J_t^f )}^-$ will have at the moment $t+1$ will decrease (with the probability ${(p^f )}^- > 0.5$), and c) the values of all the prices that the underlying assets of the futures contracts from the set ${(J_t^f)}^0$ will have at the moment $t+1$ may increase and may decrease with the same probability equaling ${(p^f )}^0=0.5$. (See the reasoning regarding this feature of the prices with respect to standard securities on page 24.) As before (see Section 5.1), it is assumed that the prices for both futures contracts and their underlying assets either change in the same direction, or do not change. 

Consider futures contracts forming the set ${(J_t^f )}^+$ at the moment $t$. Let a) ${(v_{t+1}^f)}^+ \in R_+^{|{(J_t^f )}^+ |}$ be the vector whose components are the numbers of futures contracts that the trader plans to buy at the moment $t$, b) ${(s_{t+1}^f)}^+ \in R_+^{|{(J_t^f )}^+ |}$ be the vector of the values of the market prices which futures contracts from the set ${(J_t^f )}^+$ may assume at the moment $t+1$ if the trader correctly determines (with the probability ${(p^f )}^+$) the direction of changing these values, and c) ${(u_{t+1}^f)}^+ \in R_+^{|{(J_t^f )}^+ |}$ be the vector of the values of the market  prices which futures contracts from the set ${(J_t^f )}^+$ may assume at the moment $t+1$ if the trader incorrectly determines (with the probability $1-{(p^f )}^+$) the direction of changing these values. 

For the same reasons as those presented in Section 4, one may consider that the inclusions ${(v_{t+1}^f)}^+ \in {(V_{t+1}^f)}^+ \subset R_+^{|{(J_t^f )}^+ |}$, ${(s_{t+1}^f)}^+ \in {(S_{t+1}^f)}^+ \subset R_+^{|{(J_t^f )}^+ |}$, ${(u_{t+1}^f)}^+ \in {(U_{t+1}^f)}^+ \subset R_+^{|{(J_t^f )}^+ |}$ hold. Here ${(S_{t+1}^f)}^+$ and ${(U_{t+1}^f)}^+$ are (non-empty) convex polyhedra described by compatible systems of linear equations and inequalities, and $V_{t+1}^f$ is a non-empty subset of a convex polyhedron (described by a compatible system of linear equations and inequalities) consisting of the vectors all whose coordinates are non-negative integers. As before (see Section 4), it is natural to assume that the set $V_{t+1}^f$ contains the zero vector.

The trader's best investment strategy in dealing with the futures contracts of her interest from the set ${(J_t^f )}^+$ in her game with the stock exchange can be found by solving the problem
\begin{equation*}
\begin{split}
& {(p^f )}^+ \min_{ {(s_{t+1}^f )}^+ \in {(S_{t+1}^f )}^+ }{  \left( \langle{(v_{t+1}^f )}^+,{(s_{t+1}^f )}^+ \rangle - \langle {(K_{t+1}^f )}^+,{(v_{t+1}^f )}^+ \rangle - \langle {(c_{t+1}^f )}^+,{(v_{t+1}^f)}^+ \rangle \right) }+ \\
& +\left( 1-{(p^f )}^+ \right) \min_{ {(u_{t+1}^f )}^+ \in {(U_{t+1}^f )}^+ }{  \left( \langle{(v_{t+1}^f )}^+,{(u_{t+1}^f )}^+ \rangle - \langle {(K_{t+1}^f )}^+,{(v_{t+1}^f )}^+ \rangle \right. }- \\
& - \left. \langle {(c_{t+1}^f )}^+,{(v_{t+1}^f)}^+ \rangle \right) \to \max_{ {(v_{t+1}^f )}^+ \in {(V_{t+1}^f )}^+ }, \\
\end{split}
\end{equation*}
where ${(K_{t+1}^f )}^+=\left({(K_{1,t+1}^f )}^+,...,{(K_{|{(J_t^f )}^+ |,t+1}^f )}^+  \right) \in R_+^{|(J_t^f )^+  |}$,\\ ${(v_{t+1}^f)}^+=\left({(v_{1,t+1}^f )}^+,...,{(v_{|{(J_t^f )}^+ |,t+1}^f )}^+  \right) \in {(V_{t+1}^f )}^+ \subset  R_+^{|(J_t^f )^+  |}$, \\ ${(c_{t+1}^f )}^+=\left({(c_{1,t+1}^f )}^+,...,{(c_{|{(J_t^f )}^+ |,t+1}^f )}^+  \right) \in R_+^{|(J_t^f )^+  |}$. 

One can easily be certain that the equality 
\begin{equation*}
\begin{split}
& \max_{ {(v_{t+1}^f )}^+ \in {(V_{t+1}^f )}^+ }{ \left[ {(p^f )}^+ \min_{ {(s_{t+1}^f )}^+ \in {(S_{t+1}^f )}^+ }{  \left( \langle {(v_{t+1}^f )}^+,{(s_{t+1}^f )}^+ \rangle - \langle {(K_{t+1}^f )}^+,{(v_{t+1}^f )}^+ \rangle - \right. } \right. } \\
& - \left. \langle {(c_{t+1}^f )}^+, {(v_{t+1}^f )}^+ \rangle \right) + \left. \left( 1-{(p^f )}^+ \right) \min_{ {(u_{t+1}^f )}^+ \in {(U_{t+1}^f )}^+ }{  \left( \langle{(v_{t+1}^f )}^+,{(u_{t+1}^f )}^+ \rangle - \right. }\right.\\
& - \left. \left. \langle {(K_{t+1}^f )}^+,{(v_{t+1}^f )}^+ \rangle - \langle {(c_{t+1}^f )}^+,{(v_{t+1}^f )}^+ \rangle \right) \right]= \\
& = \max_{ {(v_{t+1}^f )}^+ \in {(V_{t+1}^f )}^+ }{ \left[ \min_{ {(s_{t+1}^f )}^+ \in {(S_{t+1}^f )}^+ }{  \left( \langle{(v_{t+1}^f )}^+, H^{|{(J_t^f )}^+ |} \left( {(p^f )}^+ \right) {(s_{t+1}^f )}^+ \rangle -\right. } \right. } \\
& \left. -  {(p^f )}^+ \langle {(K_{t+1}^f )}^+,{(v_{t+1}^f )}^+ \rangle  - {(p^f )}^+ \langle {(c_{t+1}^f )}^+, {(v_{t+1}^f )}^+ \rangle \right) + \\
& + \left.  \min_{ {(u_{t+1}^f )}^+ \in {(U_{t+1}^f )}^+ }{  \left( \langle{(v_{t+1}^f )}^+, H^{|{(J_t^f )}^+ |} \left( 1-{(p^f )}^+ \right) {(u_{t+1}^f )}^+ \rangle  \right. } \right. \\
& \left. \left. - \left( 1-{(p^f )}^+ \right) \langle {(K_{t+1}^f )}^+,{(v_{t+1}^f )}^+ \rangle - \left( 1-{(p^f )}^+ \right) \langle {(c_{t+1}^f )}^+, {(v_{t+1}^f )}^+ \rangle \right) \right] \\
\end{split}
\end{equation*}
holds, and since the vectors ${(s_{t+1}^f )}^+$ from the set ${(S_{t+1}^f )}^+$ and the vectors ${(u_{t+1}^f )}^+$ from the set ${(U_{t+1}^f )}^+$ are chosen independently of each other, the equality
\begin{equation*}
\begin{split}
& \max_{ {(v_{t+1}^f )}^+ \in {(V_{t+1}^f )}^+ }{ \left[ \min_{ {(s_{t+1}^f )}^+ \in {(S_{t+1}^f )}^+ }{  \left( \langle{(v_{t+1}^f )}^+, H^{|{(J_t^f )}^+ |} \left( {(p^f )}^+ \right) {(s_{t+1}^f )}^+ \rangle -  \right. } \right. } \\
& \left. -{(p^f )}^+ \langle {(K_{t+1}^f )}^+,{(v_{t+1}^f )}^+ \rangle -{(p^f )}^+ \langle {(c_{t+1}^f )}^+,{(v_{t+1}^f)}^+ \rangle  \right) +\\
& + \left.  \min_{ {(u_{t+1}^f )}^+ \in {(U_{t+1}^f )}^+ }{  \left( \langle{(v_{t+1}^f )}^+, H^{|{(J_t^f )}^+ |}  \left( 1-{(p^f )}^+ \right) {(u_{t+1}^f )}^+ \rangle  \right. } \right. \\
& \left. \left. - \left( 1-{(p^f )}^+ \right) \langle {(K_{t+1}^f )}^+,{(v_{t+1}^f )}^+ \rangle - \left( 1-{(p^f )}^+ \right) \langle {(c_{t+1}^f )}^+,{(v_{t+1}^f)}^+ \rangle \right) \right]= \\
& = \max_{ {(v_{t+1}^f )}^+ \in {(V_{t+1}^f )}^+ }{ \left[ \min_{ {(w_{t+1}^f )}^+ \in {(\theta_{t+1}^f )}^+ }{ \langle {(v_{t+1}^f )}^+, H^{2|(J_t^f )^+ |} \left( {(p^f )}^+ , \left( 1-{(p^f )}^+ \right) \right) {(w_{t+1}^f )}^+ \rangle } \right. } \\
& \left. - \langle {(K_{t+1}^f )}^+,{(v_{t+1}^f )}^+ \rangle  - \langle {(c_{t+1}^f )}^+,{(v_{t+1}^f)}^+ \rangle\right] ,\\
\end{split}
\end{equation*}
where ${(w_{t+1}^f )}^+=({(s_{t+1}^f )}^+,{(u_{t+1}^f )}^+)$, ${(\theta_{t+1}^f )}^+={(S_{t+1}^f )}^+ \times {(U_{t+1}^f )}^+$, \\ $H^{2|{(J_t^f )}^+ |} \left( {(p^f )}^+ , \left( 1-{(p^f )}^+ \right) \right)$ is the matrix formed by ascribing the matrix \\ $H^{|{(J_t^f )}^+ |} \left( 1-{(p^f )}^+ \right)  $ to the matrix $H^{|{(J_t^f )}^+ |} \left( {(p^f )}^+ \right)$ from the right so that \\ $H^{2|{(J_t^f )}^+ |} \left( {(p^f )}^+ , \left( 1-{(p^f )}^+ \right) \right) = H^{|{(J_t^f )}^+ |} \left( {(p^f )}^+ \right) H^{|{(J_t^f )}^+ |}  \left( 1-{(p^f )}^+ \right) $.

Thus, the trader's best investment strategy with respect to the futures contract of her interest from the set ${(J_t^f) }^+$ in the game with the stock exchange can be found by solving the problem 
\begin{equation*}
\begin{split}
& \min_{ {(w_{t+1}^f )}^+ \in {(\theta_{t+1}^f )}^+ }{ \left[ \langle {(v_{t+1}^f )}^+, H^{2|(J_t^f )^+ |} \left( {(p^f )}^+ , \left( 1-{(p^f )}^+ \right) \right) {(w_{t+1}^f )}^+ \rangle \right.} - \\
& - \left. \langle {(K_{t+1}^f )}^+,{(v_{t+1}^f )}^+ \rangle  - \langle {(c_{t+1}^f )}^+,{(v_{t+1}^f)}^+ \rangle \right]\to \max_{ {(v_{t+1}^f )}^+ \in {(V_{t+1}^f )}^+ }. \\
\end{split}
\end{equation*}

Analogously, the trader's best investment strategy with respect to the futures contracts of her interest from the sets ${(J_t^f) }^-$ and ${(J_t^f )}^0$ in the game with the stock exchange can be found by solving the problems
\begin{equation*}
\begin{split}
& \min_{ {(w_{t+1}^f )}^- \in {(\theta_{t+1}^f )}^- }{ \left[ \langle {(v_{t+1}^f )}^-, H^{2|(J_t^f )^- |} \left( {(p^f )}^- , \left( 1-{(p^f )}^- \right) \right) {(w_{t+1}^f )}^- \rangle \right.} - \\
& - \left. \langle {(K_{t+1}^f )}^-,{(v_{t+1}^f )}^- \rangle  - \langle {(c_{t+1}^f )}^-,{(v_{t+1}^f )}^- \rangle \right]\to \max_{ {(v_{t+1}^f )}^- \in {(V_{t+1}^f )}^- }, \\
\end{split}
\end{equation*}
and 
\begin{equation*}
\begin{split}
& \min_{ {(w_{t+1}^f )}^0 \in {(\theta_{t+1}^f )}^0 }{ \left[ \langle {(v_{t+1}^f )}^0, H^{2|(J_t^f )^0 |} \left( {(p^f )}^0 , \left( 1-{(p^f )}^0 \right) \right) {(w_{t+1}^f )}^0 \rangle \right.} - \\
& - \left. \langle {(K_{t+1}^f )}^0,{(v_{t+1}^f )}^0 \rangle  - \langle {(c_{t+1}^f )}^0,{(v_{t+1}^f )}^0 \rangle\right] \to \max_{ {(v_{t+1}^f )}^0 \in {(V_{t+1}^f )}^0 }, \\
\end{split}
\end{equation*}
respectively, where ${(K_{t+1}^f )}^-=\left({(K_{1,t+1}^f )}^-,...,{(K_{|{(J_t^f )}^- |,t+1}^f )}^-  \right) \in R_+^{|(J_t^f )^-  |}$, ${(c_{t+1}^f )}^-=\left({(c_{1,t+1}^f )}^-,...,{(c_{|{(J_t^f )}^- |,t+1}^f )}^-  \right) \in R_+^{|(J_t^f )^-  |}$, ${(v_{t+1}^f)}^-=\left({(v_{1,t+1}^f )}^-,...,{(v_{|{(J_t^f )}^- |,t+1}^f )}^-  \right) \in {(V_{t+1}^f )}^-\subset R_+^{|(J_t^f )^-  |}$,
 ${(w_{t+1}^f )}^-=({(s_{t+1}^f )}^-,{(u_{t+1}^f )}^-)$, ${(\theta_{t+1}^f )}^-={(S_{t+1}^f )}^- \times {(U_{t+1}^f )}^-$, \\ $H^{2|{(J_t^f )}^- |} \left( {(p^f )}^- , \left( 1-{(p^f )}^- \right) \right)=H^{|{(J_t^f )}^- |} \left( {(p^f )}^- \right) H^{|{(J_t^f )}^- |} \left( \left( 1-{(p^f )}^- \right) \right) $, \\ ${(K_{t+1}^f )}^0=\left({(K_{1,t+1}^f )}^0,...,{(K_{|{(J_t^f )}^0 |,t+1}^f )}^0  \right) \in R_+^{|(J_t^f )^0  |}$, ${(c_{t+1}^f )}^0=\left({(c_{1,t+1}^f )}^0,...,{(c_{|{(J_t^f )}^0 |,t+1}^f )}^0  \right) \\ \in R_+^{|(J_t^f )^0  |}$, ${(v_{t+1}^f)}^0=\left({(v_{1,t+1}^f )}^0,...,{(v_{|{(J_t^f )}^0 |,t+1}^f )}^0  \right) \in {(V_{t+1}^f )}^0\subset R_+^{|(J_t^f )^0  |}$, \\  ${(w_{t+1}^f )}^0=({(s_{t+1}^f )}^0,{(u_{t+1}^f )}^0)$, ${(\theta_{t+1}^f )}^0={(S_{t+1}^f )}^0 \times {(U_{t+1}^f )}^0$, \\ $H^{2|{(J_t^f )}^0 |} \left( {(p^f )}^0 , \left( 1-{(p^f )}^0 \right) \right)=H^{|{(J_t^f )}^0 |} \left( {(p^f )}^0 \right) H^{|{(J_t^f )}^0 |}  \left( 1-{(p^f )}^0 \right) $.

Here, the probabilities $(p^f )^-$  and $(p^f )^0$  have the meaning similar to that of $(p^f )^+$ with respect to the sets of the futures contracts  $(J_t^f )^-$ and $(J_t^f )^0$, respectively, and the matrices $H^{2|{(J_t^f )}^- |} \left( {(p^f )}^- , \left( 1-{(p^f )}^- \right) \right)$   and  $H^{2|{(J_t^f )}^0 |} \left( {(p^f )}^0 , \left( 1-{(p^f )}^0 \right) \right)$ are formed by ascribing the matrices $ H^{|{(J_t^f )}^- |} \left( 1-{(p^f )}^- \right)$ and $ H^{|{(J_t^f )}^0 |} \left( 1-{(p^f )}^0 \right)$ to the matrices $H^{|{(J_t^f )}^- |} \left( {(p^f )}^- \right)$ and $H^{|{(J_t^f )}^0 |} \left( {(p^f )}^0 \right)$ from the right, respectively. 
Also, similar to the assumptions on the set ${(V_{t+1}^f )}^+$, it is assumed that a) ${(V_{t+1}^f )}^-$ and ${(V_{t+1}^f )}^0$ are non-empty subsets of convex polyhedra (each of which is described by a compatible system of linear equations and inequalities) consisting of the vectors all whose coordinates are non-negative integers, and b) each of these two sets contains the zero vector. One should bear in mind that similar to the assumptions made in Sections 4 and 5, to deal with the futures contracts from the sets $(J_t^f)^+$, $(J_t^f)^-$,  and $(J_t^f)^0$, the trader should either buy them with her own cash or borrow them from a broker (or from brokers) to open short positions. It is assumed that the financial possibilities of the trader to borrow these derivative financial instruments (and then sell them short) are reflected in the above-mentioned equations and inequalities describing the sets $(V_{t+1}^f)^+$, $(V_{t+1}^f)^-$, and $(V_{t+1}^f)^0$.

Let now $v_{t+1}^f=({(v_{t+1}^f )}^+,{(v_{t+1}^f )}^-,{(v_{t+1}^f )}^0 ) \in M_t^f\subseteq {(V_{t+1}^f )}^+ \times {(V_{t+1}^f )}^- \times {(V_{t+1}^f )}^0$, $w_{t+1}^f=({(w_{t+1}^f )}^+,{(w_{t+1}^f )}^-,{(w_{t+1}^f )}^0 ) \in \theta_t^f\subseteq {(\theta_{t+1}^f )}^+ \times {(\theta_{t+1}^f )}^- \times {(\theta_{t+1}^f )}^0$, and let the matrix $D_t^f$ have the form 
$$
{\scriptsize D_t^f=\begin{pmatrix} H^{2|{(J_t^f )}^+ |} \left( {(p^f )}^+ , \left( 1-{(p^f )}^+ \right) \right) & 0 & 0 \\ 0 & H^{2|{(J_t^f )}^- |} \left( {(p^f )}^- , \left( 1-{(p^f )}^- \right) \right) & 0 \\ 0 & 0 & H^{2|{(J_t^f )}^0 |} \left( {(p^f )}^0 , \left( 1-{(p^f )}^0 \right) \right)   \end{pmatrix}} .
$$

Then the trader's best investment strategy with respect to the futures contracts of her interest consists of choosing the vector $v_{t+1}^f$ at which
$$
\max_{v_{t+1}^f \in M_t^f}{ \left [\min_{ w_{t+1}^f  \in \theta_{t+1}^f  }{ \left[ \langle v_{t+1}^f , D_t^f w_{t+1}^f  \rangle - \langle K_{t+1}^f ,v_{t+1}^f  \rangle   - \langle c_{t+1}^f ,v_{t+1}^f  \rangle \right] }  \right]}
$$
is attained, where the equalities $K_{t+1}^f=({(K_{t+1}^f )}^+,{(K_{t+1}^f )}^-,{(K_{t+1}^f )}^0 )$, and $c_{t+1}^f=({(c_{t+1}^f )}^+,{(c_{t+1}^f )}^-,{(c_{t+1}^f )}^0 )$ hold.

Let us now assume that at the moment $t$, the trader already possesses  futures contracts in her portfolio that form the vector $v_t^f=((v_t^f)^+, (v_t^f)^-, (v_t^f)^0)$.  In just the same way this was done in Section 4, one can easily  show that the trader's best investment strategy (with respect to the futures contracts) consists of choosing the vector $v_{t+1}^f$  at which 
$$
\max_{v_{t+1}^f \in M_t^f}{ \left [\min_{ w_{t+1}^f  \in \theta_{t+1}^f  }{ \left[ \langle v_{t+1}^f , D_t^f w_{t+1}^f  \rangle - \langle K_{t+1}^f ,v_{t+1}^f  \rangle   - \langle c_{t+1}^f ,v_{t+1}^f  \rangle + \langle q^f ,w_{t+1}^f  \rangle  \right] }  \right]}
$$
where $q^f=\big( (p^f)^+(v_t^f)^+, \  ((1-(p^f)^+)(v_t^f)^+,  \ ( p^f)^-(v_t^f)^-, \  ((1-(p^f)^-)(v_t^f)^-$,      
$(p^f)^0(v_t^f)^0, \  ((1-(p^f)^0)(v_t^f)^0\big)$, is attained.

The same reasoning is applicable to the options contracts. Particularly, let at the moment $t$, the trader have contracts in her portfolio that form the vector $v_t^{op}=\big((v_t^{op})^+, (v_t^{op})^-, (v_t^{op})^0   \big)$. Further,  let $(p^{op})^+$, $(p^{op})^-$, and $(p^{op})^0$ be the probabilities with which the trader assumes in which direction the values of the prices that the underlying assets of the options contracts may change at the moment $t+1$ will change at the moment $t+1$. Then, one can show  that the trader's best investment strategy in the options contracts of her interest consists of choosing the vector $v_{t+1}^{op}$ at which 
$$
\max_{v_{t+1}^{op} \in M_t^{op}}{ \left [\min_{ w_{t+1}^{op}  \in \theta_{t+1}^{op}  }{ \left[ \langle v_{t+1}^{op} , D_t^{op} w_{t+1}^{op}  \rangle - \langle K_{t+1}^{op} ,v_{t+1}^{op}  \rangle - \langle \gamma_{t+1}^{op} ,v_{t+1}^{op} \rangle  + \langle q^{op} ,w_{t+1}^{op}  \rangle  \right] }  \right]}
$$
is attained, where $\gamma_{t+1}^{op}$ is a vector whose components are the values of the premiums for the corresponding options contracts (put options contracts and call options contracts). Here, for the options contracts, the vectors $v_{t+1}^{op}$, $w_{t+1}^{op}$, $K_{t+1}^{op}$, $\gamma_{t+1}^{op}$, $v_t^{op}$, $q^{op}$,  the matrix $D_t^{op}$, and the sets $M_t^{op}$ and $\theta_{t+1}^{op}$ have the meaning completely identical to that of the vectors $v_{t+1}^f$, $w_{t+1}^f$,  $K_{t+1}^{f}$, $c_{t+1}^f$, $v_t^{f}$, $q^{f}$, the matrix $D_t^f$, and the sets $M_t^f$ and $\theta_{t+1}^f$, respectively. To be certain about this one should bear in mind  that in determining the best strategy of dealing with the options contracts the trader should buy and keep only those ones for which the expectation of the profit of executing the contract is not negative, which makes this strategy completely identical to the strategy of dealing with the futures contracts.

Finally, let $x=\left( x_t, v_{t+1}^f,v_{t+1}^{op} \right)$,  $w=\left( w_{t+1},w_{t+1}^f,w_{t+1}^{op} \right)$, and

$$
K=\left( 0,K_{t+1}^f+c_{t+1}^f,K_{t+1}^{op}+\gamma_{t+1}^{op} \right), \  {D=\begin{pmatrix} D_t& 0 & 0 \\ 0 & D_t^f& 0 \\ 0 & 0 & D_t^{op} \end{pmatrix}} .
$$
Then finding the best investment strategies of the trader dealing with all the securities and derivative financial instruments (in the form of futures contracts and options contracts) that she has in her portfolio at the moment $t$ is reducible to finding 

\begin{equation}
\max_{x \in M(t)} {\left[ \min_{w \in W(t+1)}{\langle x,D w \rangle} -  \langle K,x \rangle + \langle \pi,w \rangle \right]},
\end{equation} 
where $M(t)\subseteq M_t \times M_t^f \times M_t^{op}$, $W(t+1)\subseteq \theta_{t+1} \times \theta_{t+1}^f \times \theta_{t+1}^{op}$, and ${M}(t)$ and $W(t+1)$ are (non-empty) convex polyhedra described by compatible systems of linear equations and inequalities in the spaces of corresponding dimensions, $\pi=(q,q^f,q^{op})$,  $M(t)=\{x \ge 0: G(t) x \ge g(t), x \in Q_+^n \times Q_+^{|J_t^f|} \times Q_+^{|J_t^{op}|} \}$, $W(t+1)=\{w \ge 0: F(t+1) w \ge f(t+1)\}$, $|J_t^f|=|{(J_t^f)}^+|+|{(J_t^f)}^-|+|{(J_t^f)}^0|$, $|J_t^{op}|=|{(J_t^{op})}^+|+|{(J_t^{op})}^-|+|{(J_t^{op})}^0|$, and $Q_+^{|J_t^f|}$, $Q_+^{|J_t^{op}|}$ are direct products of $|J_t^f|$ and $|J_t^{op}|$ sets $Q_+$, respectively. 

In just the same way it was shown for problem (3), finding this maximin is reducible to solving the following mixed programming problem:
$$ 
\langle f(t+1),z \rangle -\langle K,x \rangle \to \max_{ \{(x,z) \ge 0 : \ G(t)x \ge g(t), \ z F(t+1) \le x D + \pi, \ x \in Q_+^n \times Q_+^{|J_t^f|} \times Q_+^{|J_t^{op}|} \}},
$$
where  $z$ is a vector of the corresponding size (equaling the number of rows in the matrix $F(t+1)$), whereas finding an upper bound of this maximin (corresponding to the case in which all the components of the vector $x$ are considered to be non-negative, real numbers) is reducible to solving linear programming problems forming a dual pair
$$
\langle \pi,w \rangle + \langle -g(t),\tau \rangle \to \min_{(w,\tau) \in  Q'},
$$
$$
\langle f(t+1),z \rangle - \langle K,x \rangle \to \max_{(z,x) \in P'},
$$
where $Q'=\{(w,\tau) \ge 0: \tau G(t) \le K - D w, F(t+1) w \ge f(t+1) \}$, $P'=\{(z ,x) \ge 0: z F(t+1) \le x D+\pi, G(t) x \ge g(t) \}$.

Remark 6. One should bear in mind that, generally, the set $M(t)$ is not a direct product of the sets $M_t$, $M_{t}^f$, and  $M_{t}^{op}$, since there may be at least one constraint binding all the vector variables $x_t$, $v_{t+1}^f$, and $v_{t+1}^{op}$ , which reflects the fact that the trader may consider all the securities and derivative financial instruments to have an equal potential to affect the value of her portfolio. However, in just the same way this was done in the course of proving the Theorem (see Section 4) with respect to the set $M_t$, one can be certain that if each of the sets $M_t, \ M_t^f$, and $M_t^{op}$ or/and the set $M(t)$ have at least one constraint binding all the variables from these sets, finding the best trader's investment strategy in both standard securities and derivative financial instruments is still reducible to solving problem (5), and finding an upper bound of the maximin in (5) is reducible to finding saddle points in the above antagonistic game. 

Remark 7. As before (see Remark 3), one should notice that solving problem (5) determines only the trader's best investment strategy at the moment $t$, and it does not determine the financial result of applying this strategy, since the goal function in maximin problem (5) does not take into consideration certain components of the trader's welfare at the moment $t+1$. Besides such components mentioned in Remark 3, there could be additional ones associated with dealing with futures and options contracts.

\section{Illustrative example }
\label{sec:6}

The aim of this section is to illustrate how a price-taking trader may make decisions on forming her portfolio out of standard securities when at the moment $t$, she can make no assumptions on probability distributions of the values of the share prices that (standard) securities of her interest may have at the moment $t +1$. As shown in Section 4, if, nevertheless, the trader can estimate the areas in which the values of the share prices of these securities may change at the moment $t +1$, game models of a special kind may help the trader calculate her optimal investment strategies at the moment $t$ (aimed at increasing the value of her portfolio at the moment $t +1$). Particularly, the present section illustrates how game models described in Section 4 are formed, and how linear programming problems are solved to find saddle points in one of these games with the use of standard software packages for solving linear programming problems. To this end, a numerical example is considered in one of the situations mentioned in the text of Section 4. In this example, a trader possessing only a certain amount of cash forms a new portfolio out of (standard) securities of her interest that are traded in a stock exchange at the moment $t$ based upon her expectations and beliefs of how the values of the share prices of these securities will change at the moment $t+1$. In the description of this example, the notation from Section 4 is used.

As was pointed out earlier (see Section 4), solving the above-mentioned linear programming problems lets the trader determine only an upper bound of the expected increment value of her portfolio. However, as mentioned in Remark 2 (see Section 4), considering volumes (numbers of shares) of securities to be bought and sold as non-negative, real numbers is in line with traditional approaches exercised in theoretical studies of stock exchanges. Moreover, even from a practical viewpoint-- when the number of different securities that interest the trader is large--solving mixed programming problems to calculate the exact (integer) numbers of shares for each (standard) security to buy and to sell to secure the exact value of the expected increment of the trader portfolio's value may present substantial difficulties. If this is the case, finding the exact numbers of shares of the above-mentioned standard securities will hardly interest the traders in making decisions on managing their investment portfolios.

In just the same way as in Section 4, in the illustrative example to follow, the optimality of the trader's investment strategy is considered in the sense of the value of her portfolio at the moment $t+1$.
 
{\it Illustrative Example.}

Consider a trader who plans to interact with a stock exchange by forming a portfolio of financial instruments. Let us assume that at the moment $t$, the trader  a) is interested in only two particular standard securities that are traded in the stock exchange (so that $N=\{1,2\}$ for this trader), b) does not have a portfolio of financial instruments traded in this stock exchange (so that $v_1=v_2=0$ for this trader), c) has the amount  of cash equaling $m_t=10,000.00$ financial units, for instance, US dollars, and d) has a broker who is ready to provide her with a credit. It is assumed that  a) the credit leverage equals $k_t=0.5$ for borrowing standard securities from the broker  to let the trader open short positions there, and b) the broker is ready to offer the trader securities from the set $N$ (which are the only securities that interest the trader at the moment $t$) to borrow. 

Let at the moment $t$, the values of the share prices equal $s_{1,t}=100$ US dollars for security 1 and  $s_{2,t}=50$ US dollars for security 2. Further,  let the trader believe that the value of the share price of security 1 will increase at the moment $t+1$, whereas the value of the share price of security 2 will decrease at the moment $t+1$ so that $I_t^+=\{1\}$, $I_t^-=\{2\}$ and $I_t^0=\emptyset$. Moreover, let the trader be confident that the price values of the above two securities will change the way she believes they will with the probabilities $p^+=0.6$ and  $p^-=0.7$, respectively. Finally, let the trader adhere to Approach 3 to the understanding of what should be viewed as the set  $X_t^-$ (see Section 4). 

The first step in finding the trader's best investment strategy is to find out how much of additional cash  she can have as a result of borrowing securities from the broker and selling them short  at the moment $t$.  Further, since security 2 is the only one that the trader should be interested in borrowing from the broker (hoping that the share price value of this security will decrease in the future), the trader should  determine how many shares of security 2 she should borrow to sell them at the moment $t$. It is obvious that since the total cost of the shares of security 2 that the trader can borrow from the broker at the moment $t$ cannot exceed 5,000,00, and the share price value of one share of security 2 equals  50.00 US dollars at the moment $t$, the maximum number of shares of  security 2 that the trader can borrow equals 100.

Let $x_1^+$ and $x_2^-$ be the numbers of shares of security 1 and security 2, respectively, that the trader plans to have in her portfolio at the moment $t+1$, which means that the trader plans to buy $x_1^+$ shares of security 1 and $x_2^-$ shares of security 2 at the moment $t$. According to the description of the trader's actions in forming her portfolio at the moment $t$, presented in Section 4, the trader should estimate how many shares and  of which securities from the set $N$ she should have at the moment $t+1$ that would maximize the value of her portfolio at the moment $t+1$. It is clear that in this particular example, one should expect the trader not to buy any shares of security 2. However, one should bear in mind that, generally, despite the fact that at the moment $t$, the trader borrows (from the broker) at least some securities from the set $X_t^-$  to receive additional cash, it may happen that the portfolio with the maximum value at the moment $t+1$ may include at least some of the securities that were borrowed  at the moment $t$ (security 2 in the example under consideration). Thus, for the purpose of illustrating the trader's actions in the general case, buying both shares of security 1 and shares of security 2 are considered.

As mentioned in Section 4, the trader determines the description of the sets $X_t^+$ and $X_t^-$ at her own discretion, so  let the trader describe them with the following system of linear inequalities (proceeding from her financial abilities at the moment $t$):
$$ 
x_1^+ \ge 0;
$$
$$
 x_2^- \ge 0;
$$
$$
s_{1,t} x_1^++ s_{2,t}  x_2^-\le m_ t+5000.
$$
Here, the first two of the above three inequalities reflect the condition of non-negativity of the transaction volumes, whereas the third one  puts the limit on the volume of securities 1 and 2  that the trader can buy with her own money and with the money to be received from selling at the moment $t$  shares of security 2 (borrowed from the broker).

Thus,  $M_t=\{x_t \in R_+^2: B_t x_t \ge d_t \}$, the set of the volumes of securities 1 and 2 that the trader can buy at the moment $t$, where $x_t=(x_{1,t}^+, x_{2,t}^-)=(x_1^+, x_2^-)$,  is such that 
$$
B_t=\begin{pmatrix} 1 &  0 \\ 0 & 1 \\ -s_{1,t} & -s_{2,t}  \end{pmatrix} =\begin{pmatrix} 1 &  0 \\ 0 & 1 \\ -100 & -50  \end{pmatrix},   
 d_t=\begin{pmatrix}   0 \\ 0  \\ -m_t-5000  \end{pmatrix} =\begin{pmatrix}   0 \\ 0  \\ -15000   \end{pmatrix},
$$
and the inequality
$$
 \begin{pmatrix} 1 &  0 \\ 0 & 1  \\ -100 & -50  \end{pmatrix}(x_1^+,x_2^-)\ge \begin{pmatrix}   0 \\ 0  \\ -15000   \end{pmatrix}
$$
holds (see Section 4). To simplify the notation in the description of the illustrative example to follow, let also
$$
 y_{1,t+1}^+=y_1^+, \ y_{2,t+1}^-=y_2^-, \ z_{1,t+1}^+=z_1^+, \ z_{2,t+1}^-=z_2^-.
$$

While $x_t=(x_1^+,x_2^- )$ is the vector of the trader's strategies in her game with the stock exchange (see Section 4), the strategies of the stock exchange can be represented by the vector $w_{t+1}=(y_1^+,z_1^+,y_2^-,z_2^- )$ whose components are the (expected) values of the share prices of securities 1 and 2 at the moment $t+1$. Here, $y_1^+$, $y_2^-$ are the (expected) values of the share prices of securities 1 and 2 at the moment $t+1$, respectively,  if the trader has correctly predicted directions in which the values of these two securities will change, and $z_1^+$, $z_2^-$ are the (expected) values of the share prices of securities 1 and 2 at the moment $t+1$, respectively,  if the trader has failed to predict these directions correctly.  

Let the trader believe that the maximum and the minimum values of the share prices of securities 1 and 2 at the moment $t+1$ will be $s_{1,t+1}^{max}=115$, $s_{2,t+1}^{max}=65$, $s_{1,t+1}^{min}=75$, $s_{2,t+1}^{min}=35$ US dollars, respectively. Further, let the trader put stop orders on the above maximum and minimum price values of securities 2 and 1 at the moment $t+1$ to avoid unexpected financial losses associated with increasing the value of the share price of security 2 beyond $s_{2,t+1}^{max}$ and with decreasing the value of the share price of security 1 below $s_{1,t+1}^{min}$, respectively. Then, $\theta_{t+1}=\{w_{t+1} \in R_+^4:A_t w_{t+1} \ge b_t\}$, the set of possible strategies of the stock exchange in the game, can be described by the system of inequalities
$$
s_{1,t} \le y_1^+ \le s_{1,t+1}^{max},
$$
$$
s_{1,t+1}^{min} \le z_1^+ \le s_{1,t},
$$
$$
s_{2,t+1}^{min} \le y_2^- \le s_{2,t},
$$
$$
s_{2,t} \le z_2^- \le s_{2,t+1}^{max},
$$
which takes the following vector-matrix form: 
$$
A_t=\begin{pmatrix} 1 & 0 & 0 & 0 \\ -1 & 0 & 0 & 0 \\ 0 & 1 & 0 & 0 \\ 0 & -1 & 0 & 0 \\ 0 & 0 & 1 & 0 \\ 0 & 0 & -1 & 0 \\ 0 & 0 & 0 & 1 \\ 0 & 0 & 0 & -1 \\   \end{pmatrix}, b_t=\begin{pmatrix}   s_{1,t} \\ -s_{1,t+1}^{max} \\ s_{1,t+1}^{min} \\ -s_{1,t} \\ s_{2,t+1}^{min} \\ -s_{2,t} \\ s_{2,t} \\ -s_{2,t+1}^{max} \end{pmatrix} =\begin{pmatrix}   100 \\ -115  \\ 75 \\ -100 \\ 35 \\ -50 \\ 50  \\ -65  \end{pmatrix},
$$ 
$$
\begin{pmatrix} 1 & 0 & 0 & 0 \\ -1 & 0 & 0 & 0 \\ 0 & 1 & 0 & 0 \\ 0 & -1 & 0 & 0 \\ 0 & 0 & 1 & 0 \\ 0 & 0 & -1 & 0 \\ 0 & 0 & 0 & 1 \\ 0 & 0 & 0 & -1 \\   \end{pmatrix}(y_1^+, z_1^+,y_2^-,z_2^-)\ge \begin{pmatrix}   100 \\ -115  \\ 75 \\ -100 \\ 35 \\ -50 \\ 50  \\ -65  \end{pmatrix}.
$$
According to the Theorem (see Section 4), the payoff function of the game between the trader and the stock exchange takes the form $\langle x_t,D_t w_{t+1} \rangle$, where 
$$
D_t=\begin{pmatrix} p^+ & 1-p^+ & 0 & 0 \\ 0 & 0 & p^- & 1-p^- \end{pmatrix} =\begin{pmatrix} 0.6 & 0.4 & 0 & 0 \\ 0 & 0 & 0.7 & 0.3 \end{pmatrix} .
$$ 
To simplify the notation further, let 
$$
h_{1,t}=h_1, \ h_{2,t}=h_2, \ h_{3,t}=h_3,  \ h_{4,t}=h_4,  \ h_{5,t}=h_5,  \ h_{6,t}=h_6,  \ h_{7,t}=h_7,  \ h_{8,t}=h_8,
$$
and let 
$$
\pi_{1, t+1}=u_1, \  \pi_{2, t+1}=u_2, \ \pi_{3, t+1}=u_3.
$$
As shown in Section 4, saddle points in the game under consideration can be found by solving linear programming problems 
\begin{equation}
\begin{split}
100h_1-115h_2+75h_3-100h_4+35h_5-50h_6+50h_7-65h_8 \to \\
\to \max_{(h_1, h_2,h_3,h_4,h_5,h_6,h_7,h_8;x_1^+,  x_2^-)}⁡,
\end{split}
\end{equation}
$$
h_1-h_2 \le 0.6x_1^+,
$$
$$
h_3-h_4 \le 0.4x_1^+,
$$
$$
h_5-h_6 \le 0.7x_2^-,
$$
$$
h_7-h_8 \le 0.3x_2^-,
$$
$$
-100x_1^+-50x_2^- \ge -15000,
$$
$$
h_i \ge 0,i=\overline{1,8} , 
$$
$$
x_1^+ \ge 0,x_2^- \ge 0,
$$
and 
\begin{equation}
15000u_3\to \min_{(u_1,u_2, u_3;y_1^+,z_1^+, y_2^-,z_2^-)},
\end{equation}
$$
u_1-100u_3 \le -0.6y_1^+-0.4z_1^+,
$$
$$
u_2-50u_3\le -0.7y_2^--0.3z_2^-,
$$
$$
100 \le y_1^+ \le 115,
$$
$$
75 \le z_1^+ \le 100,
$$
$$
35\le y_2^- \le 50,
$$
$$
50 \le z_2^- \le 65,
$$
$$
u_i \ge 0,i=\overline{1,3},
$$
forming a dual pair. 

Solutions to problems (6) and (7) were found with the use of a computer program implemented on the Maple 7 computing platform, which includes software for solving linear programming problems. These solutions are 
$$
x_1^+ = 150, x_2^- =  0, 
$$
$$
h_1= 90, h_2= 0, h_3= 60, h_4= 0, h_5= 0, h_6= 0, h_7= 0, h_ 8= 0, 
$$
for problem (6), and
$$
u_1=0, u_2=0, u_3=0.9,
$$
$$
y_1^+ = 100 ,z_1^+ = 75 ,y_2^- =35  ,z_2^- =50,
$$
for problem (7). 

Thus, the trader's optimal strategy consists of a) borrowing from a broker 100 units of security 2 and selling them at the moment $t$,  and b) buying  150  units of security 1 at the moment $t$.  As a result of the deployment of this optimal strategy, the expectation of the value of the trader's portfolio at the moment $t+1$  equals  13500.

\section{Concluding remarks}
\label{sec:7}

1. Studying the financial behavior of small and medium price-taking traders in their interaction with a stock exchange presents both scientific and practical interest. As a result, both researchers of stock markets and successful stock market players offer their viewpoints on how the stock exchange functions and their explanations of why the market players act as they do. They also offer their recommendations on how the market players should act to succeed, and what decision-making models can be viewed as those adequately describing the interaction of individual market players with the stock exchange. 

The authors believe that currently, two competing viewpoints on what models should be considered adequate prevail in both scientific and mass media publications. 

Fundamental scientific approaches to mathematically modeling the interaction of a trader and a particular stock exchange, briefly surveyed, for instance,  in [Belenky \& Egorova 2015], underlie the first one.  This viewpoint is based on the belief that an adequate model is the one of the so-called representative agent, who is rational in adopting decisions on forming and managing her portfolio of securities and derivative financial instruments and tries to maximize her welfare. This belief is accompanied by the assumption that this ``rational'' agent a) knows the probability distribution of the values of future prices for every financial instrument that is of her interest and is traded in the stock exchange (with which this trader interacts), and b) makes her decisions based upon this knowledge. However, the real life does not seem to support either the above assumption or the above belief underlying this viewpoint. As mentioned earlier, deviations of the trader's financial behavior from a rational one [Barberis \& Thaler 2002; Kahneman 2011; Mullainathan \& Thaler 2000], as well as the inability of even financial analysts to make rational investment decisions and forecast directions in which the values of the share prices of particular securities (considered as random variables) will change (under any assumptions on the probability distributions of the values of these share prices), have widely been reported in scientific publications [Barber \& Odean 2000; Malkiel \& Saha 2005; Odean 1999; Penikas \& Proskurin 2013; Soderlind 2010]. 

The other viewpoint on the decision-making models adequately describing the interaction of a trader with a stock exchange is ``pushed'' by particular ``lucky traders'' who have managed to make money on adopting non-standard financial decisions. Some of them, particularly, N. Taleb [Taleb 2008], even deny the effectiveness of any economic and mathematical theories describing the functioning of the stock market for forming a trader's decision on managing her portfolio. Instead of adhering to such theories in managing the portfolio of a trader, N. Taleb suggests the trader to focus her attention exceptionally on the crises that happen in a stock exchange and in the world.  He believes that only at the time of these crises can a trader expect to attain significant financial results. However, as shown in [Aleskerov \& Egorova 2012], at least under quite natural assumptions, a price-taking trader who is capable of recognizing regular events with a probability even slightly exceeding 50\% is almost guaranteed to receive a positive average gain. It is clear that this  may or may not be the case if all the trader's activities consist of waiting for ``black swan'' events to occur. 

The authors believe that both viewpoints on the adequacy of the decision-making models are extreme, and neither reflects the practice of the interaction of a trader with a stock exchange. This state of affairs raises the following two  groups of questions:

1)  Can any alternative to the above extreme views on the adequacy of the decision-making mode be proposed?  Can  mathematical models capable of facilitating the decision-making process that small and medium price-taking traders undergo in estimating the expected financial results be proposed? Can such models be proposed  in the absence of knowledge on any probability distribution of future price values of financial instruments traded in a particular stock exchange?

2)  Can one propose mathematical models the use of which would allow a trader (with a confirmed ability to correctly estimate directions of changing the price values of financial instruments of her interest) make rational decisions on the structure of her portfolio at a particular moment $t$ in principle? Can such models be proposed  if the trader can indicate a segment within which the future values of the price of a particular financial instrument will change being uniformly distributed? Can one propose such models if the trader can  estimate only the expected areas in which the values of the prices for the groups of financial instruments forming together the whole set of the financial instruments of her interest (into which this set is divided by the trader) may change? Can one develop these models with the  use of only the simplest linear equations and inequalities of a balance type? 

The present paper offers positive answers to all the above questions. However,  the authors believe that the proposed mathematical models and approaches to finding trader's optimal investment strategies need to be tested and researched by both economists and other analysts studying financial aspects of the interaction between a trader and a stock exchange. The authors consider the tools proposed in this paper mostly as a powerful instrument allowing interested researchers to study particular aspects of the stock exchange behavior in the framework of a large-scale decision-support system. This system allows one to use the models with millions of variables and constraints, which distinguishes the authors' approach to modeling stock exchanges from those already proposed. 

2. As is well known, global optimization problems are difficult to solve, and there are no uniform approaches allowing one to find global extrema in problems mathematically formalizing many of theoretical and practical optimization problems. Thus, detecting classes of problems in which not only global extrema can be found in principle, but those in which these extrema can be found with the use of the most powerful computational techniques, linear programming being one of them, undoubtedly presents both scientific and applied interest. As shown in the paper, finding a point of the global maximum of a particular nonlinear function (the minimum function on a convex polyhedron described by a compatible system of linear equations and inequalities) on a subset of another convex polyhedron formed by vectors with all the coordinates being non-negative integers is reducible to solving a mixed programming problem. It was also shown that finding the global maximum of the above function on a convex polyhedron (described by another compatible system of linear equations and inequalities) is reducible to solving linear programming problems forming a dual pair. 

3. While there are numerous schemes for and approaches to forecasting time series, the need in tools helping a potential or an acting small or medium price-taking trader reliably estimate the ability to divine future values of the share prices of securities  remains high. Such tools can save a lot of money to private investors and even prevent personal financial tragedies. It is clear that a) a detected ability to divine future values of the share prices of particular securities by processing results of the trials according to the Bernoulli scheme,  and b) the ability  to divine the actual values of the share prices of particular securities  in dealing with these prices in real life may not be the same.  So the availability of the tool that allows one to compare both abilities seems critical at least from a practical viewpoint.

4. In two mathematical models proposed in this paper, the authors assumed that for all the securities being of interest to a trader, the trader either a) can indicate a segment within which the values of the prices of a particular financial instrument will change being uniformly distributed, or b) can only estimate the areas in which the expected values of the prices for the whole set of financial instruments that interest her may change. However, it is possible that there are two groups of securities that interest the trader, and for one group, her ability to divine future values of the share prices of particular securities corresponds to case a) from point 3 of this section, whereas for the other group, the ability to divine directions in which the price values of securities from this group will change corresponds to case b) from the same point of this section. If the trader is firm in dividing financial resources  available to her between these two groups (in dealing with securities from these groups), then both models can be used separately.  If this is the case, the trader's optimal investment strategies can be determined by solving corresponding mathematical programming problems considered in Sections 3 and 4 of this paper. Otherwise, the trader faces a complicated problem of dividing financial resources available to her at the moment $t$ between the two groups, which leads to considering models whose structure and features are completely different from those considered in the present paper. 

The authors would like to emphasize that in the models formalizing the interaction of a trader with the stock exchange in the form of  mathematical programming problems with Boolean variables, presented in Sections 3 and 5 of the paper, they did not consider some particular risks that the trader may be interested in taking into consideration in making her decision on developing or changing her portfolio of securities. Though such risks are traditionally considered in publications on modeling the behavior of traders trading securities in a stock exchange, the inclusion of the risks considered, for instance, in [Markowitz 1952], in the models proposed in this paper would lead to solving large-scale nonlinear programming problems with integer or mixed variables (formulated on the basis of these models). Such problems are difficult to solve even for relatively small problem sizes, and from the authors' viewpoint, this inclusion would hardly make corresponding models and problems an effective tool of studying stock exchanges and traders' behavior in interacting with them. At the same time, the authors would like to make it clear that their search for the models that could be considered an effective tools for studying the stock exchange behavior continues, and models of the mentioned kind presented in this paper should be viewed as no more than only the first step towards this direction.

5. Finally, only the modeling of the decision-making process that individual price-taking traders undergo in the course of their interaction with a stock exchange was the subject of this paper.  However, one should bear in mind that both small and medium price-taking traders may form coalitions and act either as one legal entity or as groups in the framework of which the interests of all the group members within each group are to be observed. Moreover, such groups are implicitly formed when some (and, possibly, quite a substantial number of) small price-taking traders exercise the strategy of following someone's decisions (for instance, those of large traders or ``lucky'' ones) independently of their (groups') sizes. Studying aspects of the financial behavior of these groups presents obvious interest in an attempt to understand the mechanisms of the interacting between individual traders and a stock exchange. However,  such studies require both a particular  use of known and the development of new mathematical tools, and the discussion of these issues, which invokes that of a set of fundamental modeling problems,   lies beyond the scope of the present paper.

\section*{Acknowledgements} 
The financial support from the Russian Federation Government within the framework of the implementation of the 5-100 Programme Roadmap of the National Research University Higher School of Economics is acknowledged. The authors are grateful for financial support of their work to DeCAn Laboratory at the National Research University Higher School of Economics, headed by Prof. Fuad Aleskerov, with whom the authors have had fruitful discussions on problems of developing tools for quantitatively analyzing large-scale systems in general and those for stock exchanges in particular. L. Egorova expresses her gratitude to LATNA Laboratory, NRU HSE, headed by Prof. Panos Pardalos, RF government grant, ag.11.G34.31.0057, and A. Belenky expresses his gratitude to the MIT Center for Engineering Systems Fundamentals, headed by Prof. Richard Larson.

\section*{References}

[Aleskerov \& Egorova 2012]	Aleskerov F. T., Egorova L. G. Is it so bad that we cannot recognize black swans? // Economics Letters. 2012. Vol. 117. No. 3. P. 563-565.

[Asratyan \& Kuzyurin 2004]	Asratyan A. S., Kuzyurin N. N.  Analysis of randomized rounding for integer programs // Discrete Mathematics and Applications. 2004. Vol. 14. No. 6. P. 543-554. (in Russian).

[Barber \& Odean 2000]	Barber B., Odean T.  Trading is hazardous to your wealth: The common stock investment performance of individual investors // Journal of Finance. 2000. Vol. 55. No. 2. P.773-806.

[Barberis \& Thaler 2002]	Barberis N., Thaler R. A survey of behavioral finance // NBER Working Paper Series. 2002. URL: http://www.nber.org/papers/w9222.

[Belenky 1981]  Minimax planning problems with linear constraints and methods of their solution // Automation and Remote Control - Vol. 42, No. 10, P. 1409-1419.

[Belenky \& Egorova 2015]   Belenky A.S., Egorova L.G. An Approach to Forming and Managing a Portfolio of Financial Securities by Small and Medium Price-Taking Traders in a Stock Exchange // Advances in Intelligent Systems and Computing. Proceedings of the 3rd International Conference on Modelling, Computation and Optimization in Information Systems and Management Sciences MCO 2015. 2015. P. 257-268.

[Belenky \& Egorova 2016] Belenky A.S., Egorova L.G. Optimization of portfolio compositions for small and medium price-taking traders // In Optimization and Its Applications in Control Sciences and Data Analysis, Springer, 2016 (to be published in June 2016).

[Crack 2014]   Crack T.F.  Basic Black-Scholes: Option Pricing and Trading,  Timothy Falcon Crack, 3rd edition, 2014.

[Feller 1991]	Feller, W.  An Introduction to Probability Theory and Its Applications, Vol. 1 and 2 // Wiley, 1991.

[Kahneman 2011]	Kahneman D.  Thinking, fast and slow. // N.Y.: Penguin, 2011.

[Lin et al. 1994] Lin W., Engle R., Ito T. Do bulls and bears move across borders? International transmission of stock returns and volatility	// Review of Financial Studies. 1994. Vol.7. No. 3. P. 507-538.

[Malkiel \& Saha 2005]	Malkiel B.G., Saha A.  Hedge Funds: Risk and Return // Financial Analysts Journal. 2005. Vol. 61. No. 6. P.80-88.

[Markowitz 1952]	Markowitz H.  Portfolio Selection // Journal of Finance. 1952. Vol. VII. No.1. P. 77-91.

[Mullainathan \& Thaler 2000]	Mullainathan S., Thaler R.H.  Behavioral Economics // NBER Working paper. 2000. URL: http://www.nber.org/papers/w7948.

[Odean 1999]	Odean T.  Do investors trade too much? // American Economic Review. 1999. Vol. 89. No. 5. P. 1279-1298.

[Penikas \& Proskurin 2013]	Penikas H., Proskurin S. How Well Do Analysts Predict Stock Prices? Evidence from Russia // Working papers by NRU Higher School of Economics. Series FE «Financial Economics», WP BRP 18/FE/2013.

[Soderlind 2010]	Soderlind P.  Predicting stock price movements: regressions versus economists // Applied Economic Letters. 2010. Vol. 17. P. 869-874.

[Taleb 2008]	Taleb N. N. The Black Swan: The Impact of The Highly Improbable // London, Penguin Books, 2008.

[Yudin \& Golshtein 1965] Yudin D., Golshtein E. Linear Programming // Israel Program of Scientific Translations, Jerusalem, 1965.

[Yudin \& Yudin 2009] Yudin D.B., Yudin, A.D. Extreme models in economics // LIBROKOM (in Russian), 2009.

\end{document}